\documentclass[12pt,reqno]{article}
\usepackage[english]{babel}
\usepackage{amsmath,dsfont, amsthm}
\usepackage{amsfonts}
\usepackage{graphicx}
\usepackage{enumerate}
\usepackage[usenames,dvipsnames]{color}
\usepackage{subfigure}
\usepackage[right,pagewise,displaymath, mathlines]{lineno}
\usepackage{epstopdf}
\usepackage{color}
\usepackage{multirow}
\usepackage[table,xcdraw]{xcolor}
\usepackage[title]{appendix}

\usepackage{authblk}
\usepackage[round]{natbib}
\usepackage{float}

\usepackage{enumitem}

\usepackage{tikz}
\usepackage{schemabloc}

\usepackage[bottom]{footmisc}

\usepackage[colorlinks = true,urlcolor = blue,citecolor = blue,breaklinks]{hyperref}

\usepackage{geometry}
\usepackage{rotating}

\numberwithin{equation}{section}
\numberwithin{figure}{section}
\numberwithin{table}{section}

\newtheorem{property}{Property}[section]

\theoremstyle{definition}

\newtheorem{example}{Example}[section]
\newtheorem{note}{Note}[section]

\numberwithin{equation}{section}

\definecolor{darkred}{rgb}{0.7, 0, 0}
\definecolor{darkbrown}{rgb}{0.55, 0.2, 0.15}
\definecolor{darkblue}{rgb}{0.1,0.1,0.6}
\definecolor{darkgreen}{rgb}{0.1,0.5,0.2}

\newcommand{\rd}{\color{darkred}}

\newcommand{\gr}{\color{darkgreen}}

\newcommand{\dd}{\mathrm{d}}

\usepackage{setspace}
\usepackage[bottom]{footmisc}

\usepackage{tkz-graph}
\usepackage{pgfplots}

\title{\vspace{-15mm}\Large\bf Measuring income inequality via percentile relativities\thefootnote\relax\footnotetext{
This research has been supported by the NSERC Alliance--MITACS Accelerate grant (ALLRP 580632-22) entitled ``New Order of Risk Management: Theory and Applications in the Era of Systemic Risk'' from the Natural Sciences and Engineering Research Council (NSERC) of Canada, and the national research organization Mathematics of Information Technology and Complex Systems (MITACS) of Canada, as well as by the individual NSERC Discovery Grant of R.~Zitikis (RGPIN-2022-04426).}}

\author[,1]{Vytaras Brazauskas\thanks{E-mail: \href{mailto:vytaras@uwm.edu}{vytaras@uwm.edu}}}
\author[,2]{Francesca Greselin\thanks{Corresponding author; e-mail: \href{mailto:francesca.greselin@unimib.it}{francesca.greselin@unimib.it}}}
\author[,3]{Ri\v{c}ardas Zitikis\thanks{E-mail: \href{mailto:rzitikis@uwo.ca}{rzitikis@uwo.ca}}}
\affil[1]{\normalsize University of Wisconsin-Milwaukee, Milwaukee, Wisconsin, U.S.A.}
\affil[2]{\normalsize Universit\`{a} degli Studi di Milano-Bicocca, Milano, Italy}
\affil[3]{\normalsize Western University, London, Ontario, Canada}

\date{}

\begin{document}

\maketitle


\noindent
\textbf{Abstract.}
``The rich are getting richer'' implies that the population income distributions are  getting more right skewed and heavily tailed. For such distributions, the mean is not the best measure of the center, but the classical indices of income inequality, including the celebrated Gini index, are all mean-based. In view of this, Professor Gastwirth sounded an alarm back in 2014 by suggesting to incorporate the median into the definition of the Gini index, although noted a few shortcomings of his proposed index. In the present paper we make a further step in the modification of classical indices and, to acknowledge the possibility of differing viewpoints, arrive at three median-based indices of inequality. They avoid the shortcomings of the previous indices and can be used even when populations are ultra heavily tailed, that is, when their first moments are infinite. The new indices are illustrated both analytically and numerically  using parametric families of income distributions, and further illustrated using capital incomes coming from 2001 and 2018 surveys of fifteen European countries. We also discuss the performance of the indices from the perspective of income transfers.

\medskip


\vfill


\section{Introduction}
\label{introduction}

Measuring income inequality has been a challenging task, as each of the indices used for the purpose  attempt to condense the complexities of populations into just one number. Among the many indices, we have the Atkinson, Bonferroni, Gini, Palma, Pietra, Theil, and Zenga indices, to name just a few associated with the names of their inventors.  Many treatises have been written on the topic, such as the handbook by \citet{AB2000,AB2015}, which also contains many references to earlier studies, and they are voluminous.

The indices are often the areas under certain income-equality curves, which are considerably more difficult to present and explain to the general audience, let alone to easily compare. For example, the Gini index of inequality is $1$ minus twice the area under the Lorenz curve. (We shall give mathematical definitions later in this paper.) The curves and thus the indices are based on comparing the mean income of the poor with other means, such as the mean income of the entire population, the mean income of the nonpoor, and the mean income of the rich, whatever the definitions of ``poor'' and  ``rich'' might be. Hence, to be well defined, the curves and the indices inevitably assume that the mean of the underlying population is finite. With the rising income inequality, and thus with the distribution of incomes becoming more skewed and heavily tailed, researchers have therefore sought other ways for measuring inequality.

\cite{g2014} proposed to use the median instead of the mean when ``normalizing'' the absolute Gini mean difference, widely known as the GMD. The author noted, however, that the proposed index might fall outside the class of normalized indices because it compares the \textit{mean} income of the poor with the \textit{median} income of the entire population. There is a natural remedy to this normalization issue: compare the \textit{median} income of the poor with the \textit{median} of the population. Even more, we can compare the median income of the poor with the median of the ``not poor'' or, for example, with the median of the rich, whatever the latter might mean. This is the path that we take in this paper to arrive at the indices to be formally introduced in the next section.

In this regard we wish to mention the study of \citet{BZ2015} where it is shown that a number of classical indices of income inequality arise naturally from a Harsanyi-inspired model of choice under risk, with persons acting as \textit{reference-dependent} expected-utility maximizers in the face of an income quantile lottery, thus giving rise to a reinterpretation of the classical indices as measures of the desirability of redistribution in society. This relativistic approach to constructing indices of income inequality was further explored by \citet{gz2018}, although more from the modeller's perspective than from the philosophical one. The present paper further advances this line of research by showing how naturally percentile-based indices arise in this relativistic context, and how they facilitate inequality measurement even in those populations whose distributions are  ultra-heavily tailed, that is, do not possess even a finite first moment.

The rest of the paper is organized as follows.
In Section~\ref{indices} we define the new inequality indices, alongside the corresponding equality curves, preceded by several known indices for comparison purposes.
In Section~\ref{parametrics} we illustrate the new indices and their curves numerically, using several popular families of distributions. In Section~\ref{empirics}, we use the indices to first analyze capital incomes of European countries using data from a 2001 survey, and then we compare the results with those obtained from a 2018 survey. In Section~\ref{est-pd} we look at the new indices from the perspective of income transfers.
Section~\ref{conclusion} concludes the paper. Proofs and other technicalities are in Appendix~\ref{technicalities}.

\section{Inequality indices and their curves}
\label{indices}

We start with technical prerequisites. Let $F$ be the cumulative distribution function of the population incomes $X$, a random variable. We assume that $F$ is non-negatively supported, that is, $F(x)=0$ for all real $x<0$. Furthermore, let $Q$ denote the (generalized) inverse of $F$, called the quantile function. That is, for each $p\in (0,1)$, $Q(p)$ is the smallest number $x$ such that $F(x)\ge p$. Hence, the population median income is
\[
m=Q(1/2)
\]
and, generally, $Q(p)$ is the $p\times 100^{\rm th}$ percentile. Furthermore, the median income of the poorest $p\times 100\%$ persons is $Q(p/2)$. Based on these quantities, we shall later describe three new ways for measuring inequality, but first, we recall the definitions of a few earlier indices that serve as benchmarks for the new ones.

\subsection{In the classical mean-based world}

The index of \citet{g1914} is the most widely-used measure of inequality. It can be expressed in a myriad of ways \citep[e.g.,][]{y1998,YS2013}. For example, the Gini index can be written in terms of the Bonferroni curve
\[
B(p)={1\over \mu p }\int_0^p Q(s)\dd s, \quad 0 \le p \le 1,
\]
as follows:
\begin{align}
G &= 2\int_0^1 \bigg (1-{{1\over p}\int_0^p Q(s)\dd s \over \mu}  \bigg )p\dd p
\notag
\\
&= 1- \int_0^1 {{1\over p}\int_0^p Q(s)\dd s \over \mu}  2p\dd p
\notag
\\
&= 1-\int_0^1 B(p) 2p\dd p,
\label{gini-0}
\end{align}
where
\[
\mu=\int_0^1 Q(s)\dd s
\]
is the mean of $X$.

\citet{z2007} argued that the mean income of those below the percentile $Q(p)$ need to be compared not with the mean of all the incomes but with the mean income of those above the percentile $Q(p)$. This point of view led the author to the index
\[
Z =1-\int_0^1 {{1\over p}\int_0^p Q(s)\dd s \over {1\over 1-p}\int_p^1 Q(s)\dd s }  \dd p.
\]

\citet{dg2019,dg2020} suggested to modify Zenga's idea by comparing the mean income of those below the percentile $Q(p)$ with the mean income of those above the percentile $Q(1-p)$. This point of view led the authors to the index
\[
D =1-\int_0^1 {{1\over p}\int_0^p Q(s)\dd s \over {1\over p}\int_{1-p}^1 Q(s)\dd s } \dd p.
\]
Of course, $1/p$ in the numerator and denominator cancel out, but in this way written $D$ facilitates an easier comparison with $Z$.

\subsection{A transition into the heavy-tailed modern world}
\label{gast}

Unlike the above three mean-based indices $G$, $Z$ and $D$, the index of \citet{g2014}  is a mean-median based index. Namely, given the well-known expression
\begin{equation}\label{gmd}
G=  {\textrm{GMD} \over 2 \mu }
\end{equation}
of the Gini index $G$ in terms of the Gini mean difference (GMD), which is often written as the expectation $\mathbb{E}(|X_1-X_2|)$, where $X_1$ and $X_2$ are two independent copies of $X$, \citet{g2014} argued that comparing the GMD with twice the median would be better than comparing with twice the mean as in equation~\eqref{gmd}. This viewpoint has given rise to the index
\begin{align*}
G_2&=  {\textrm{GMD} \over 2 m }
\\
&= \int_0^1 \bigg ( {\mu \over m} -{{1\over p}\int_0^p Q(s)ds \over m}\bigg )2p\dd p
\\
&= {\mu \over m}-\int_0^1 {{1\over p}\int_0^p Q(s)ds \over m} 2p\dd p .
\end{align*}
Note that $\mu/m$, which can be viewed as the benchmark replacing $1$ in the previous indices, is the mean-median ratio that has been used as an easy to understand -- and thus to convey to the general audience -- indicator of wealth and income distribution  \citep[e.g.,][]{g2020}. In the case of symmetric distributions, $\mu/m$ is of course equal to $1$.

\subsection{In the skewed and heavy-tailed modern world: new indices}

The above discussion naturally leads to three strategies of defining purely median-based indices of income inequality and their corresponding curves of equality, all based on percentiles and thus well defined irrespective of whether the income variable $X$ has a finite first or any other  moment.

\paragraph{Strategy 1:}
Compare the median income of the poorest $p\times 100\%$ persons
\begin{figure}[h!]
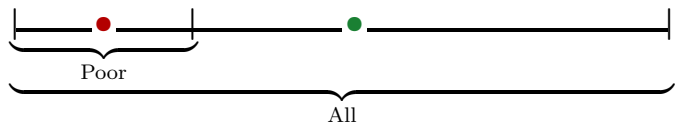

\begin{center}
\[
\underbrace{\underbrace{\boldsymbol{|}\hspace{-0.5mm}\rule{10mm}{0.5mm}\hspace{-0.5mm}
\boldsymbol{\rd\bullet}\hspace{-0.5mm}\rule{10mm}{0.5mm}\hspace{-0.5mm}
\boldsymbol{|}}_{\textrm{Poor}}\hspace{-1.5mm} \rule{20mm}{0.5mm}\hspace{-0.5mm}
\boldsymbol{\gr\bullet}\hspace{-0.5mm} \rule{40mm}{0.5mm}\hspace{-0.5mm}\boldsymbol{|}}_{\textrm{All}}
\]
\end{center}
  \caption{The median of the poor (red) and the median of all (green).}\label{fig21}
\end{figure}
with the median of the entire population (Figure~\ref{fig21}).
This leads to the equality curve
\begin{equation}\label{curve-1}
\psi_1(p) = {Q(p/2) \over Q(1/2)} , \quad 0<p<1.
\end{equation}
Averaging this curve over all $p$'s gives rise to the inequality index
\begin{equation}\label{index-1}
\Psi_1= 1-\int_{0}^{1} {Q(p/2) \over Q(1/2)}  \dd p.
\end{equation}
Note the mathematical similarity between the Bonferroni curve $b$ and the curve $\psi_1$:
\[
b(p)={{1\over p}\int_0^p Q(s)\dd s \over \int_0^1 Q(s)\dd s} ,
\quad
\psi_1(p) ={{1\over p}\int_0^p Q(p/2)\dd s \over \int_0^1 Q(1/2)\dd s}  .
\]

\paragraph{Strategy 2:}
Compare the median income of the poorest $p\times 100\%$ persons
\begin{figure}[h!]
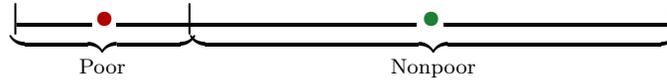

\begin{center}
\[
\underbrace{\boldsymbol{|}\hspace{-0.5mm}\rule{10mm}{0.5mm}\hspace{-0.5mm}
\boldsymbol{\rd\bullet}\hspace{-0.5mm}\rule{10mm}{0.5mm}\hspace{-1.0mm}
\boldsymbol{|}}_{\textrm{Poor}}\hspace{-1.2mm}\underbrace{\rule{30mm}{0.5mm}\hspace{-0.2mm}
\boldsymbol{\gr\bullet}\hspace{-0.5mm} \rule{30mm}{0.5mm}\hspace{-0.5mm}\boldsymbol{|}}_{\textrm{Nonpoor}}
\]
\end{center}
  \caption{The median of the poor (red) and the median of the nonpoor (green).}\label{fig22}
\end{figure}
with the median of the nonpoor (Figure~\ref{fig22}).  This leads to the equality curve
\begin{equation}\label{curve-2}
\psi_2(p) = {Q(p/2) \over Q(1/2+p/2)} , \quad 0<p<1,
\end{equation}
and, after averaging over all $p$'s, to the inequality index
\begin{equation}\label{index-2}
\Psi_2= 1-\int_{0}^{1}  {Q(p/2) \over Q(1/2+p/2)}  \dd p .
\end{equation}
Note the mathematical similarity between the Zenga curve $z$ and the curve $\psi_2$:
\[
z(p)= {{1\over p}\int_0^p Q(s)\dd s \over {1\over 1-p}\int_p^1 Q(s)\dd s } ,
\quad
\psi_2(p) = {{1\over p}\int_0^p Q(p/2)\dd s \over {1\over 1-p}\int_p^1 Q(p+(1-p)/2)\dd s }  .
\]

\paragraph{Strategy 3:}
Compare the median income of the poorest $p\times 100\%$ persons
\begin{figure}[h!]
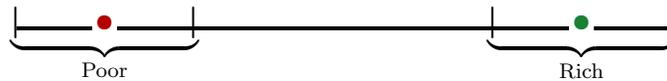

\begin{center}
\[
\underbrace{\boldsymbol{|}\hspace{-0.5mm}\rule{10mm}{0.5mm}\hspace{-0.5mm}
\boldsymbol{\rd\bullet}\hspace{-0.5mm}\rule{10mm}{0.5mm}\hspace{-0.5mm}
\boldsymbol{|}}_{\textrm{Poor}}\hspace{-1.5mm} \rule{40mm}{0.5mm}\hspace{-1.5mm}
\underbrace{\boldsymbol{|}\hspace{-0.5mm} \rule{10mm}{0.5mm}\hspace{-0.5mm}
\boldsymbol{\gr\bullet}\hspace{-0.5mm} \rule{10mm}{0.5mm}\hspace{-0.5mm}\boldsymbol{|}}_{\textrm{Rich}}
\]
\end{center}
  \caption{The median of the poor (red) and the median of the rich (green).}\label{fig23}
\end{figure}
with the median of the richest $p\times 100\%$ persons (Figure~\ref{fig23}).  This leads to the equality curve
\begin{equation}\label{curve-3}
\psi_3(p) = {Q(p/2) \over Q(1-p/2)}, \quad 0<p<1,
\end{equation}
and, after averaging over all $p$'s, to the inequality index
\begin{equation}\label{index-3}
\Psi_3= 1-\int_{0}^{1} {Q(p/2) \over Q(1-p/2)}  \dd p.
\end{equation}
Note the mathematical similarity between the Davydov-Greselin curve $d$ and the curve $\psi_3$:
\[
d(p)=  {{1\over p}\int_0^p Q(s)\dd s \over {1\over p}\int_{1-p}^1 Q(s)\dd s }  ,
\quad
\psi_3(p) =  {{1\over p}\int_0^p Q(p/2)\dd s \over {1\over p}\int_{1-p}^1 Q(1-p+p/2)\dd s }   .
\]

Summarizing the above discussion, in view of equations~\eqref{index-1}, \eqref{index-2}, and \eqref{index-3}, the three income-equality curves are connected to the corresponding income-inequality indices via the equation
\begin{equation}\label{index-curve}
\Psi_k=  1-\int_{0}^{1} \psi_k(p)\dd p.
\end{equation}
Note that the three curves $\psi_k$ take values only in the interval $[0,1]$, and so the three indices $\Psi_k$ are always normalized, that is, $\Psi_k \in [0,1]$. In this context it is useful to look at the following unrealistic but illuminating cases:
\begin{itemize}
\item
If the income-equality curve $\psi_k$ is equal to $1$ everywhere on $(0,1)$, which means perfect equality, then the income-inequality index $\Psi_k$ is equal to $0$, which means lowest inequality.
\item
If the income-equality curve $\psi_k$ is equal to $0$ everywhere on $(0,1)$, which means extreme inequality, then the income-inequality index $\Psi_k$ is equal to $1$, which means maximal inequality.
\end{itemize}
Hence, these two extreme cases serve as benchmark curves:
the one that is identically equal to $1$ is the curve of perfect equality, and
the one that is identically equal to $0$ is the curve of extreme inequality.
We can therefore say that the three indices $\Psi_k$ measure the deviation of the actual curves $\psi_k$ from the benchmark egalitarian curve $\psi_e(p)=1$, $0\le p \le 1$, by calculating the areas between them.

\section{The new indices and curves: a parametric viewpoint}
\label{parametrics}

Modelling population incomes using parametric distributions and also fitting such distributions to income data are common approaches in the area~\cite[e.g.,][]{kk2003}. From this perspective, the  inequality indices $G$, $Z$, $D$ and $G_2$ and their corresponding equality curves have been amply discussed and illustrated by their inventors and subsequent researchers. Hence, we devote this section to illustrating only the three indices $\Psi_k$ and their corresponding curves $\psi_k$.

We use nine parametric families of distributions, most of which are common in modeling incomes~\cite[e.g.,][]{kk2003}. They are right skewed and present a full spectrum of tail heaviness: some are lightly tailed (e.g., exponential),
some are heavily tailed (e.g., Pareto distributions), and others have
the right tails of intermediate heaviness (e.g., lognormal). For their specific parametrizations, next is the list of their quantile functions:
\begin{itemize}
  \item {\it Uniform\/}$(0, \theta)$:
~$Q(p) = \theta p$.
  \item {\it Exponential\/}$(0, \theta)$:
~$Q(p) = - \theta \log (1-p)$.
  \item {\it Gamma\/}$(\theta, \alpha)$:
~$Q(p) = \theta Q_0(p)$, where $Q_0(p)$ is
the quantile function of the standard gamma distribution  (i.e., $\theta=1$)
whose cumulative distribution function is $F_0(t) = \frac{1}{\Gamma (\alpha)}
\int_0^t x^{\alpha-1} e^{-x} \; \mbox{d}x$.
  \item {\it Weibull\/}$(\theta, \tau)$:
~$Q(p) = - \theta \big( \log (1-p) \big)^{1/\tau}$.
  \item {\it Lognormal\/}$(\mu, \sigma)$:
~$Q(p) = \exp{ \left\{ \mu + \sigma \Phi^{-1}(p) \right\} }$,
where $\Phi^{-1}(p)$ is the quantile function of the standard normal distribution (i.e., $\mu = 0$
and $\sigma = 1$).
  \item {\it Log-Cauchy\/}$(\mu, \sigma)$:
~$Q(p) = \exp{ \left\{ \mu + \sigma \tan ( \pi (p-1/2) ) \right\} }$.
  \item {\it Pareto-II\/}$(\sigma, \alpha)$:
~$Q(p) = \sigma \big( (1-p)^{-1/\alpha} - 1 \big)$.
  \item {\it Pareto-III\/}$(\sigma, \gamma)$:
~$Q(p) = \sigma \big( (1-p)^{-1} - 1 \big)^{\gamma}$.
  \item {\it Pareto-IV\/}$(\sigma, \alpha, \gamma)$:
~$Q(p) = \sigma \big( (1-p)^{-1/\alpha} - 1 \big)^{\gamma}$.
\end{itemize}

We have computed the inequality indices $\Psi_k$ for these distributions under various parameter choices. The results are in Table~\ref{tab-1},
\begin{table}[h!]
\centering
\begin{tabular}{l|ccc|rrr}
\hline
\hline
\multicolumn{1}{l|}{Distributions} &
\multicolumn{3}{|c|}{Inequality indices} &
\multicolumn{3}{|c}{Ranks based on} \\
 & $\Psi_1$ & $\Psi_2$ & $\Psi_3$ & $\Psi_1$ & $~~\Psi_2$ & $\Psi_3$ \\
\hline
{\it Uniform\/}$(0, \theta)$          & 0.5010 & 0.6936 & 0.6147 &  6 &  2 & 3-4 \\
{\it Exponential\/}$(0, \theta)$      & 0.5583 & 0.8327 & 0.7026 &  7 &  7 &  7 \\
{\it Gamma\/}$(\theta, \alpha = 0.5)$ & 0.6874 & 0.9378 & 0.8020 & 12 & 10 & 11 \\
{\it Gamma\/}$(\theta, \alpha = 2)$   & 0.4360 & 0.6974 & 0.5956 &  3 &  3 &  2 \\
\hline
{\it Weibull\/}$(\theta, \tau = 0.5)$ & 0.7237 & 0.9681 & 0.8358 & 13 & 13 & 13 \\
{\it Weibull\/}$(\theta, \tau = 2)$   & 0.3810 & 0.6022 & 0.5239 &  1 &  1 &  1 \\
{\it Lognormal\/}$(\mu, \sigma = 1)$  & 0.4779 & 0.7886 & 0.6648 &  4 &  5 &  5 \\
{\it Lognormal\/}$(\mu, \sigma = 2)$  & 0.6648 & 0.9527 & 0.8122 & 11 & 12 & 12 \\
\hline
{\it Log-Cauchy\/}$(\mu, \sigma = 1)$   & 0.6054 & 0.9382 & 0.7470 &  9 & 11 &  9 \\
{\it Log-Cauchy\/}$(\mu, \sigma = 2)$   & 0.7470 & 0.9935 & 0.8551 & 14 & 16 & 14 \\
{\it Pareto-II\/}$(\sigma, \alpha = 1)$ & 0.6147 & 0.9242 & 0.7736 & 10 &  9 & 10 \\
{\it Pareto-II\/}$(\sigma, \alpha = 2)$ & 0.5868 & 0.8863 & 0.7407 &  8 &  8 &  8 \\
\hline
{\it Pareto-III\/}$(\sigma, \gamma = 0.5)$ & 0.4302 & 0.7344 & 0.6147 &  2 &  4 & 3-4 \\
{\it Pareto-III\/}$(\sigma, \gamma = 2)$   & 0.7736 & 0.9932 & 0.8795 & 16 & 15 & 16 \\
{\it Pareto-IV\/}$(\sigma, \alpha = 0.5, \gamma = 0.5)$ &
0.4803 & 0.8288 & 0.6887 &  5 &  6 &  6 \\
{\it Pareto-IV\/}$(\sigma, \alpha = 2, \gamma = 2)$     &
0.7495 & 0.9852 & 0.8598 & 15 & 14 & 15 \\
\hline
\end{tabular}
\caption{The inequality indices $\Psi_k$ for various parametric  distributions and the rankings of these distributions based on the indices.}
\label{tab-1}
\end{table}
where we also report the rankings of the distributions based on the new indices: rank 1 corresponds to the lowest inequality and rank 16 to the highest inequality. It is encouraging to see that while the magnitudes of the indices differ, the rankings induced by them are fairly similar.

\begin{figure}[h!]
\centering
    \resizebox{175mm}{175mm}{\includegraphics{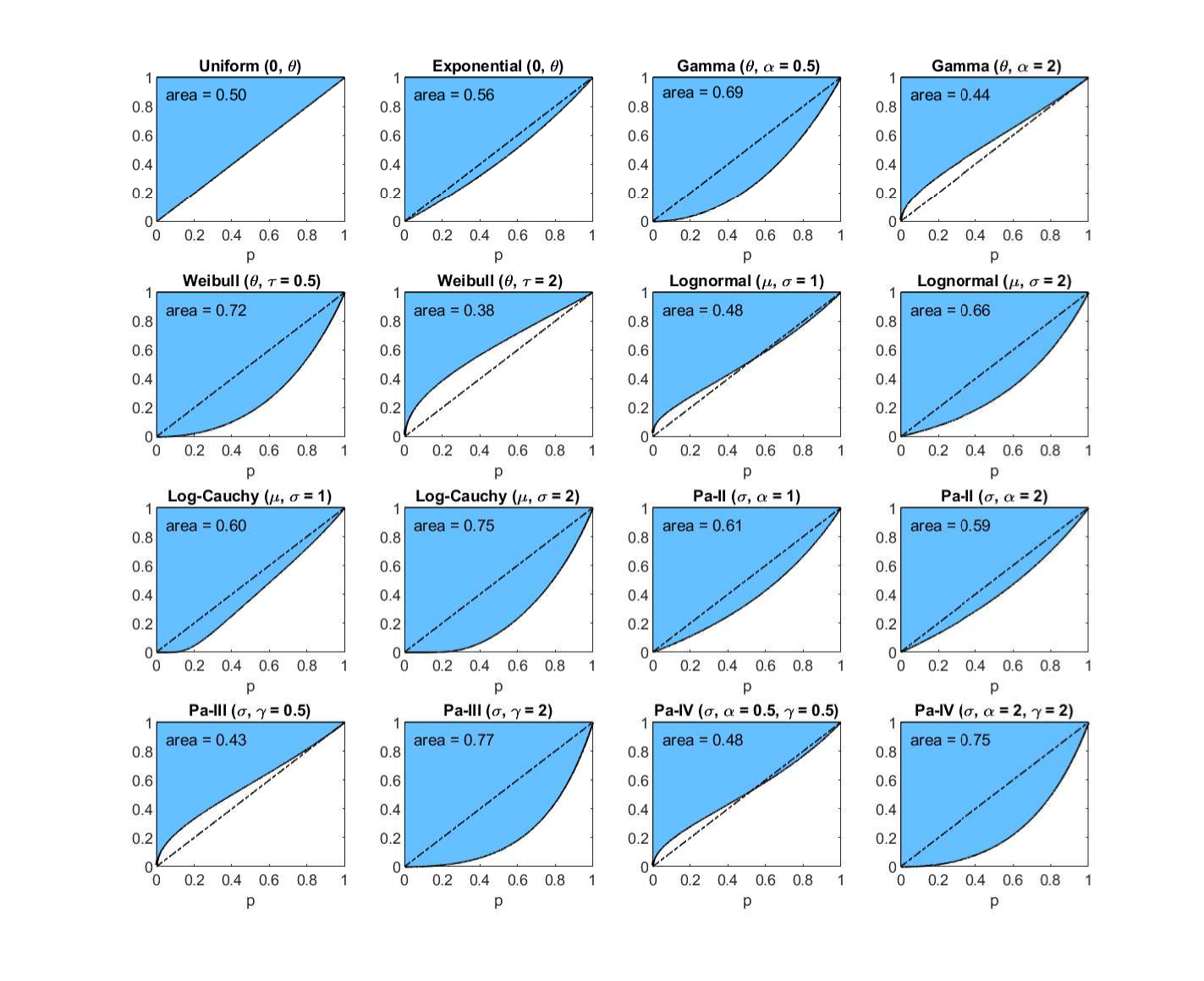}}
    \caption{The income-equality curve $\psi_1$ and the shaded-in area (i.e., $\Psi_1$) above it for the distributions of Table~\ref{tab-1}, with the dash-dotted line depicting $\psi_1$ of the uniform distribution.}
    \label{fig-1}
\end{figure}

\begin{figure}[h!]
\centering
    \resizebox{175mm}{175mm}{\includegraphics{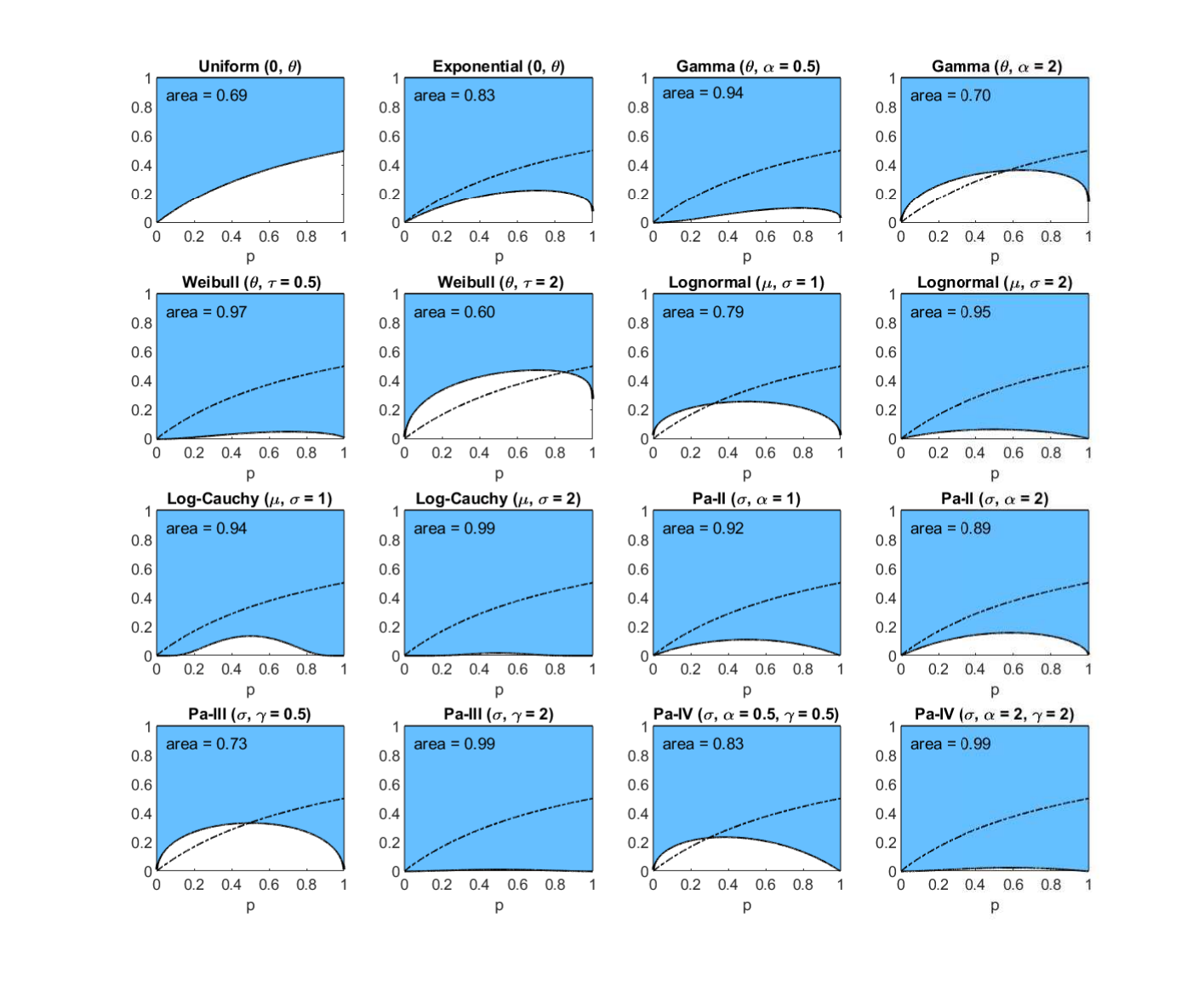}}
    \caption{The income-equality curve $\psi_2$ and the shaded-in area (i.e., $\Psi_2$) above it for the distributions of Table~\ref{tab-1}, with the dash-dotted line depicting $\psi_2$ of the uniform distribution.}
    \label{fig-2}
\end{figure}

\begin{figure}[h!]
\centering
    \resizebox{175mm}{175mm}{\includegraphics{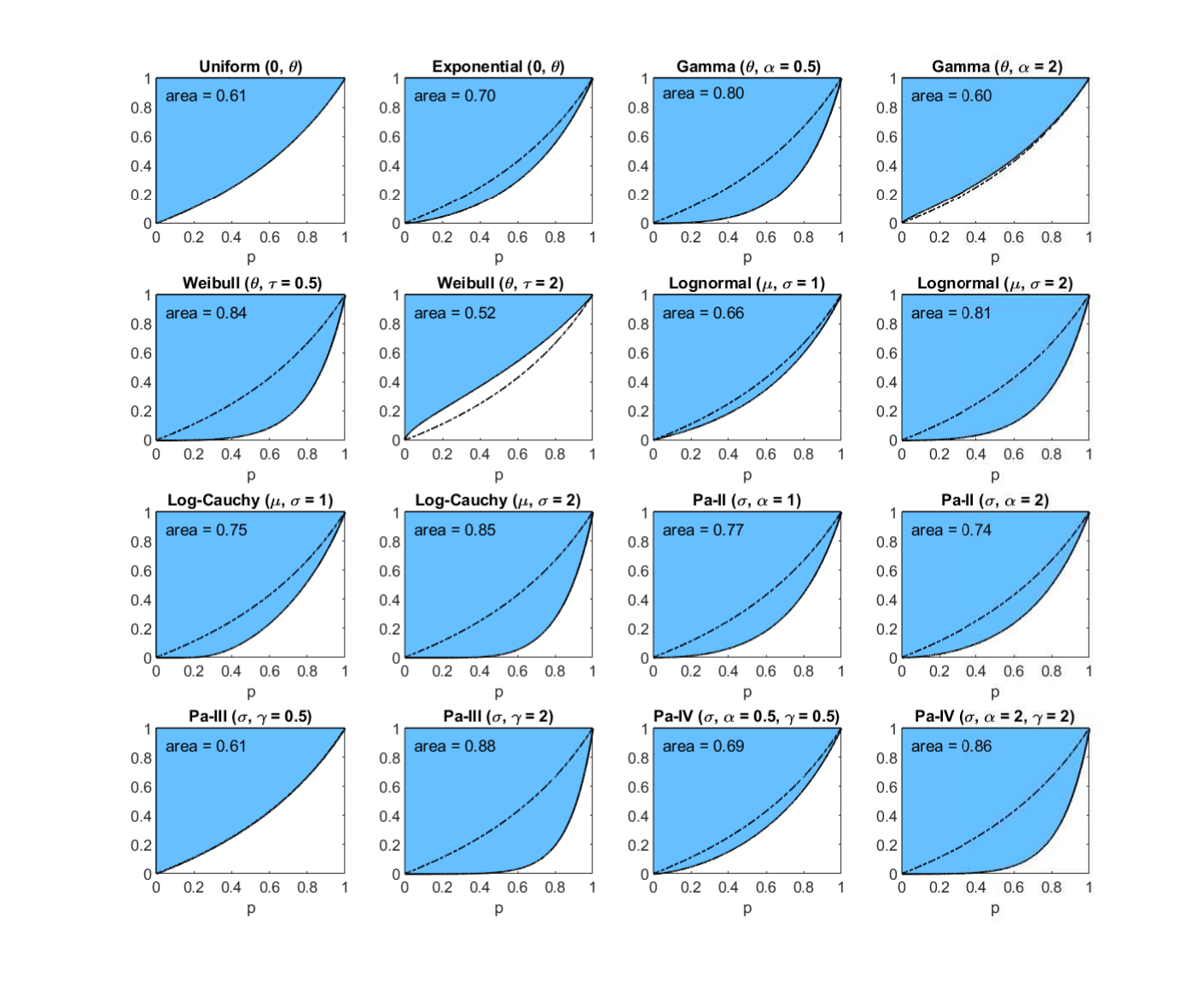}}
    \caption{The income-equality curve $\psi_3$ and the shaded-in area (i.e., $\Psi_3$) above it for the distributions of Table~\ref{tab-1}, with the dash-dotted line depicting $\psi_3$ of the uniform distribution.}
    \label{fig-3}
\end{figure}

In Table~\ref{tab-1} we have four groups consisting of four distributions. The groups reflect the fact that in Figures~\ref{fig-1}--\ref{fig-3},
the distributions are grouped into four rows each containing four panels. The figures depict the three income-equality curves $\psi_k$ for the distributions specified in Table~\ref{tab-1}.

Since the curves are ratios of percentiles, the scale parameter of each distribution has no effect on the inequality indices. The same is true for the log-location
parameter ($e^{\mu}$) of the lognormal and log-Cauchy distributions.
However, the shape ($\alpha$, $\gamma$) and the log-scale ($e^{\sigma}$)
parameters are the primary drivers of the underlying inequality.
To explore this effect, we choose a couple of values of each of these
parameters for plotting. In the plots of Figures~\ref{fig-1}--\ref{fig-3},
the uniform distribution serves as a benchmark for comparing the curves.
In each plot, the dash-dotted line (invisible in the top left panels of the figures) marks the curve $\psi_k$ in the case of the uniform distribution. Numerical evaluations labeled `area' represent the areas of the corresponding shaded regions above the curves $\psi_k$, which are the values of the inequality indices.

From Table~\ref{tab-1} and Figures~\ref{fig-1}--\ref{fig-3} we observe several facts, which can also be verified mathematically:
\begin{itemize}
\item
$\psi_1$ for {\it Pareto-III\/}$(\sigma, \gamma = 2)$ and
$\psi_3$ for {\it Pareto-II\/}$(\sigma, \alpha = 1)$
coincide, thus giving identical inequality indices $0.7736$.
\item
$\psi_1$ for {\it Pareto-II\/}$(\sigma, \alpha = 1)$
and $\psi_3$ for both {\it Uniform\/}$(0, \theta)$
and {\it Pareto-III\/}$(\sigma, \gamma = 0.5)$ coincide,
thus giving identical inequality indices $0.6147$.
\item
$\psi_1$ for {\it Lognormal\/}$(\mu, \sigma = 2)$ and
$\psi_3$ for {\it Lognormal\/}$(\mu, \sigma = 1)$
coincide, thus giving identical inequality indices $0.6648$.
\item
The curves $\psi_1$ for {\it Log-Cauchy\/}$(\mu, \sigma = 2)$ and
$\psi_3$ for {\it Log-Cauchy\/}$(\mu, \sigma = 1)$
coincide, thus giving identical inequality indices $0.7470$.
\end{itemize}

We conclude this section with the note that there are, of course, many other parametric distributions for modelling incomes \cite[see, e.g.,][]{kk2003}, and one of them is the Dagum distribution. For statistical inference for the ratio of any two quantiles of this distribution --  and the three equality curves $\psi_k(p)$ are such ratios -- we refer to \cite{jpz2023}.

\section{A nonparametric viewpoint}
\label{empirics}

We now consider nonparametric ways for estimating all the aforementioned indices of inequality and their corresponding equality curves, with an analysis of real data.

\subsection{Estimators of the new indices}

Let $X_{1},\dots , X_{n} $ denote incomes of randomly selected persons, with $X_{1:n}\le \cdots \le X_{n:n}$ denoting the ordered incomes. The empirical counterparts of the three indices $\Psi_k$ are (see their justifications in Appendix~\ref{technicalities})
\begin{align}
\Psi_{1,n}&= 1- {1\over \lfloor n/2 \rfloor} \sum_{k=1}^{\lfloor n/2 \rfloor}
{X_{k:n} \over X_{\lceil n/2 \rceil :n}} ,
\label{def-psi1}
\\
\Psi_{2,n}&= 1- {1\over \lfloor n/2 \rfloor} \sum_{k=1}^{\lfloor n/2 \rfloor}
{X_{k:n} \over X_{\lceil n/2 \rceil  +k :n}} ,
\label{def-psi2}
\\
\Psi_{3,n}&= 1- {1\over \lfloor n/2 \rfloor} \sum_{k=1}^{\lfloor n/2 \rfloor}
{X_{k:n} \over X_{n-k+1 :n}},
\label{def-psi3}
\end{align}
where, for every real $x\ge 0$, $\lfloor x \rfloor $ is the largest integer that does not exceed $x$, and $\lceil x \rceil $ is the smallest integer that is not below $x$. (These are the classical floor and ceiling functions.) When it is desirable to emphasize the dependence of the indices on incomes, we do so by writing them as $\Psi_{k,n}(\mathbf{X})$, where $\mathbf{X}=(X_{1:n}, \dots , X_{n:n})$ is the vector of all the (ordered) incomes in the sample. Next are a few immediate consequences of definitions~\eqref{def-psi1}--\eqref{def-psi3}.

\begin{property}
For every real $c\ge 0$, we have $\Psi_{k,n}(c\mathbf{X})=\Psi_{k,n}(\mathbf{X})$.
\end{property}

This property implies, for example, that changing the currency with which the incomes are reported does not affect the values of the three inequality indices.

\begin{property}\label{prop42}
We have the inequality
$\Psi_{k,n}(\mathbf{X})\ge \Psi_{k,n}(\mathbf{X}+c )$ for every real $c\ge 0$.
The inequality is strict under the following two conditions: first, $c>0$, and second, there is at least one ratio inside the sum of the definition of $\Psi_{k,n}$ that is not equal to $1$. (Note that none of the ratios exceeds $1$.)
\end{property}

This property implies that adding the same amount of income to everybody does not increase inequality and, under a minor caveat specified in the property, the index even decreases. To see the necessity of the assumption, consider the case when all $X$'s are equal, which gives  $\Psi_{k,n}(\mathbf{X})=0$ and also $\Psi_{k,n}(\mathbf{X}+c )=0$ irrespective of the value of $c$. For a proof of Property~\ref{prop42}, as well as for proofs of other properties (see Appendix~\ref{technicalities}).

\begin{property}\label{prop43}
When $c\to \infty $, we have $\Psi_{k,n}(\mathbf{X}+c ) \to 0$.
\end{property}

Intuitively, this property says that if we keep adding the same positive amount of income to everyone, all else being equal, then we shall eventually eliminate the inequality.

\subsection{Estimators of the earlier indices}

Next we report the definitions of the empirical estimators of $Z$, $D$, $G$ and $G_2$ obtained  by replacing the population quantile function $Q$ by the empirical quantile function $Q_n$, which is given by the equation
\begin{equation}\label{equantile function}
Q_n(p)=X_{\lceil np \rceil :n}
\end{equation}
for every $p\in (0,1]$. Slightly modifying the obtained expression in an asymptotically equivalent way to make it intuitively and computationally more appealing, we arrive at the estimator
\[
Z_n=1-{1\over n}\sum_{i=1}^{n-1}
{{1\over i}\sum_{k=1}^{i}X_{k:n} \over {1\over n-i}\sum_{k=i+1}^{n} X_{k:n}}
\]
of $Z$, which appears in \citet{gp2009}. Likewise, we arrive at
\[
D_n=1-{1\over n}\sum_{i=1}^{n}
{{1\over i}\sum_{k=1}^{i}X_{k:n} \over {1\over i}\sum_{k=n-i+1}^{n} X_{k:n}},
\]
which is an empirical estimator of $D$ that appeared in \citet{dg2020}. (Of course, $1/i$ in the numerator and denominator cancel out.) The same reasoning leads to the empirical Gini index
\begin{align*}
G_n
&=1-{2\over n}\sum_{i=1}^{n}
{\sum_{k=1}^{i}X_{k:n} \over \sum_{k=1}^{n} X_{k:n}} +{1\over n}
\\
&= 1-{1\over \bar{X}  n^2}\sum _{i=1}^n \big(2(n-i)+1\big) X_{i:n},
\end{align*}
where the last equation follows from simple algebra, with $\bar{X}$ denoting the mean of $X_{1},\dots , X_{n} $. Note that the last expression for $G_n$ is the one that places the empirical Gini index into the family of $S$-Gini indices introduced by \citet{dw1980} and \citet{w1980/81}; see also \citet{zg2002} for further references and statistical inference.

\begin{note}
The asymptotically negligible term $1/n$ on the right-hand side of the first equation of $G_n$ ensures that $G_n$ makes sense for all sample sizes. Without this term we may get counterintuitive values. For example, when the `incomes' are $X_1=1$, $X_2=2$ and $X_3=3$, we have $G_n=2/9$, whereas $G_n$ without the added $1/n=1/3$ would give the negative value $-1/9$, which is incompatible with the meaning of the index.
\end{note}

Finally, using the same arguments as above but now with the right-most expression for $G_{2}$ given in Section~\ref{gast} as our starting point, we arrive at
\[
G_{2,n}= {\bar{X}\over X_{\lceil n/2 \rceil :n}}-{2\over n^2}\sum_{i=1}^{n}
{\sum_{k=1}^{i}X_{k:n} \over X_{\lceil n/2 \rceil :n}}
\]
as an empirical estimator of $G_{2}$. As before, $\bar{X}$ stands for the mean of $X_{1},\dots , X_{n} $.

\subsection{An analysis of capital incomes from the ECHP 2001 survey}
\label{survey2001}

Using the formulas for calculating the aforementioned indices from data, we now analyze capital incomes reported in the European Community Household Panel survey \citep[][]{echp2001} that was  conducted by Eurostat in 2001, which is the last of the eight waves of the survey.

\begin{figure}[h!]
    \centering
    \includegraphics[width=\textwidth]{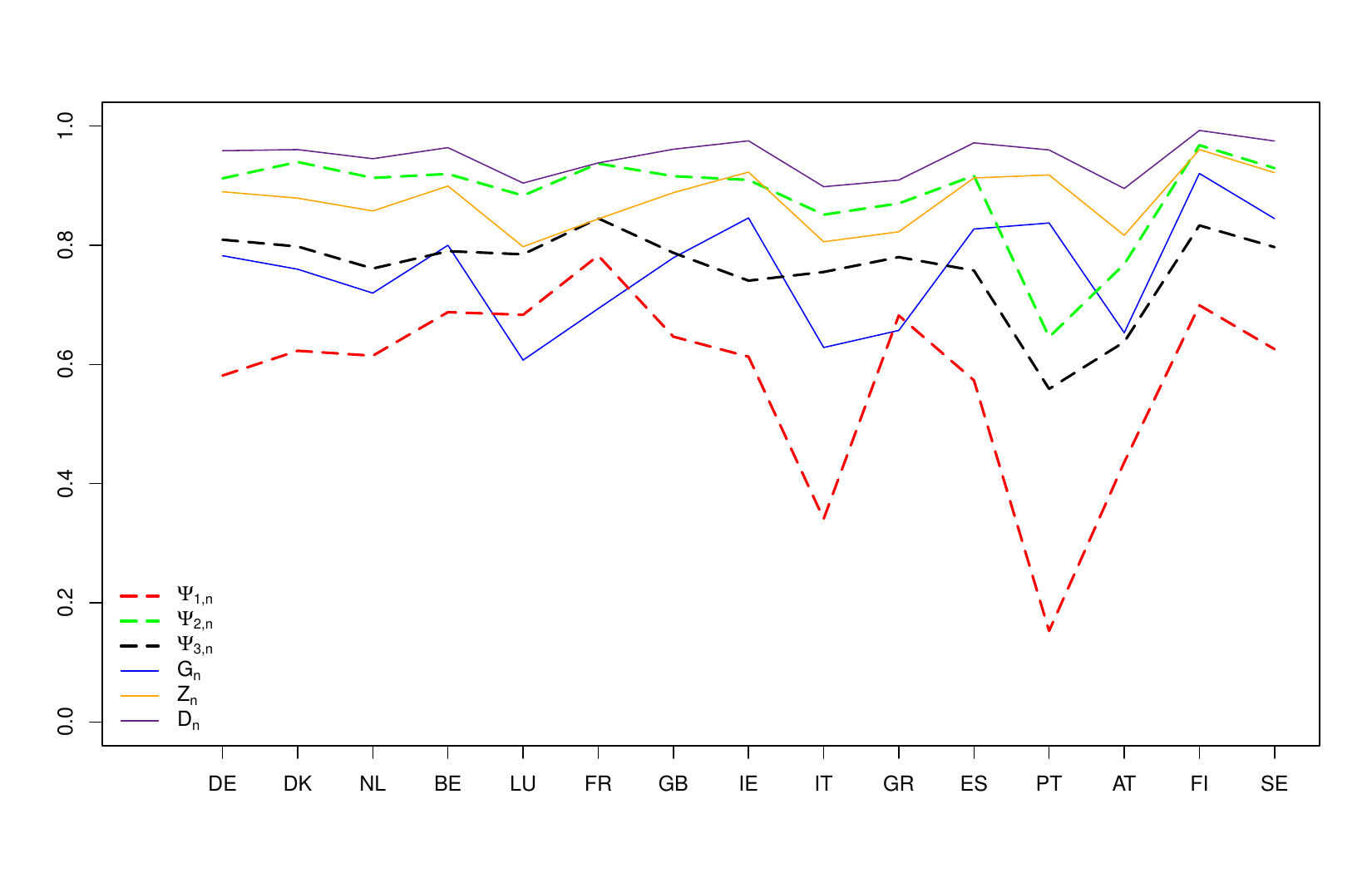}
    \caption{The income-inequality indices  $G_{n}$, $Z_{n}$,  $D_{n}$, and the new indices $\Psi_{k,n}$ for the fifteen European countries with $n=n_P$ specified in Table~\ref{tab-real} \citep[based on][]{echp2001}.}
    \label{fig-real}
\end{figure}

Specifically, the data come from 59,750 households with 121,122 persons from the fifteen European countries specified in Table~\ref{tab-real}
\begin{sidewaystable}
\centering
\begin{tabular}{l|rr|rr|cccrccc|rrr}
  \hline
   Countries & Means & Medians  &\multicolumn{2}{c|}{Sample sizes} & \multicolumn{7}{c|}{Inequality indices} & \multicolumn{3}{c}{Ranks based on}
   \\
  &&& $n_{T}~~$ & $n_{P}~~$ & $G_{n}$ & $Z_{n}$ & $D_{n}$ & $G_{2,n}$ & $\Psi_{1,n}$ & $\Psi_{2,n}$ & $\Psi_{3,n}$ & $\Psi_{1,n}$ & $\Psi_{2,n}$ & $\Psi_{3,n}$ \\

  \hline
DE & 948.373 & 186.622 & 10,624 & 4,861 & 0.782 & 0.890 & 0.959 & 3.975 & 0.581 & 0.912 & 0.809 &    5 &    7 &   13 \\
DK & 1,071.062 & 231.417 & 3,789 & 1,135 & 0.760 & 0.879 & 0.961 & 3.512 & 0.623 & 0.940 & 0.798 &    8 &   14 &   12 \\
NL & 660.744 & 214.184 & 8,608 & 2,863 & 0.720 & 0.858 & 0.945 & 2.219 & 0.615 & 0.913 & 0.761 &    7 &    8 &    6 \\
BE & 5,309.168 & 1,374.805 & 4,299 &  690 & 0.800 & 0.899 & 0.964 & 3.091 & 0.688 & 0.920 & 0.790 &   13 &   11 &   10 \\
LU & 1,982.621 & 1,214.678 & 4,916 &  769 & 0.607 & 0.798 & 0.904 & 0.989 & 0.683 & 0.883 & 0.785 &   12 &    5 &    8 \\
FR & 716.679 & 359.932 & 10,119 & 4,347 & 0.694 & 0.844 & 0.938 & 1.381 & 0.783 & 0.937 & 0.845 &   15 &   13 &   15 \\
GB & 1,522.177 & 368.826 & 8,521 & 3,477 & 0.779 & 0.888 & 0.961 & 3.214 & 0.647 & 0.916 & 0.787 &   10 &    9 &    9 \\
IE & 604.580 & 99.040 & 4,023 &  949 & 0.846 & 0.923 & 0.975 & 5.157 & 0.613 & 0.910 & 0.741 &    6 &    6 &    3 \\
IT & 1.762 & 0.480 & 13,392 & 1,111 & 0.628 & 0.806 & 0.898 & 2.303 & 0.341 & 0.851 & 0.755 &    2 &    3 &    4 \\
GR & 2,256.554 & 1,232.575 & 9,419 &  335 & 0.657 & 0.823 & 0.909 & 1.197 & 0.682 & 0.870 & 0.780 &   11 &    4 &    7 \\
ES & 240.838 & 37.431 & 11,964 & 6,541 & 0.827 & 0.913 & 0.972 & 5.322 & 0.573 & 0.917 & 0.758 &    4 &   10 &    5 \\
PT & 1,232.674 & 116.260 & 10,915 &  600 & 0.837 & 0.918 & 0.960 & 8.862 & 0.153 & 0.646 & 0.559 &    1 &    1 &    1 \\
AT & 323.822 & 133.500 & 5,605 & 2,834 & 0.653 & 0.817 & 0.895 & 1.585 & 0.436 & 0.768 & 0.638 &    3 &    2 &    2 \\
FI & 3,662.567 & 180.634 & 5,637 & 1,509 & 0.921 & 0.961 & 0.993 & 18.651 & 0.699 & 0.968 & 0.833 &   14 &   15 &   14 \\
SE & 601.528 & 84.495 & 9,291 & 5,637 & 0.845 & 0.922 & 0.975 & 6.013 & 0.626 & 0.929 & 0.797 &    9 &   12 &   11 \\
   \hline
\end{tabular}
    \caption{The income-inequality indices  $G_{n}$, $Z_{n}$,  $D_{n}$, $G_{2,n}$, and the new indices $\Psi_{1,n}$, $\Psi_{2,n}$, $\Psi_{3,n}$ for the fifteen European countries with $n=n_P$,  where $n_P$ is the number of people in the sample who possess capital incomes, and $n_T$ is the total sample size of the given country \citep[based on][]{echp2001}.}
    \label{tab-real}
\end{sidewaystable}
using the ISO 3166-1 alpha-2 (two-letter) codes.
By looking at the means and medians in Table~\ref{tab-real},
we see how skewed to the right the distributions of the countries are.
Figure~\ref{fig-real} (with $G_{2,n}$ excluded due to its large values)
visualizes the index values calculated using formulas~\eqref{def-psi1}--\eqref{def-psi3} and reported in Table~\ref{tab-real}. For a more detailed description of the data and relevant references, we refer to \citet[Section~1]{gpz2014}. Next are several observations based on Table~\ref{tab-real} and Figure~\ref{fig-real}.

Portugal has the lowest value of $\Psi_{1,n}$, with the median income of the poorest $p \times 100 \%$ persons equal, after averaging over all $p \in (0,1)$, to $84.7\%$ of the median income of the entire population.

The opposite happens in France, which provides the highest contrast among
the countries when comparing the median income of the poorest $p\times 100 \%$ persons with the overall median income: after averaging such ratios over all $p \in (0,1)$, we obtain $21.7\%$.

For France, we also observe the largest value of $\Psi_{3,n}$.
The median income of the poorest $p \times 100 \%$ people is equal, after averaging over all $p \in
(0,1)$, to only $15.5\%$ of the median income of the richest $p \times 100 \%$ persons in the population.

When we are interested in comparing the median income of the poorest $p\times 100 \%$
persons with the median income of the remaining $(1-p)\times 100\%$ part of
the population, the index $\Psi_{2,n}$ tells us that Finland is the country
in which such a contrast, after averaging over all $p \in
(0,1)$, is the largest.

\begin{figure}[h!]
\begin{center}
\includegraphics[width=0.237\linewidth]{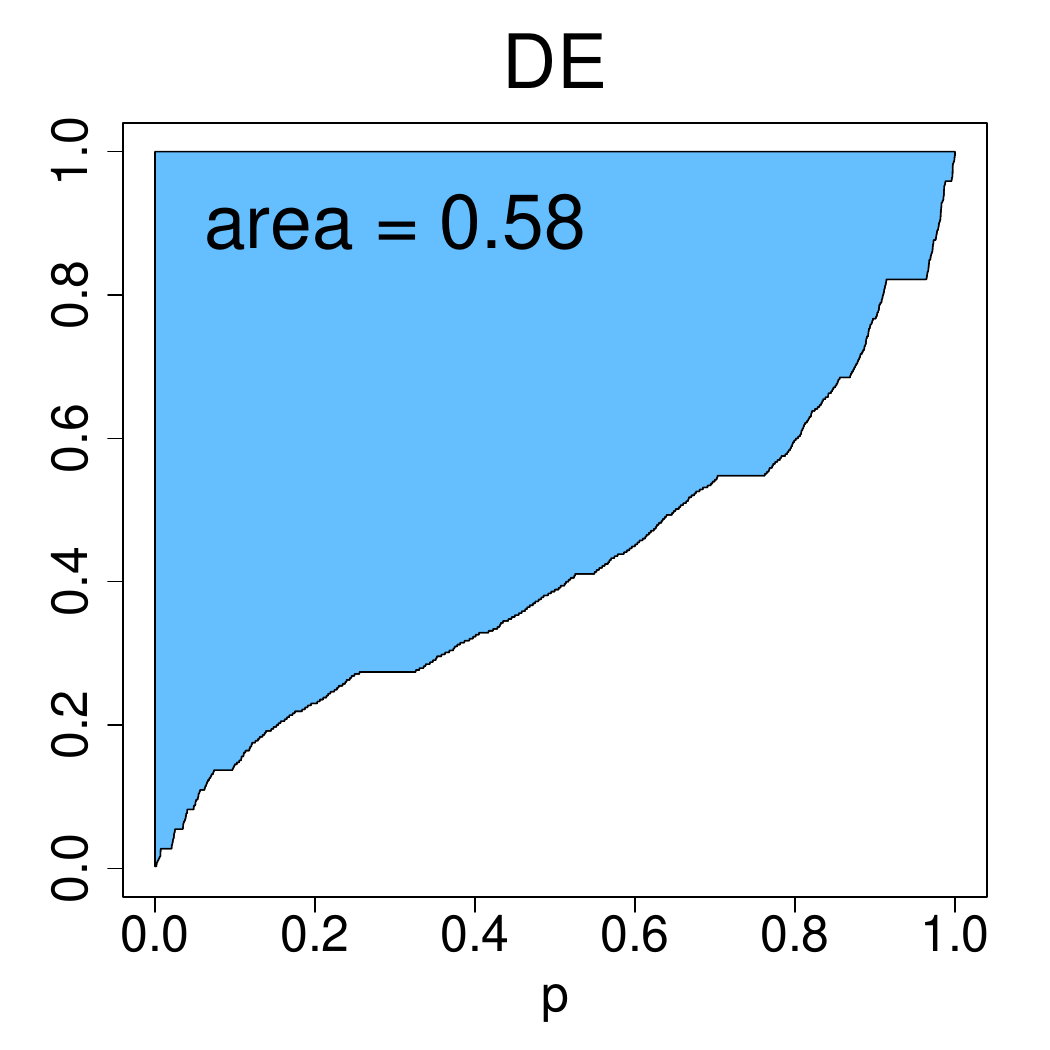}\qquad
\includegraphics[width=0.237\linewidth]{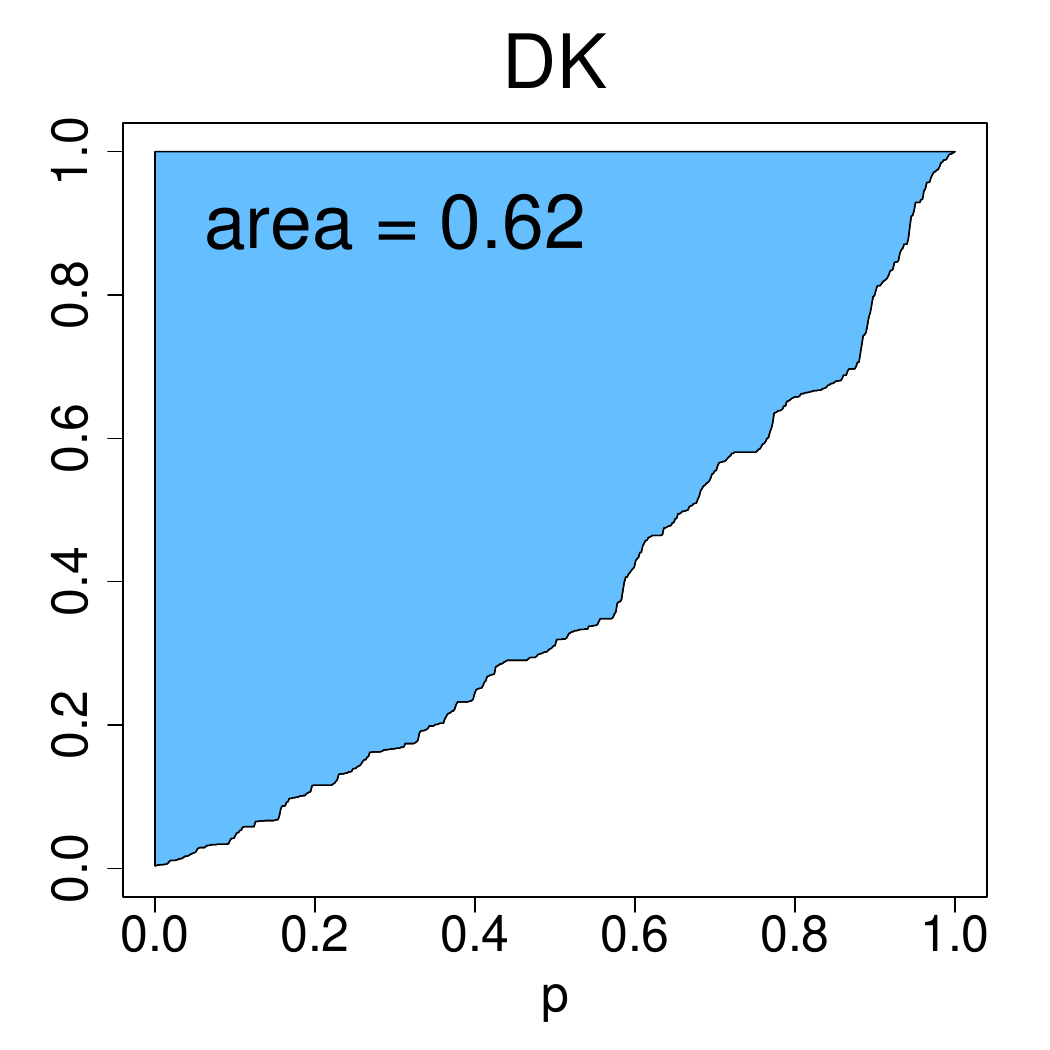}\qquad
\includegraphics[width=0.237\linewidth]{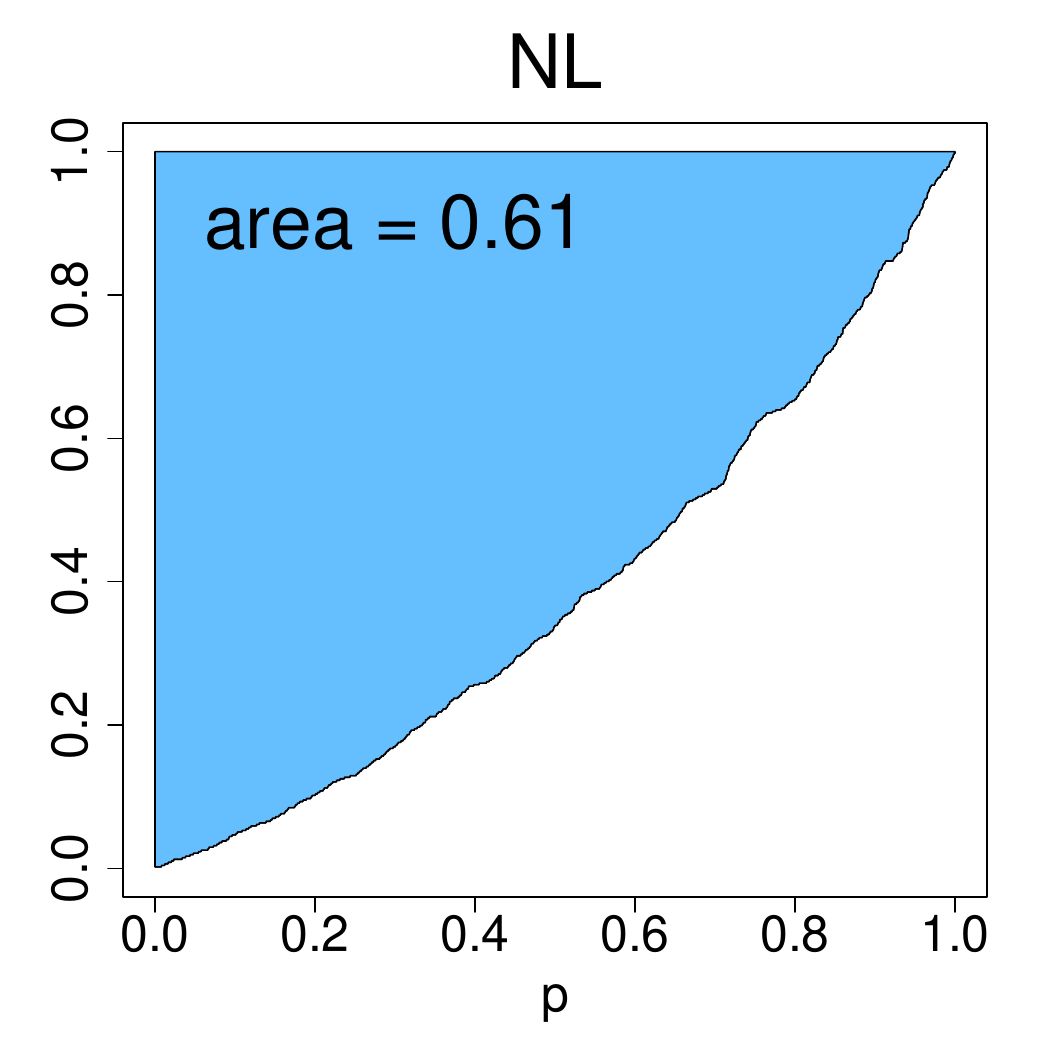}\qquad
\\
\includegraphics[width=0.237\linewidth]{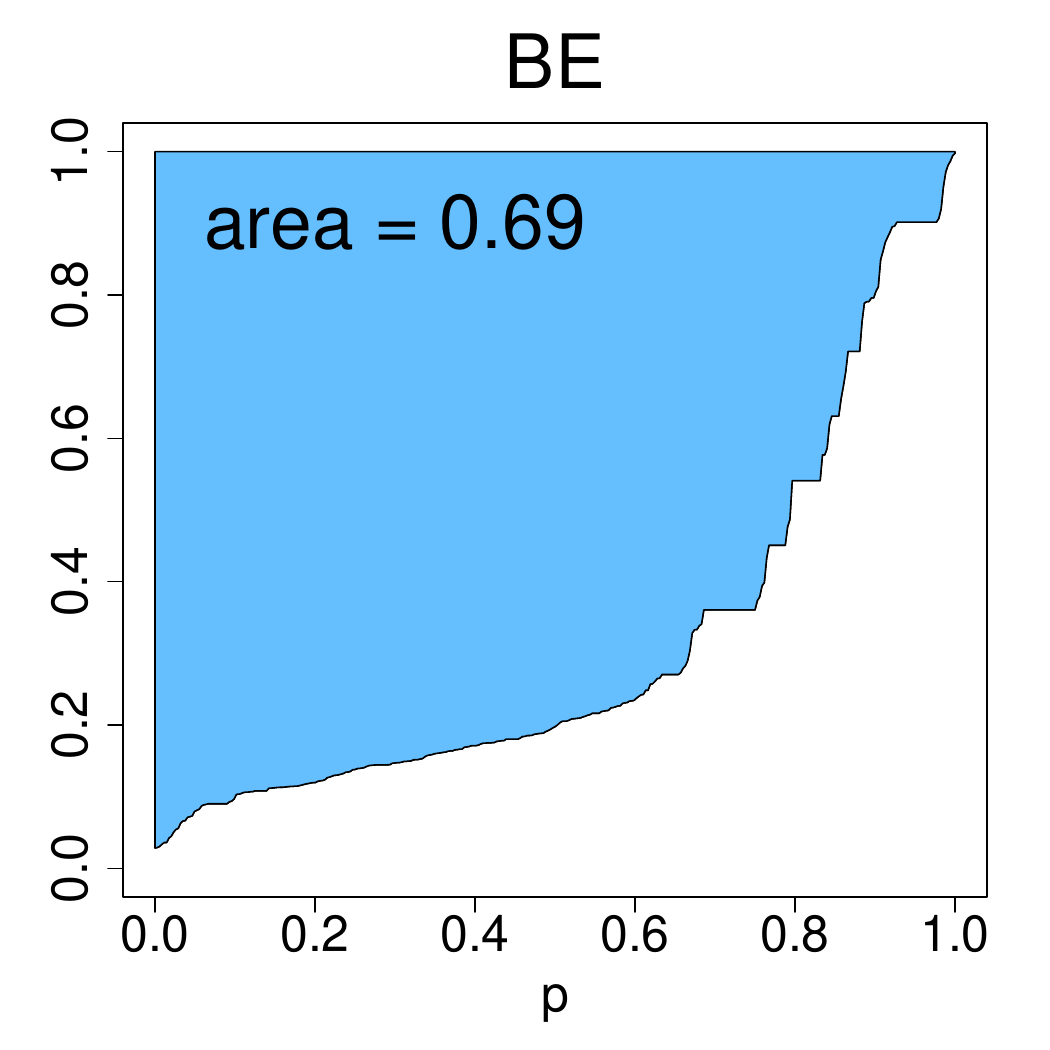}\qquad
\includegraphics[width=0.237\linewidth]{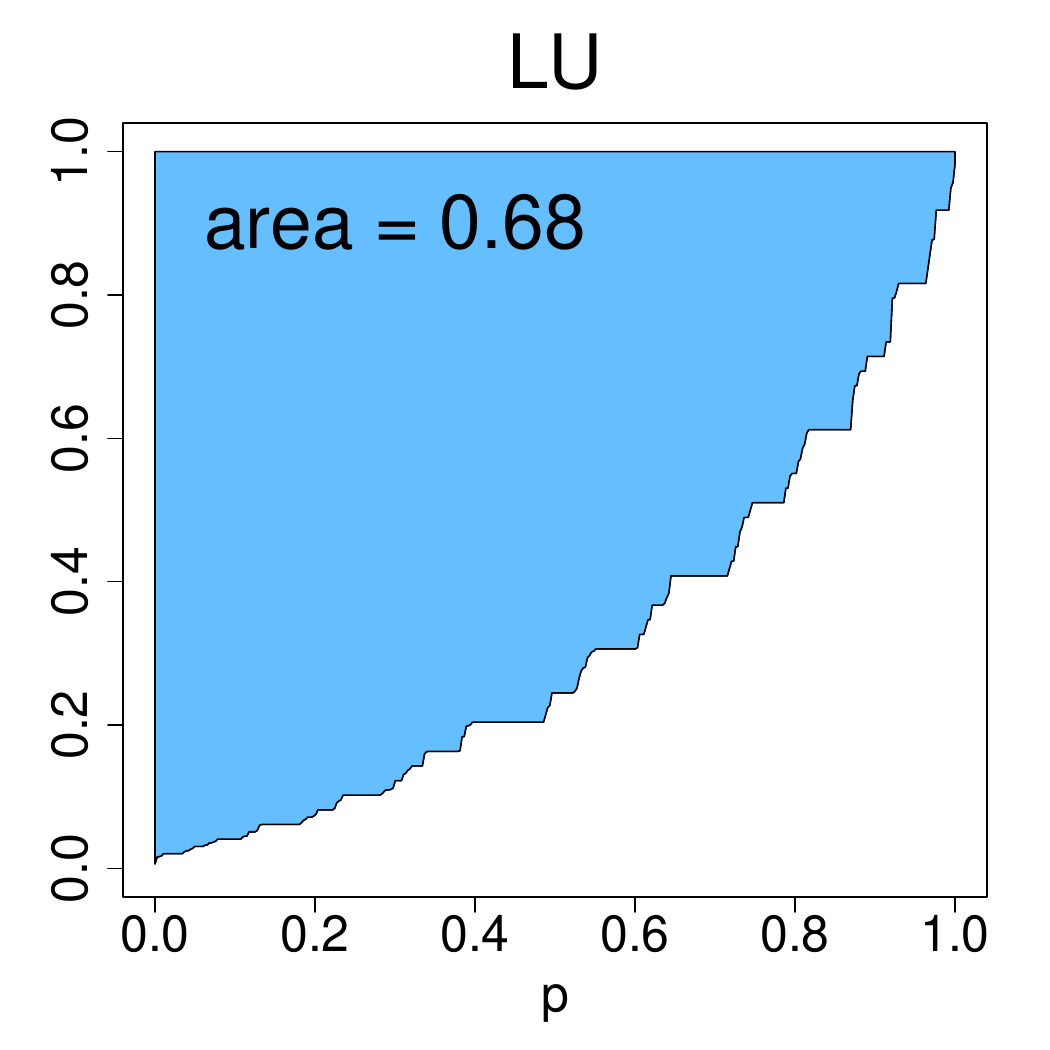}\qquad
\includegraphics[width=0.237\linewidth]{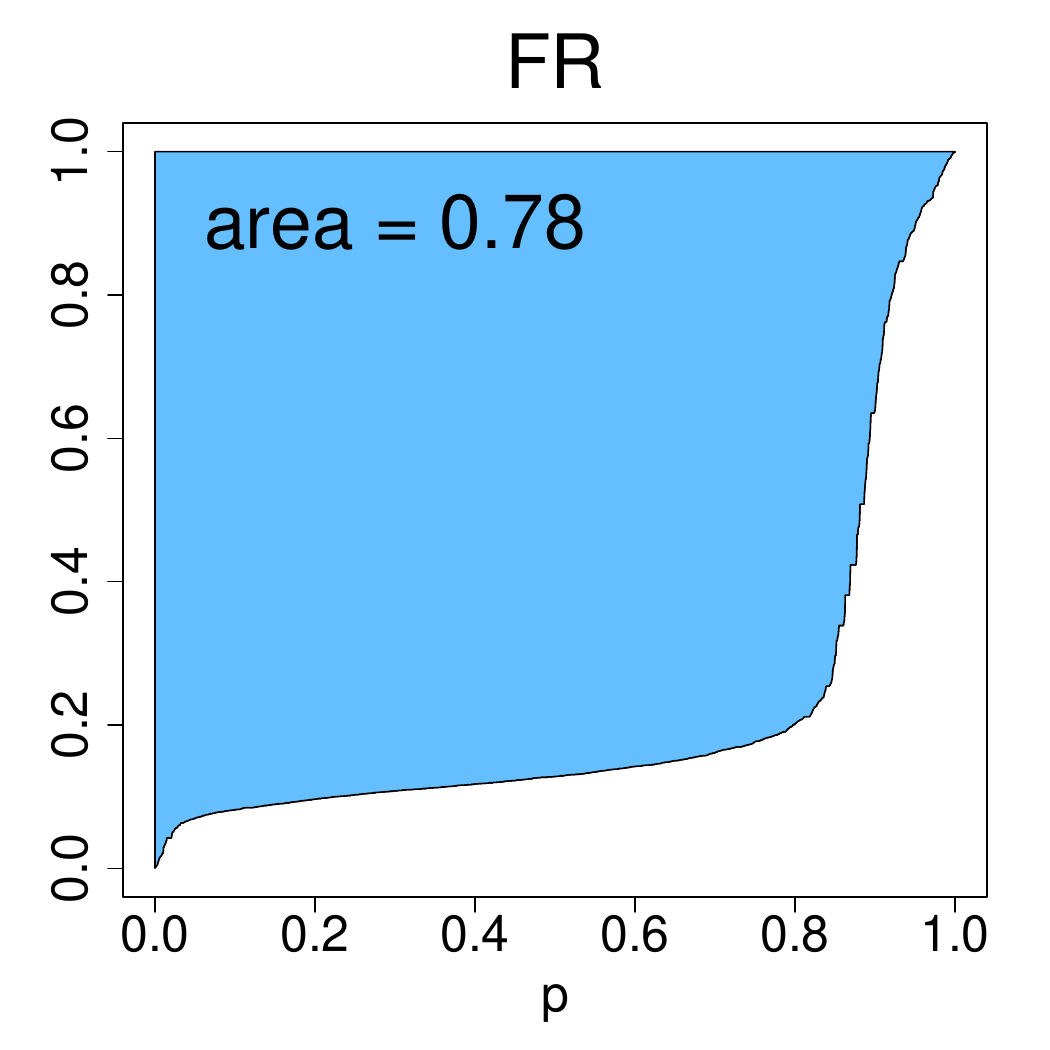}\qquad
\\
\includegraphics[width=0.237\linewidth]{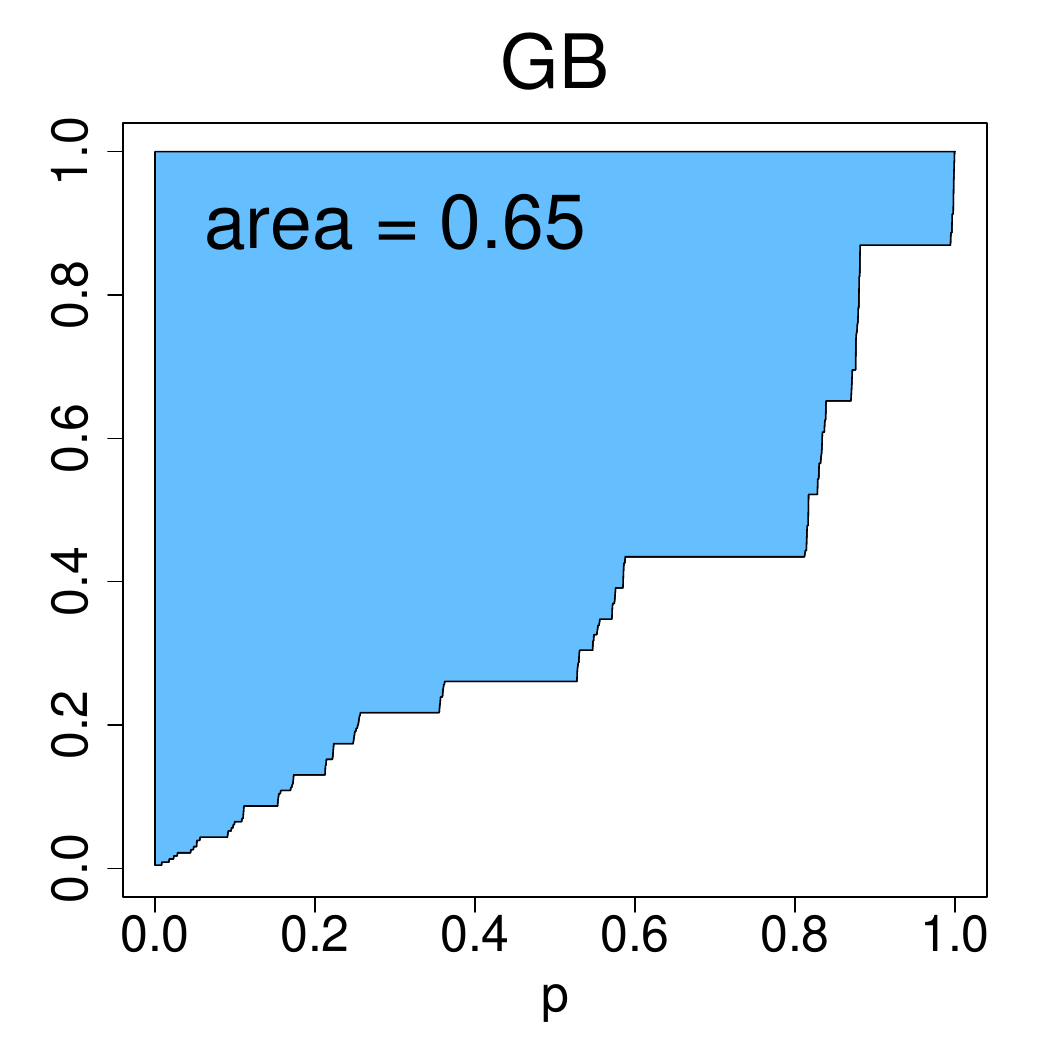}\qquad
\includegraphics[width=0.237\linewidth]{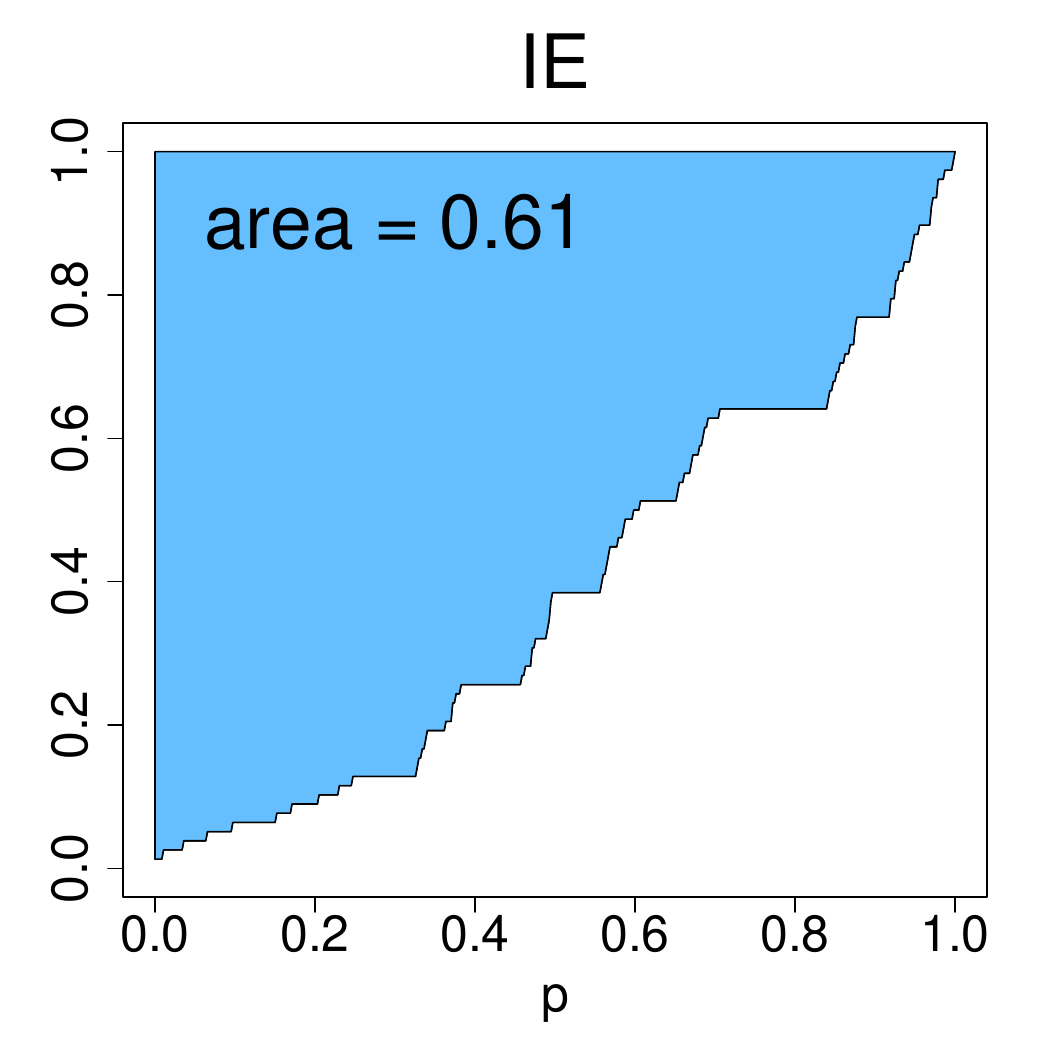}\qquad
\includegraphics[width=0.237\linewidth]{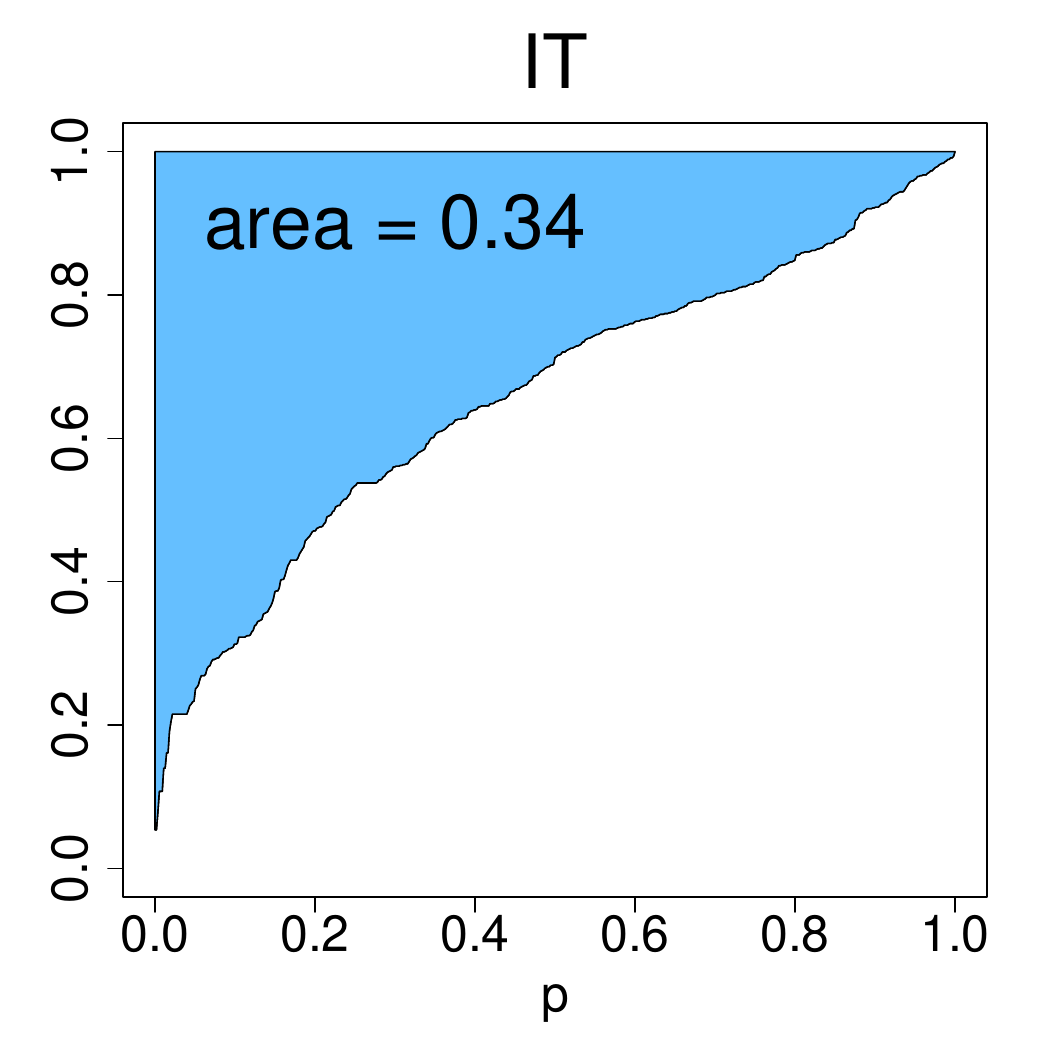}\qquad
\\
\includegraphics[width=0.237\linewidth]{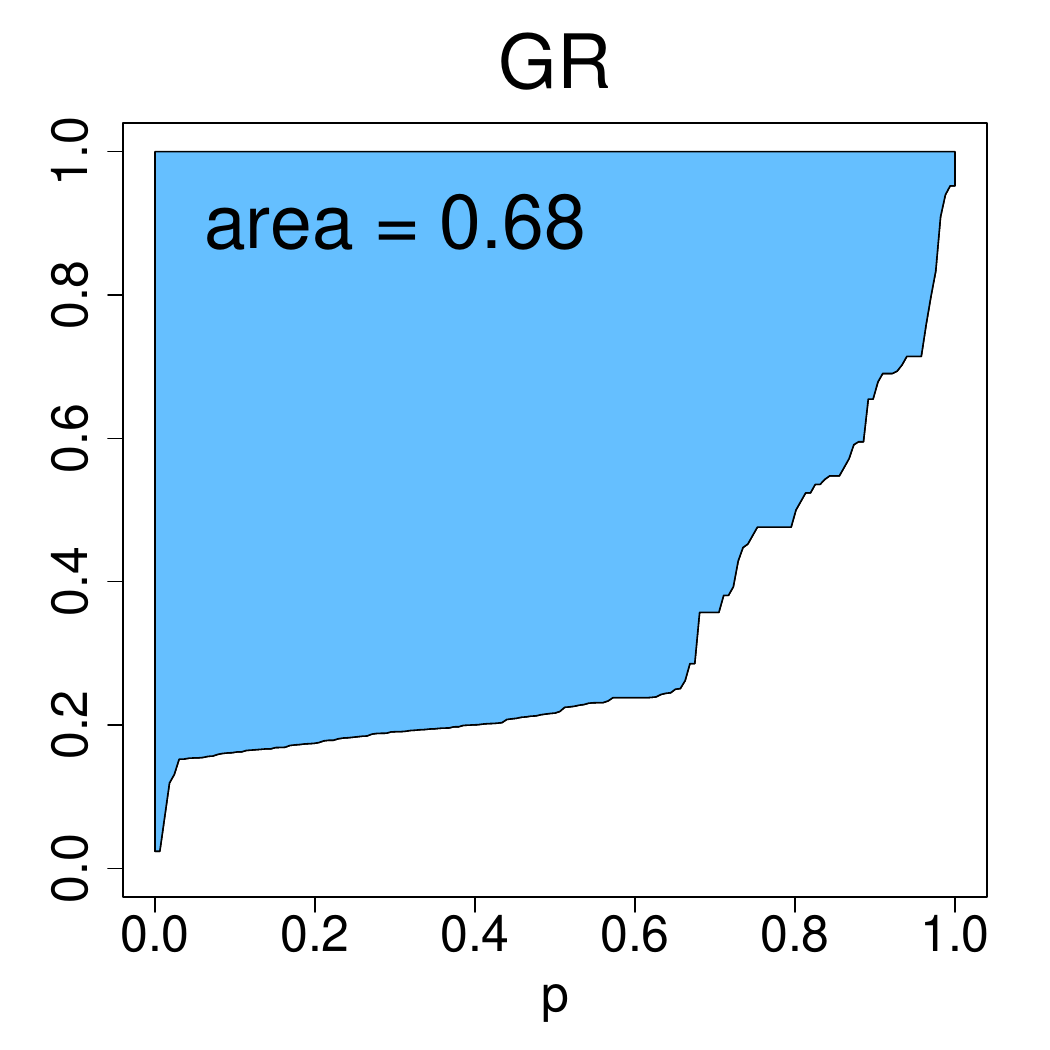}\qquad
\includegraphics[width=0.237\linewidth]{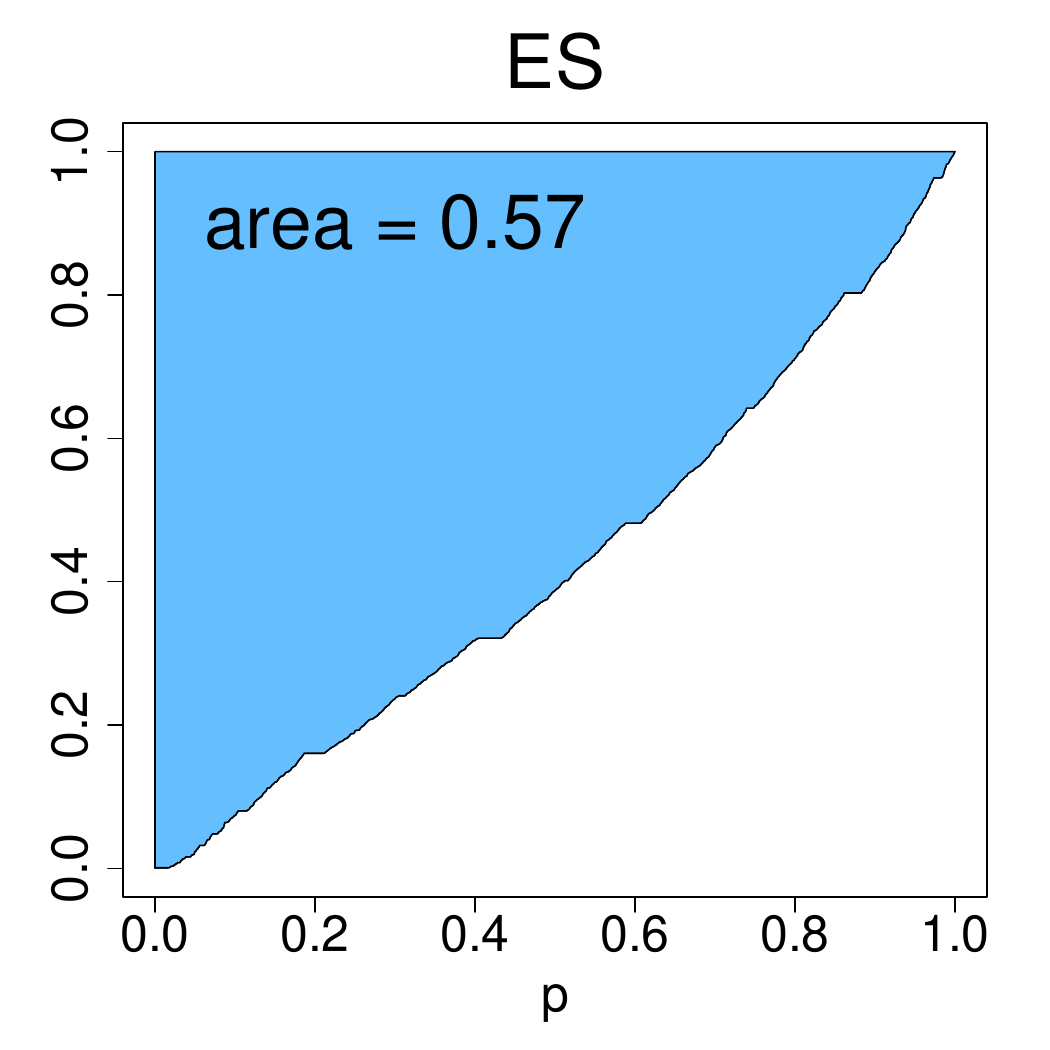}\qquad
\includegraphics[width=0.237\linewidth]{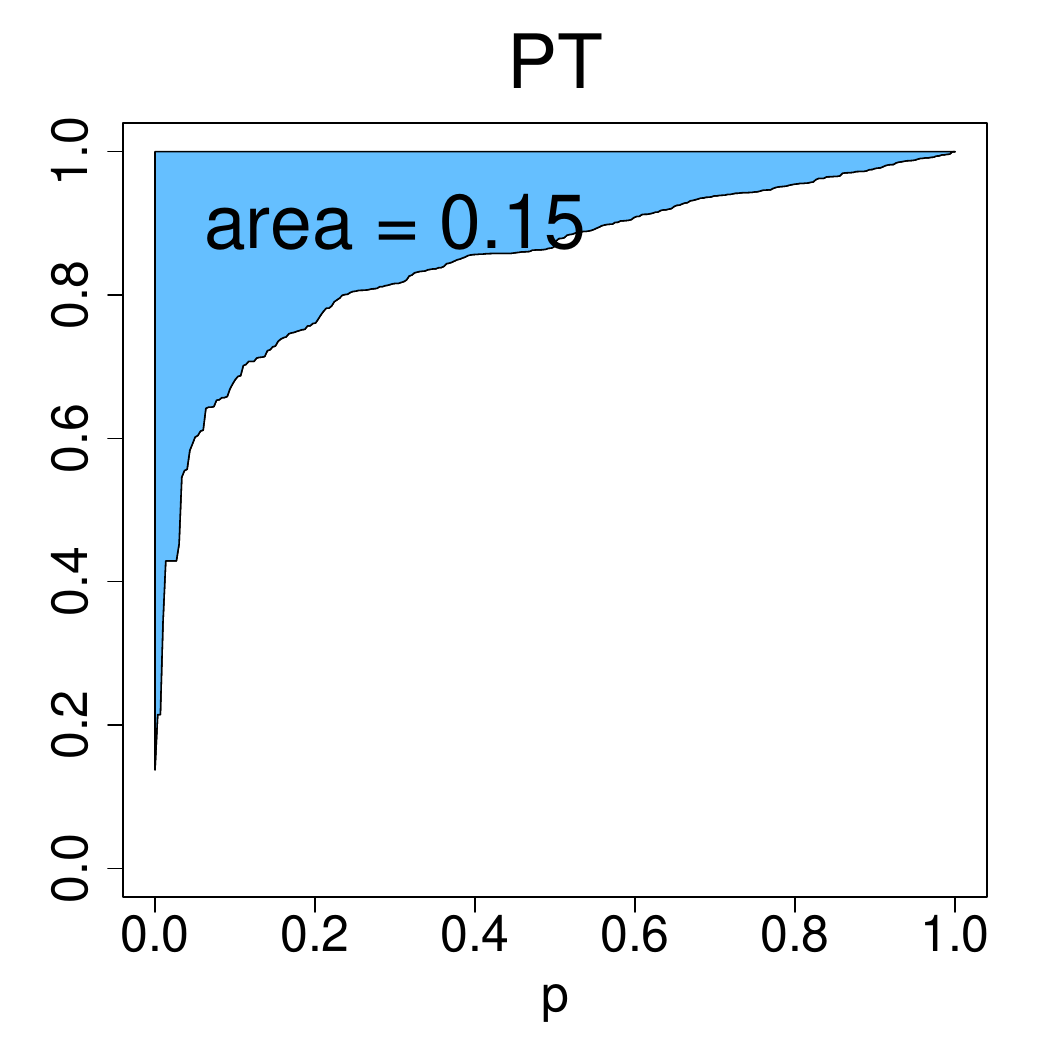}\qquad
\\
\includegraphics[width=0.237\linewidth]{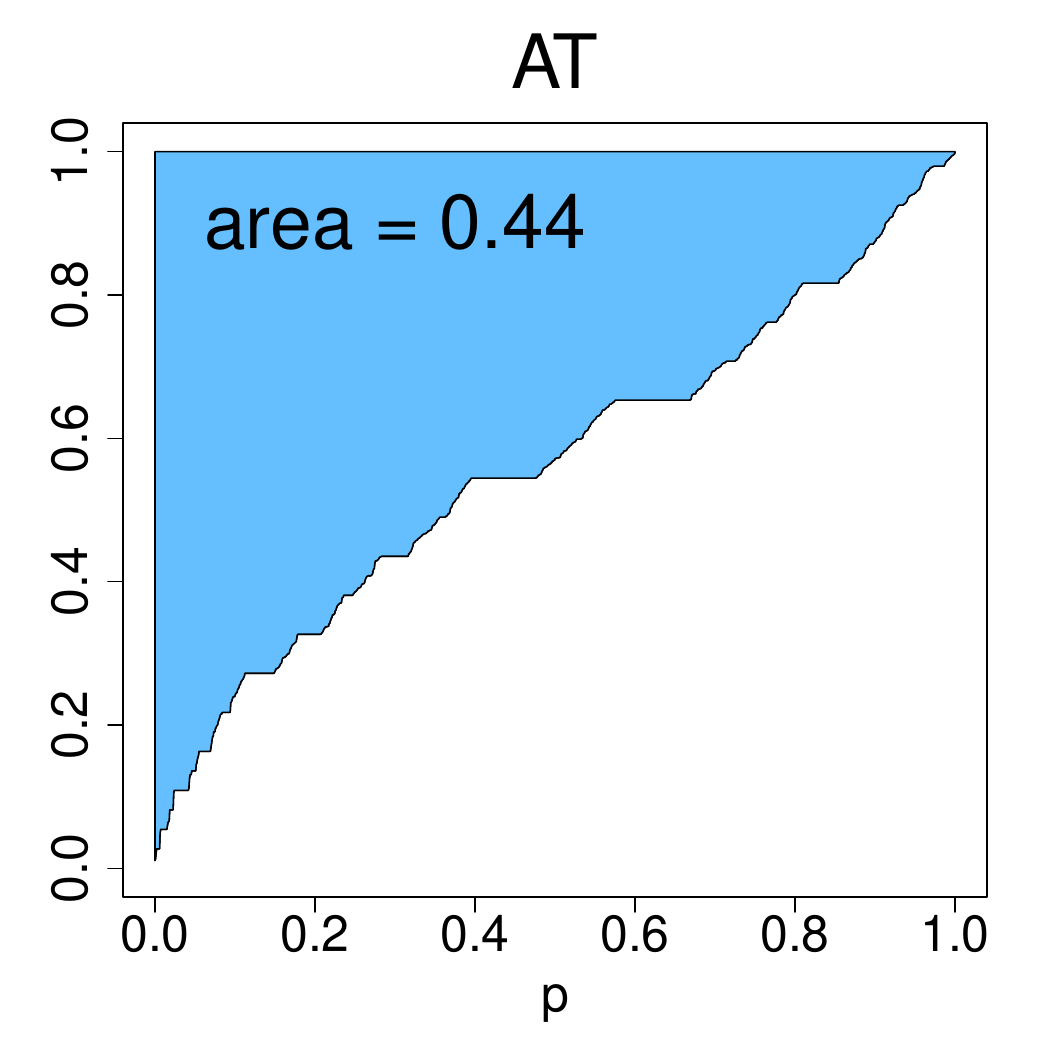}\qquad
\includegraphics[width=0.237\linewidth]{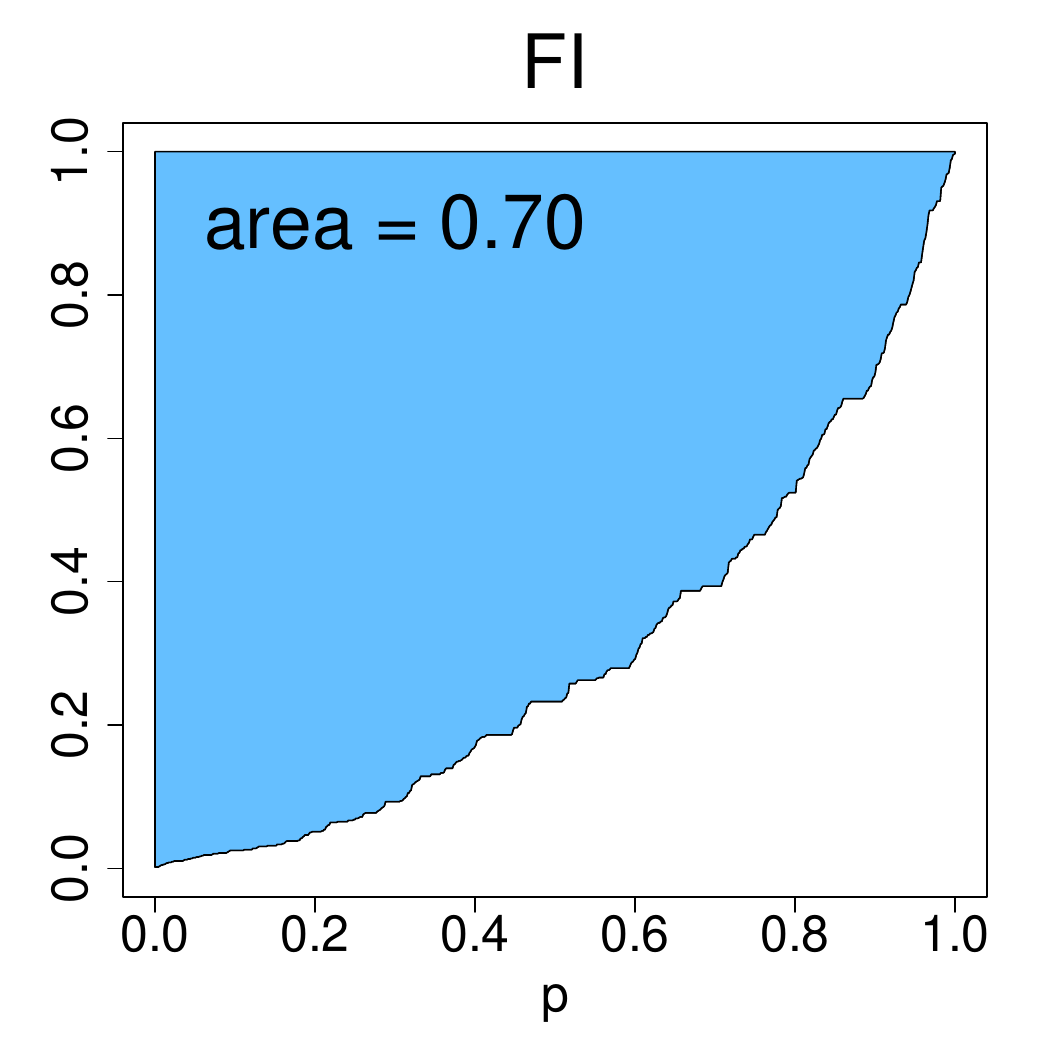}\qquad
\includegraphics[width=0.237\linewidth]{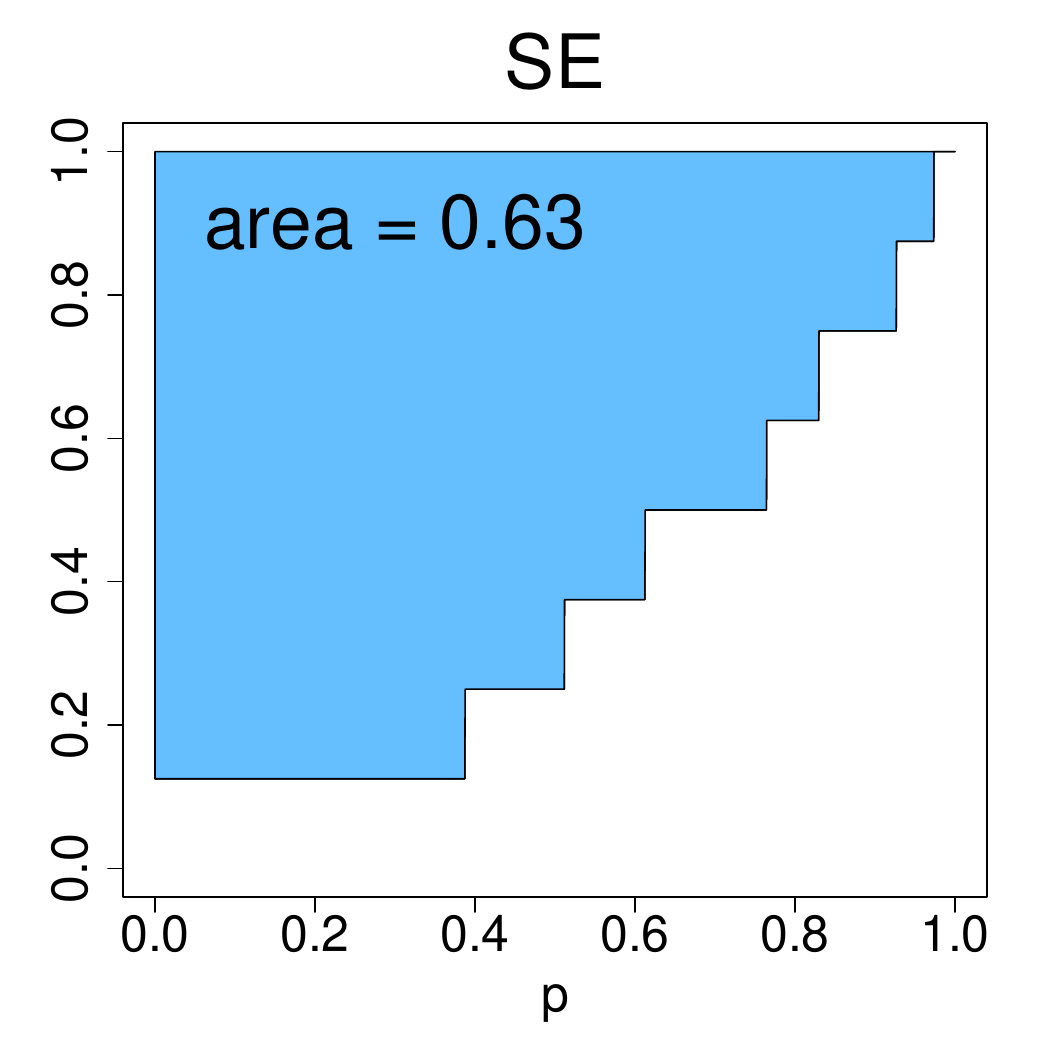}\qquad
\caption{The income-equality curve $\psi_{1,n}$ and the shaded-in area (i.e., $\Psi_{1,n}$) above it for the fifteen European countries, where $n=n_P$ is specified in Table~\ref{tab-real} \citep[based on][]{echp2001}.}
\label{fig:1}
\end{center}
\end{figure}

\begin{figure}[h!]
\begin{center}
\includegraphics[width=0.237\linewidth]{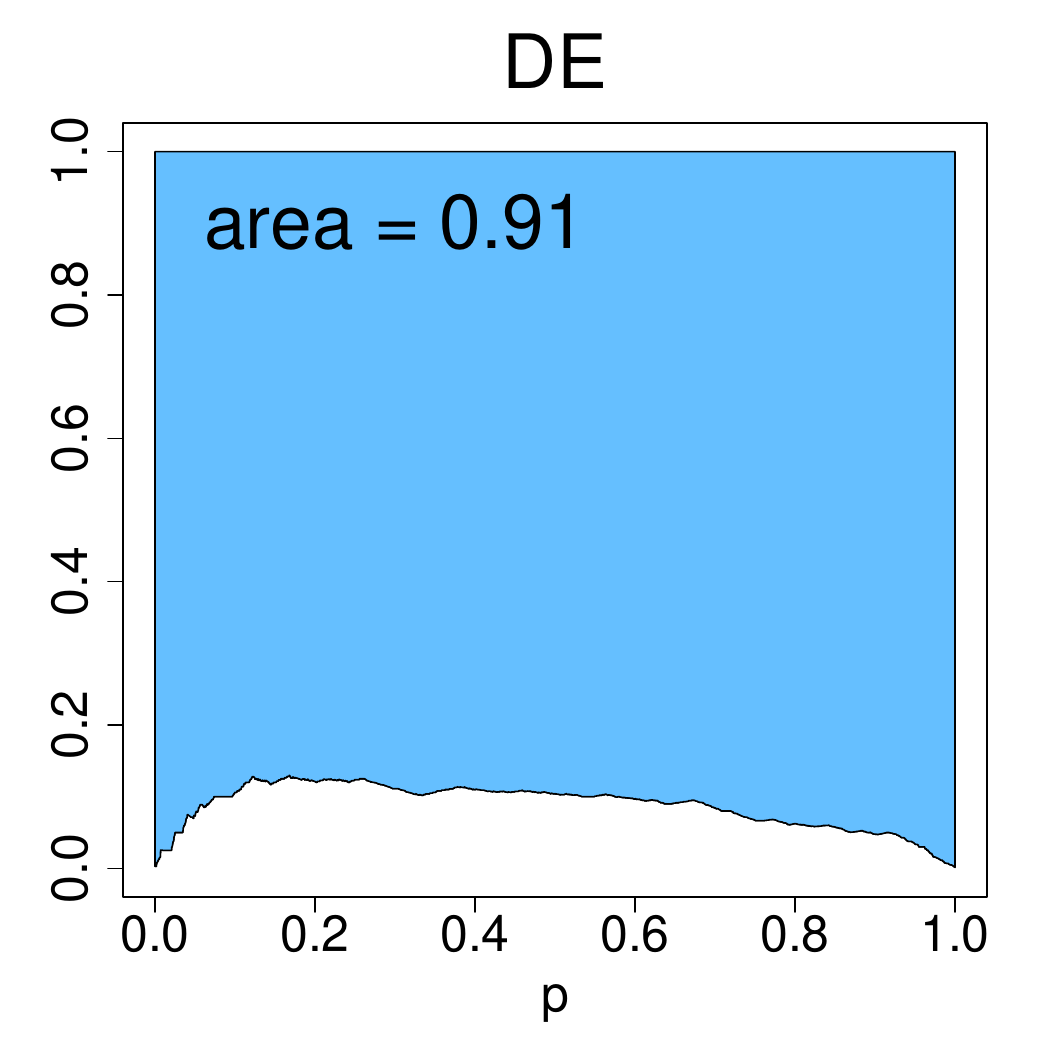}\qquad
\includegraphics[width=0.237\linewidth]{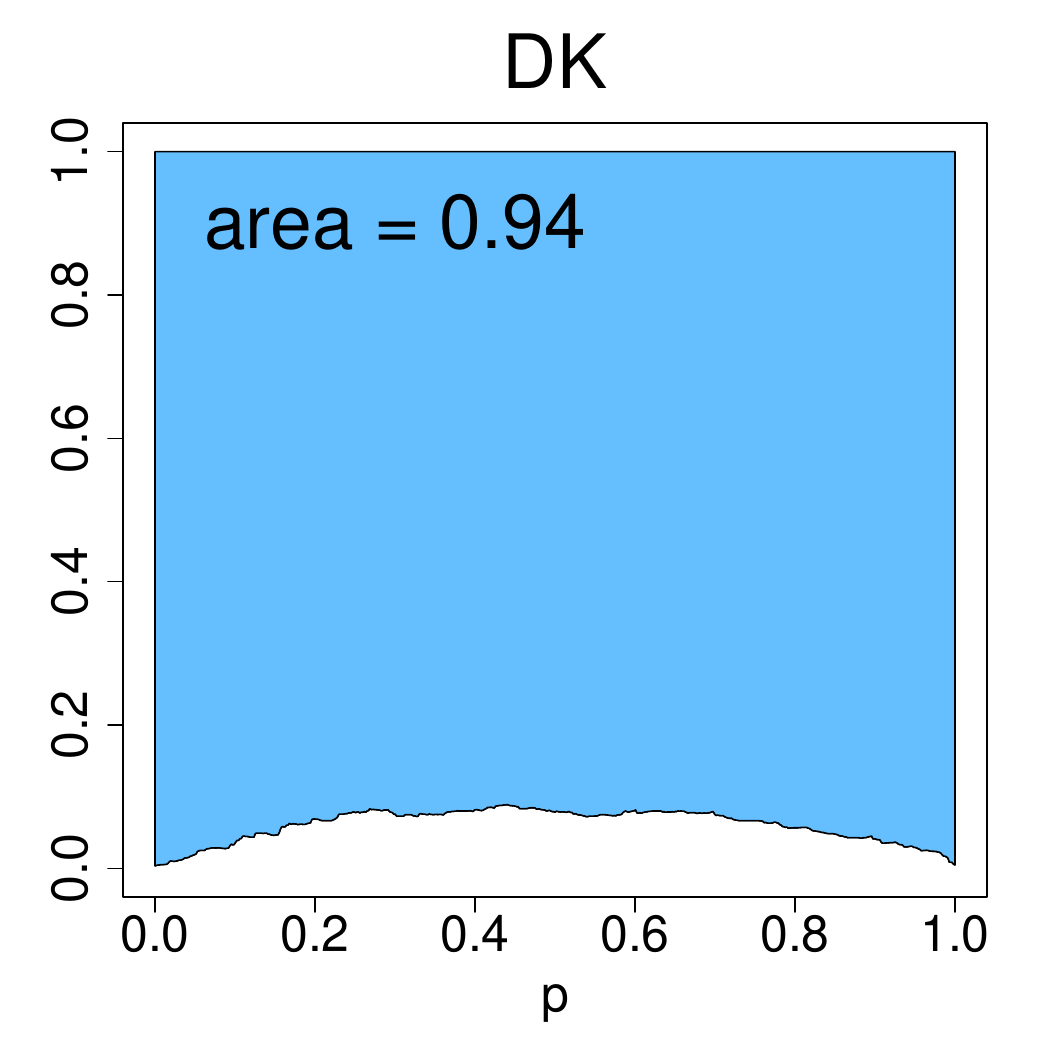}\qquad
\includegraphics[width=0.237\linewidth]{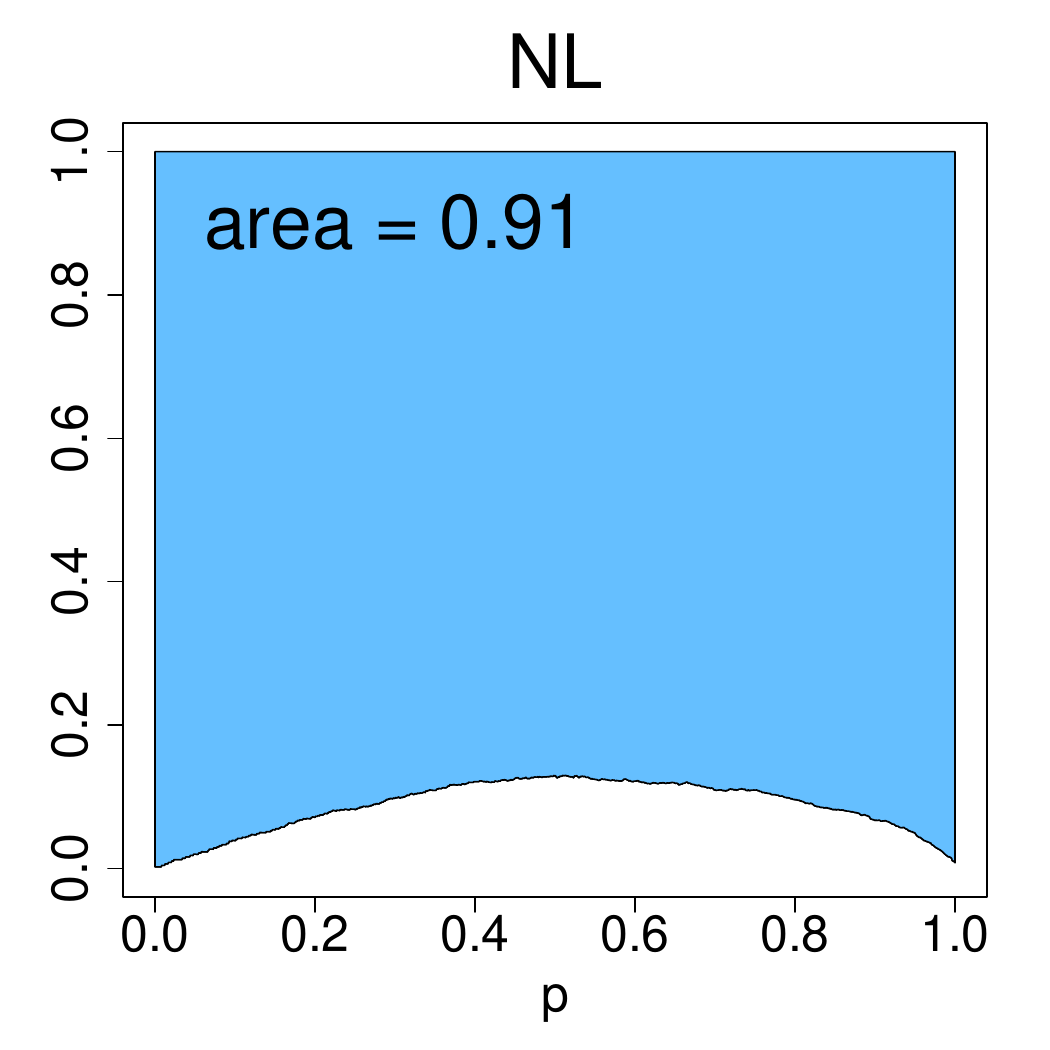}\qquad
\\
\includegraphics[width=0.237\linewidth]{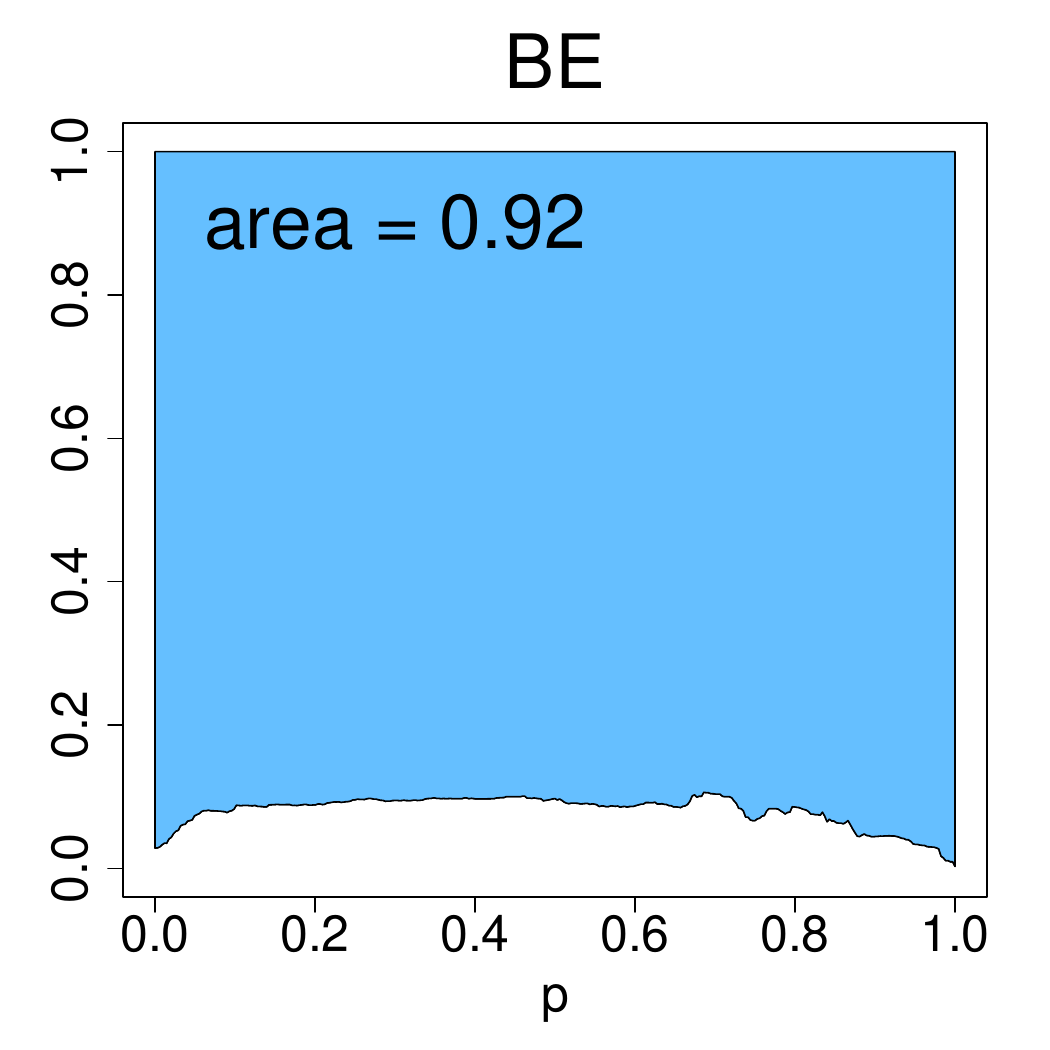}\qquad
\includegraphics[width=0.237\linewidth]{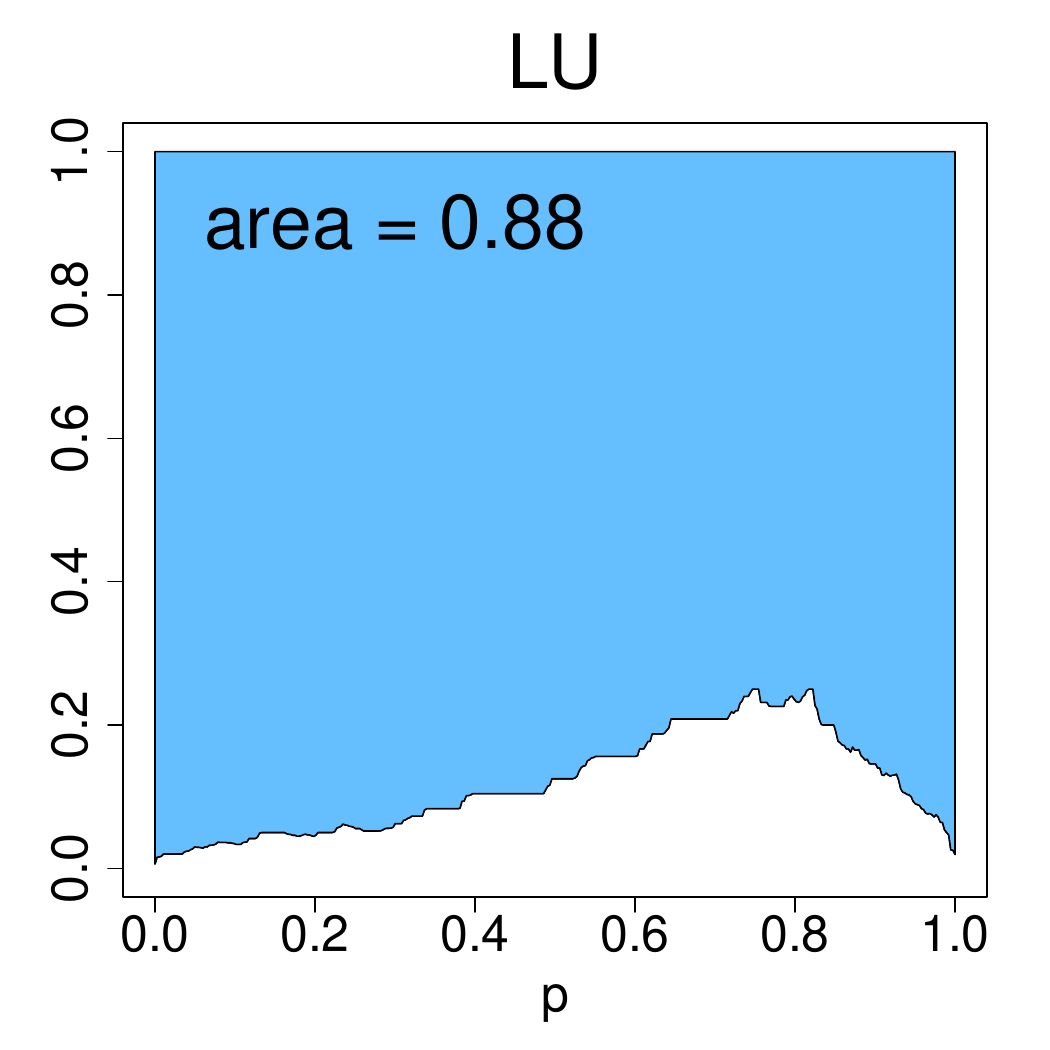}\qquad
\includegraphics[width=0.237\linewidth]{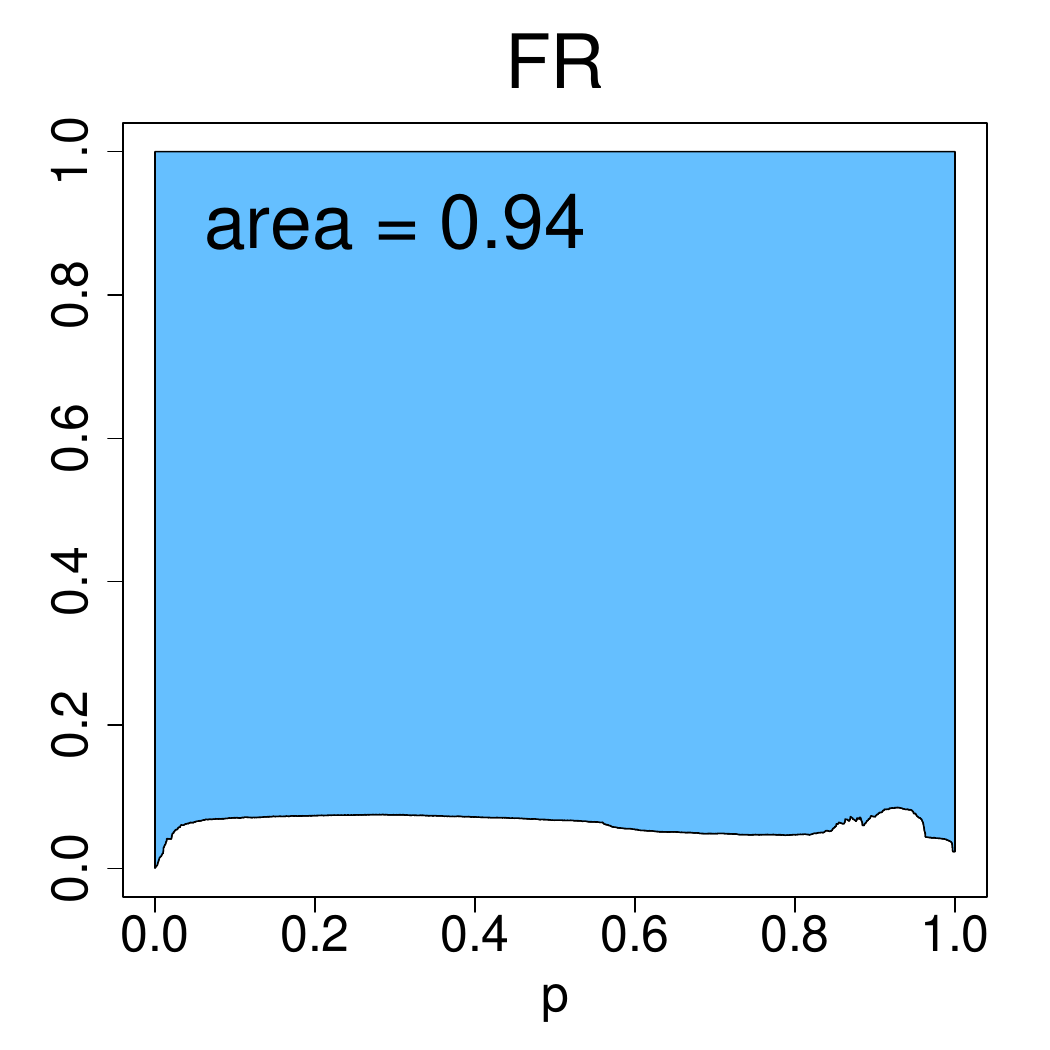}\qquad
\\
\includegraphics[width=0.237\linewidth]{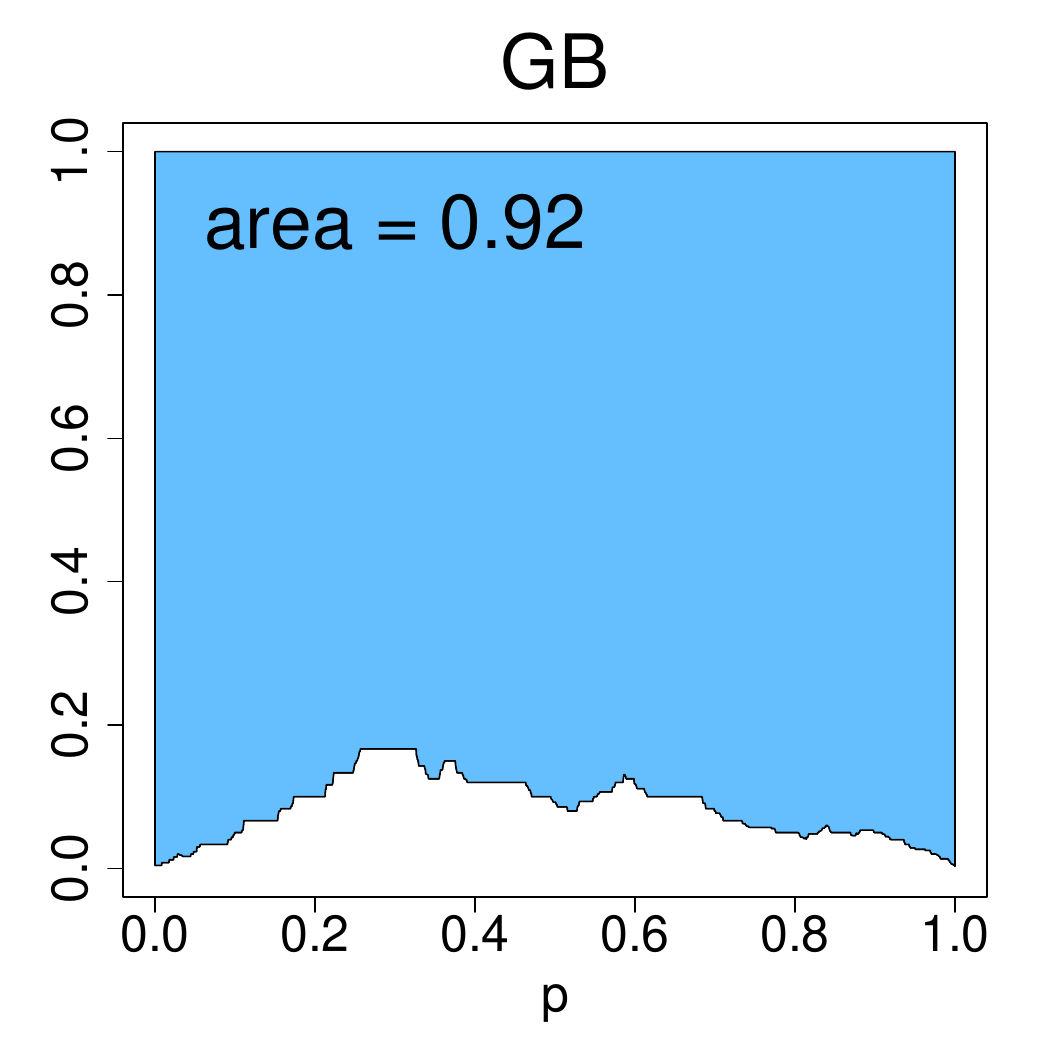}\qquad
\includegraphics[width=0.237\linewidth]{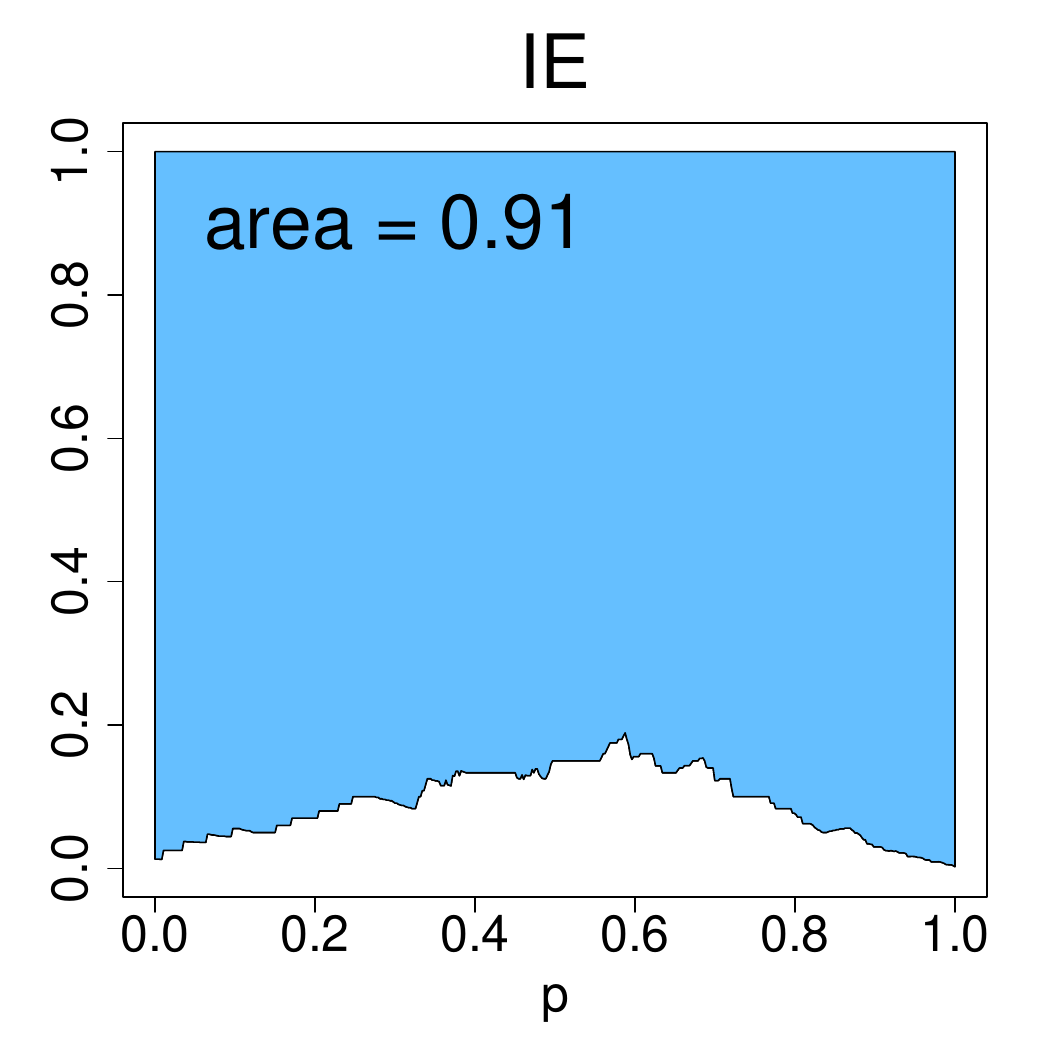}\qquad
\includegraphics[width=0.237\linewidth]{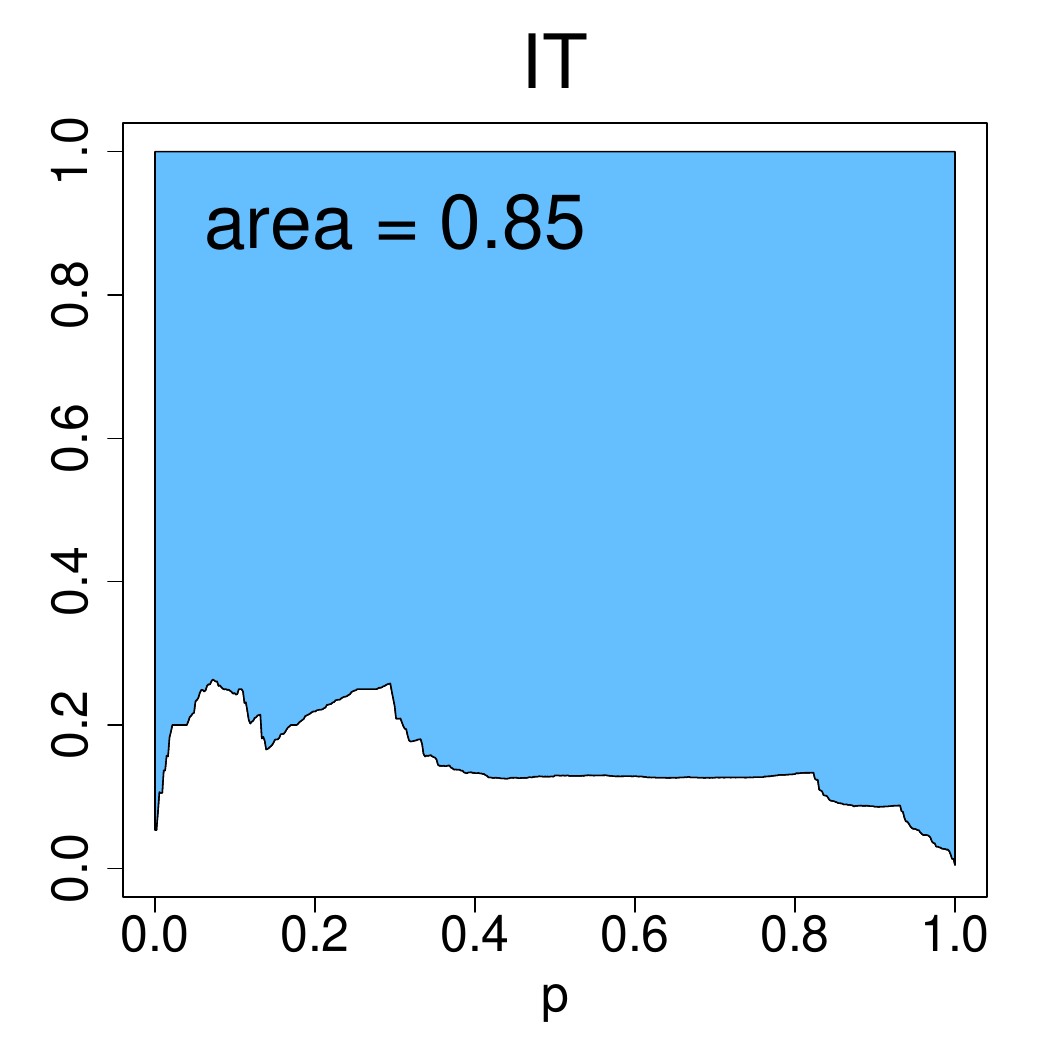}\qquad
\\
\includegraphics[width=0.237\linewidth]{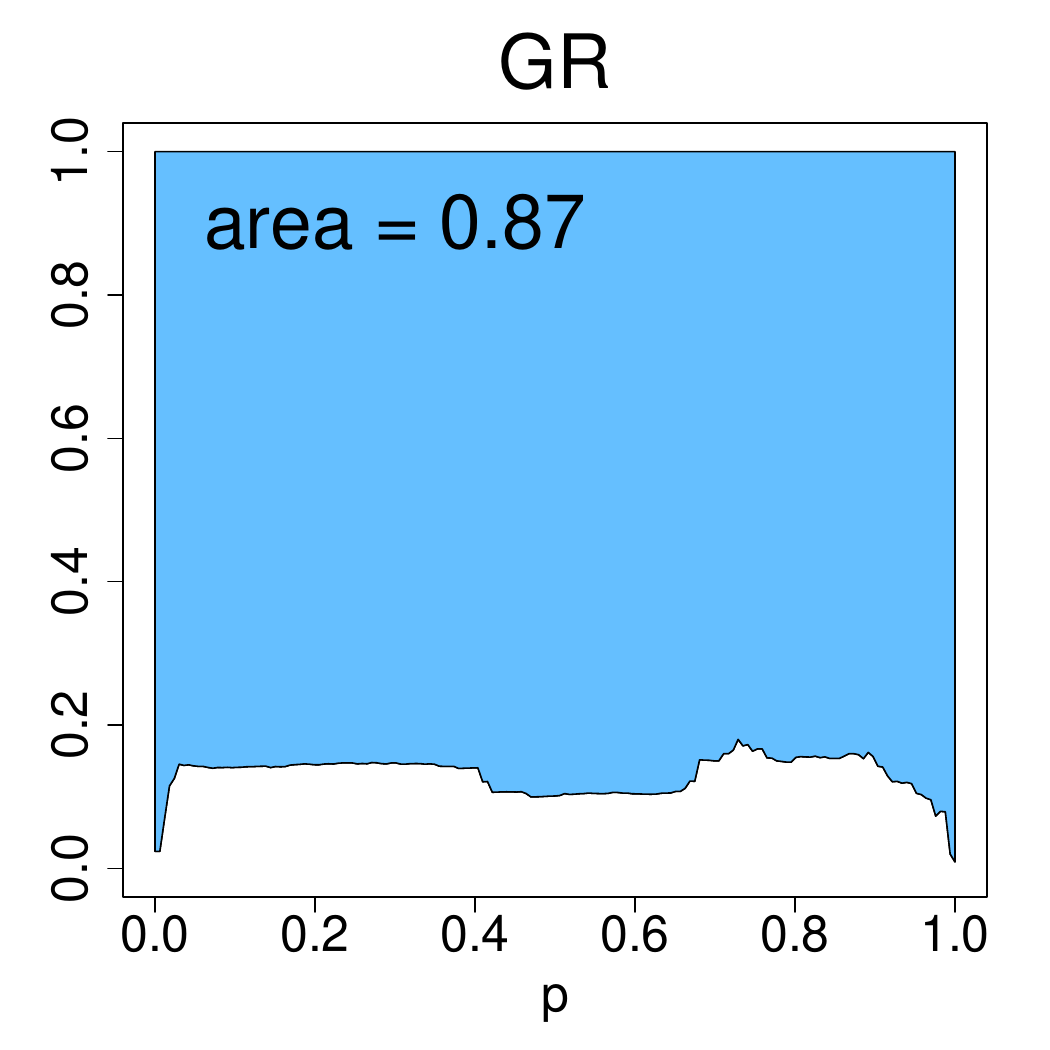}\qquad
\includegraphics[width=0.237\linewidth]{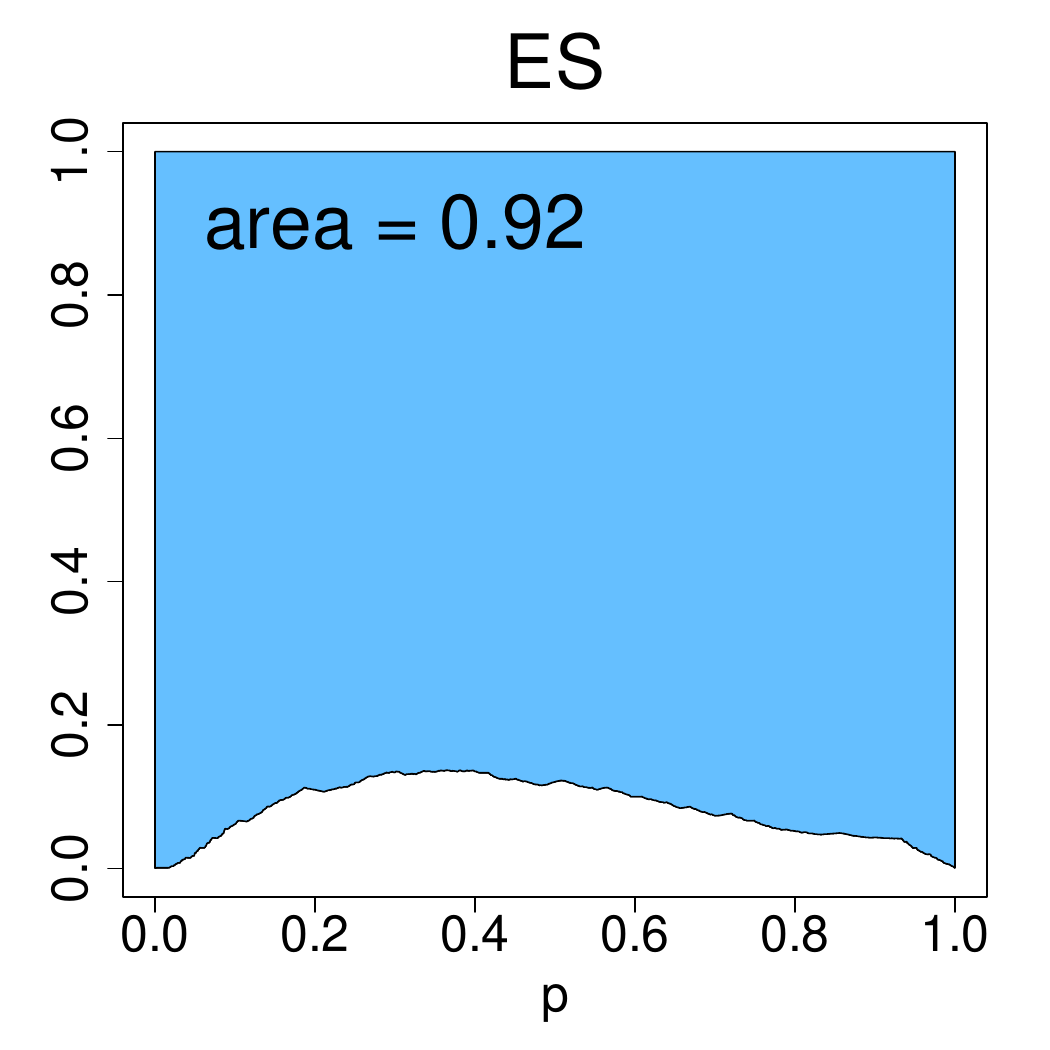}\qquad
\includegraphics[width=0.237\linewidth]{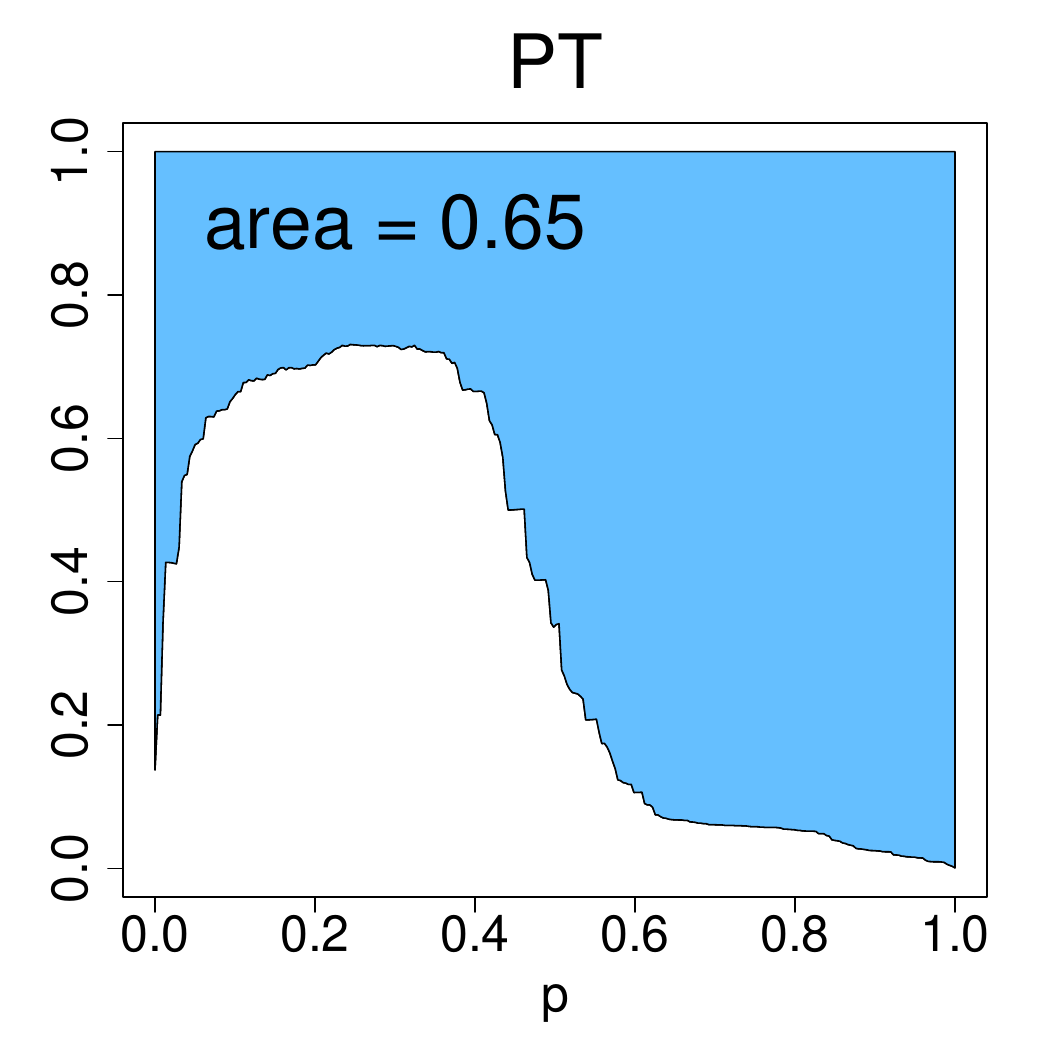}\qquad
\\
\includegraphics[width=0.237\linewidth]{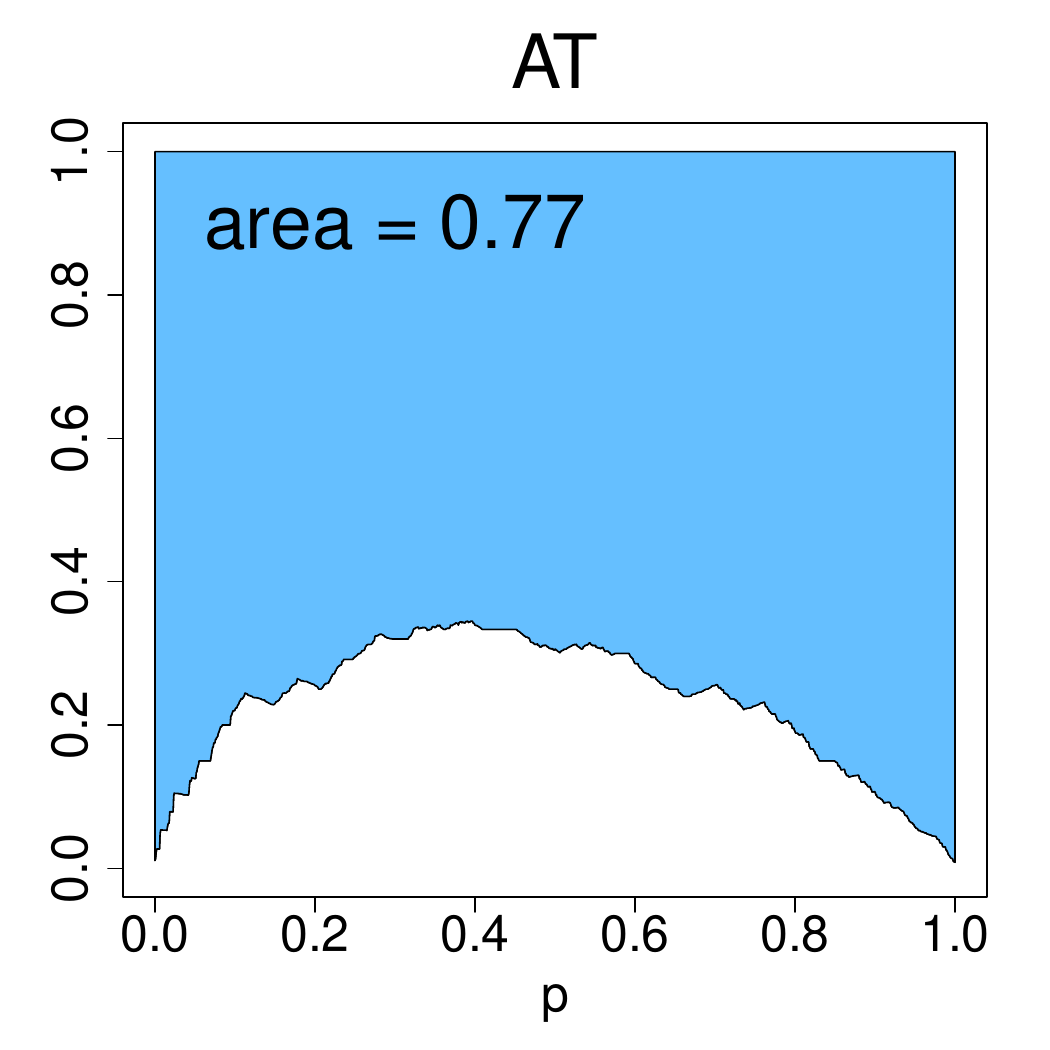}\qquad
\includegraphics[width=0.237\linewidth]{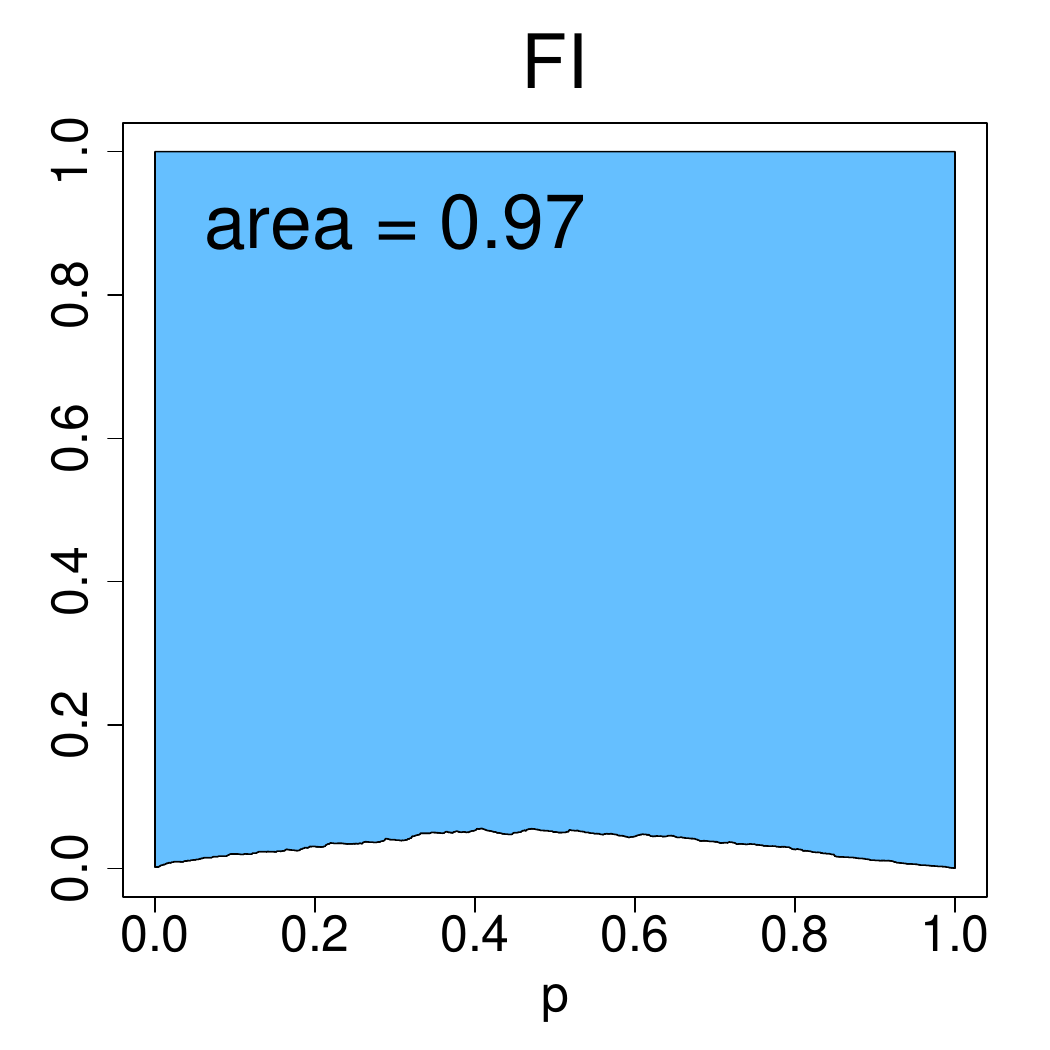}\qquad
\includegraphics[width=0.237\linewidth]{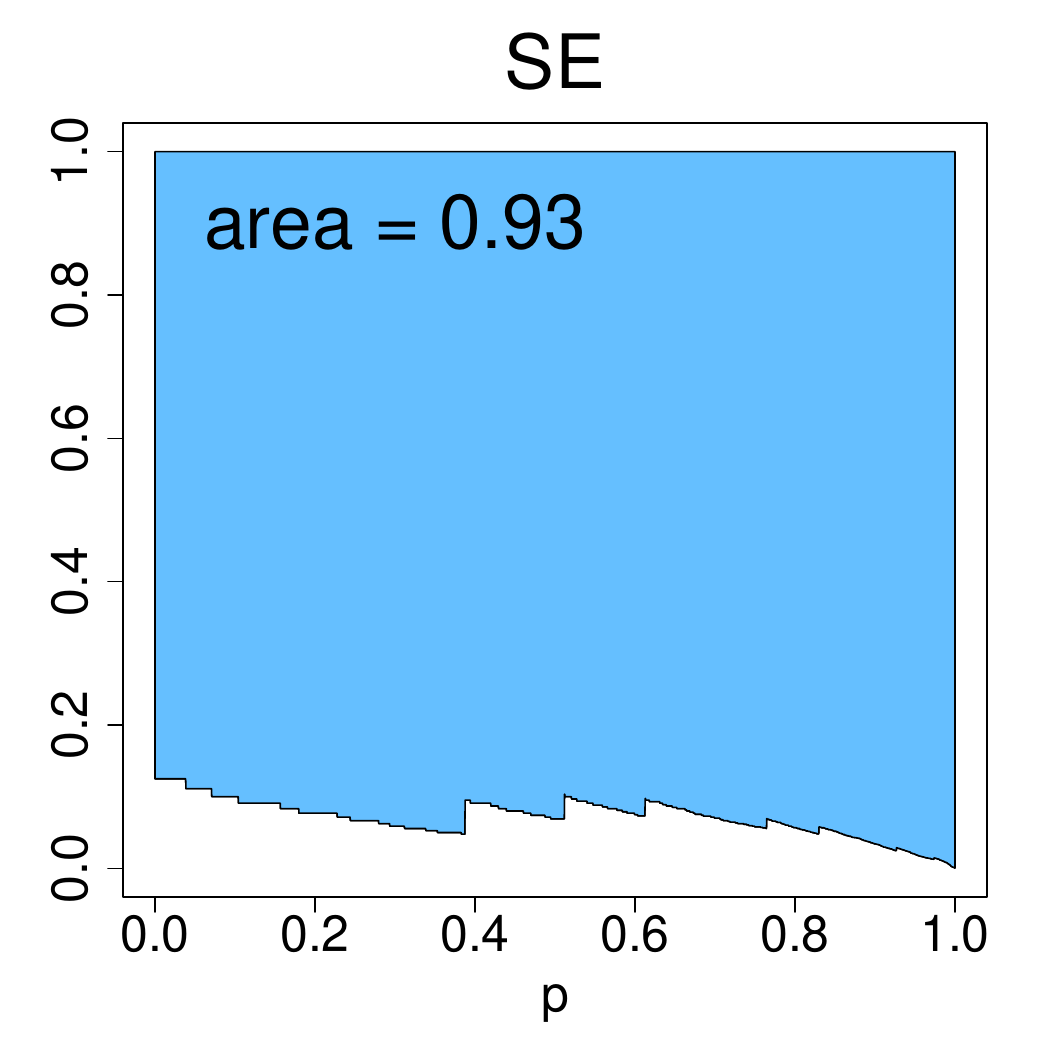}\qquad
\caption{The income-equality curve $\psi_{2,n}$ and the shaded-in area (i.e., $\Psi_{2,n}$) above it for the fifteen European countries, where $n=n_P$ is specified in Table~\ref{tab-real} \citep[based on][]{echp2001}.}
\label{fig:2}
\end{center}
\end{figure}

\begin{figure}[h!]
\begin{center}
\includegraphics[width=0.237\linewidth]{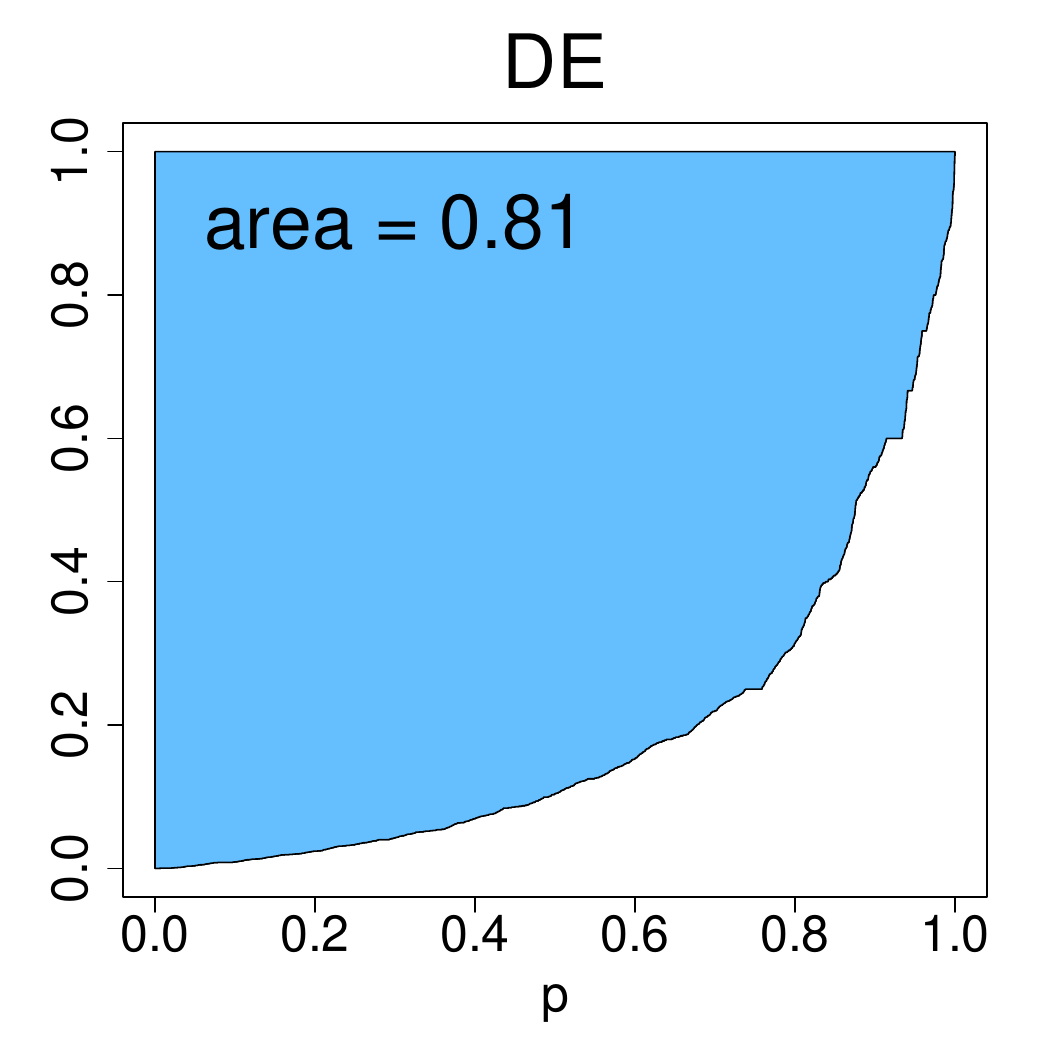}\qquad
\includegraphics[width=0.237\linewidth]{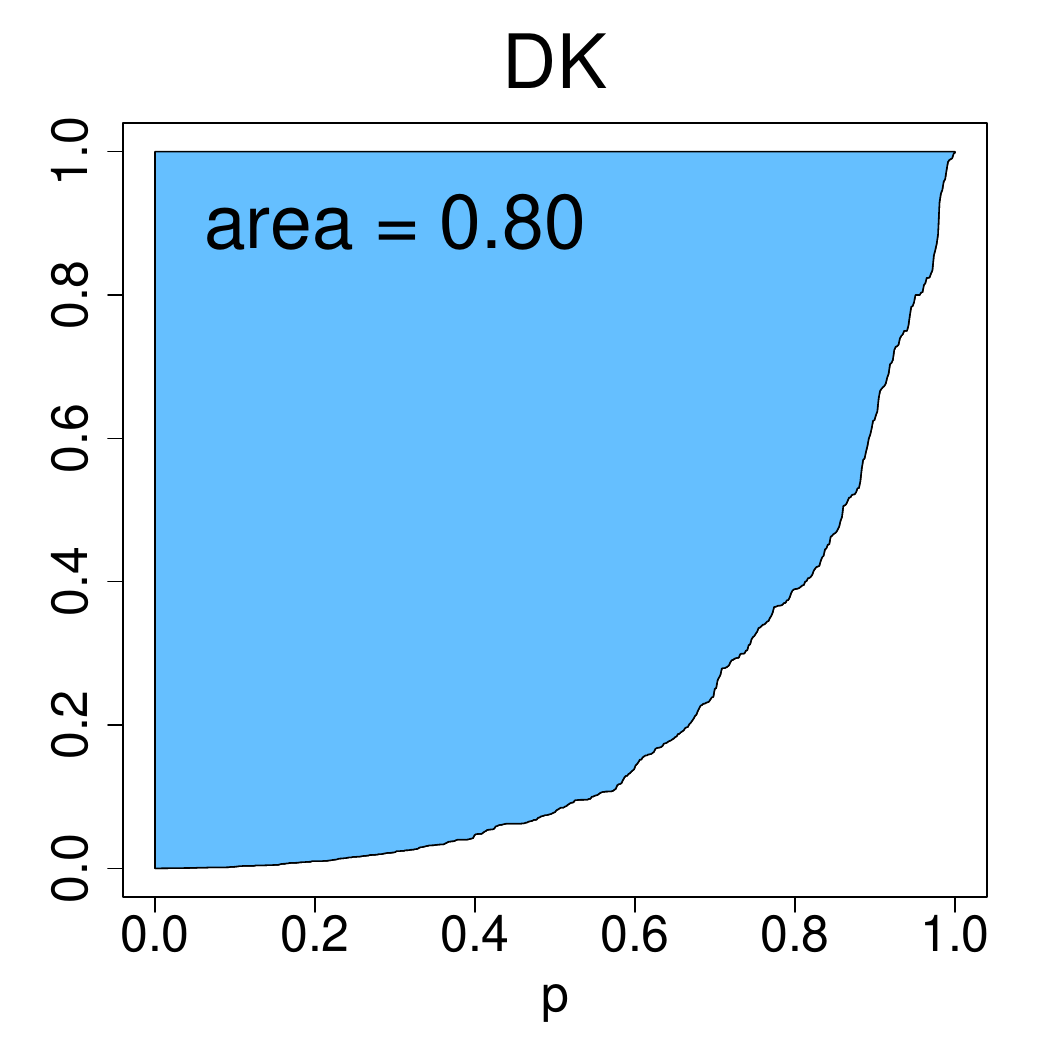}\qquad
\includegraphics[width=0.237\linewidth]{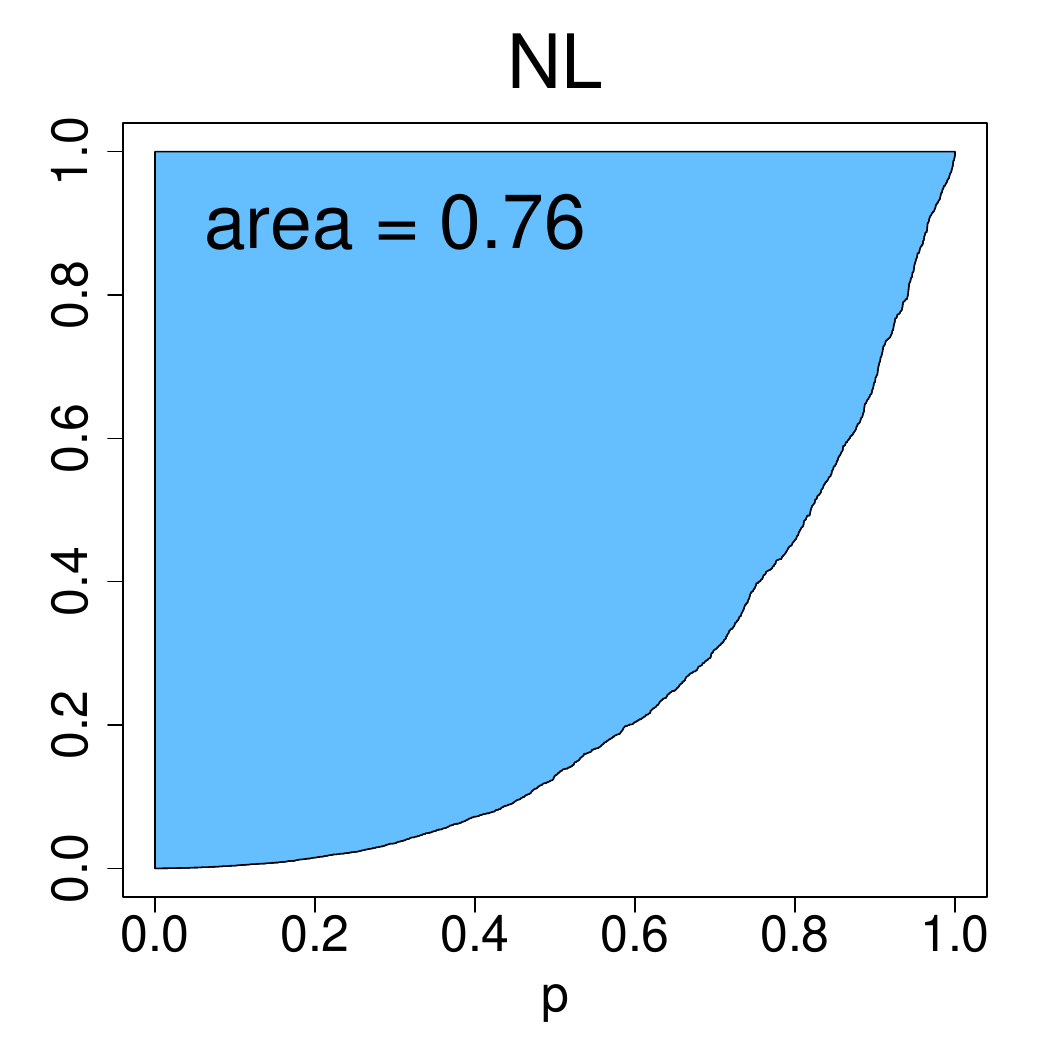}\qquad
\\
\includegraphics[width=0.237\linewidth]{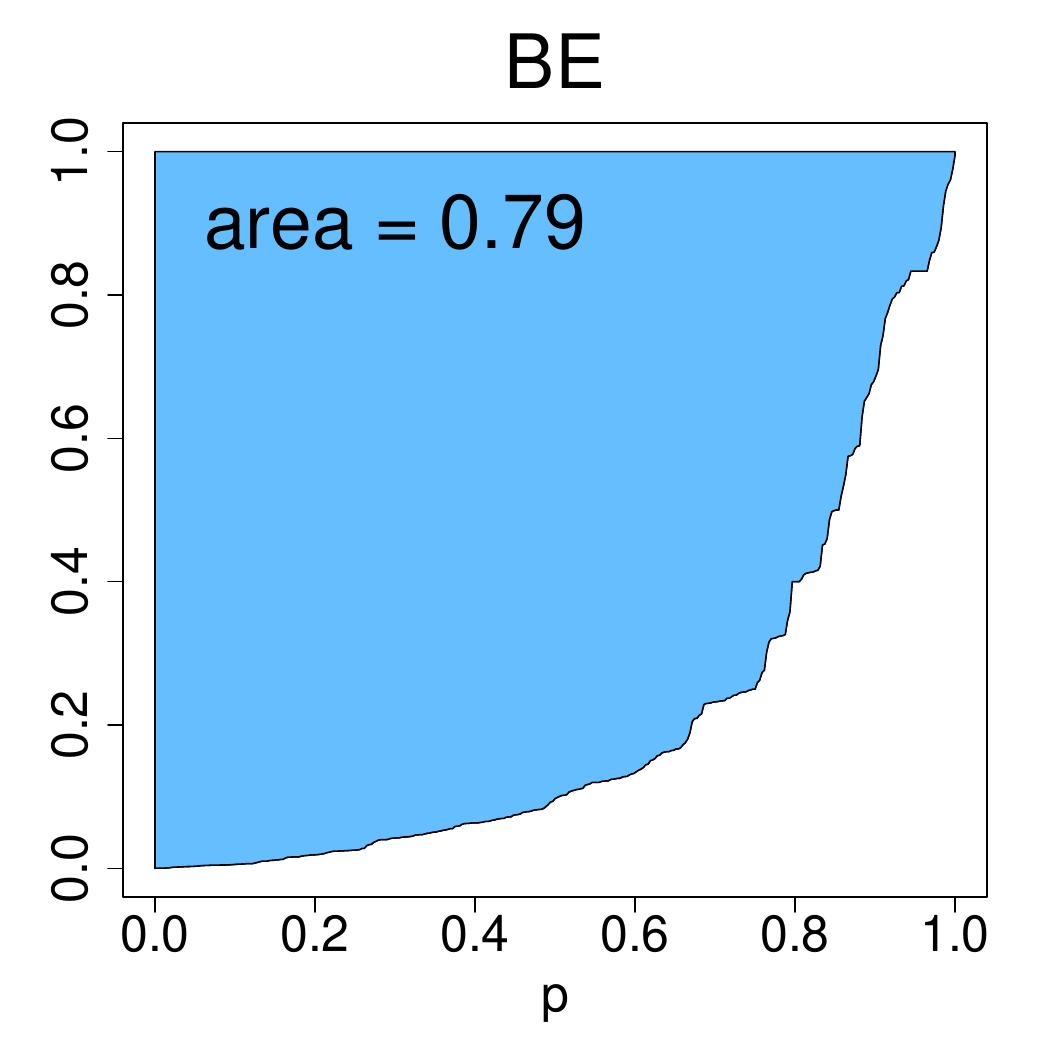}\qquad
\includegraphics[width=0.237\linewidth]{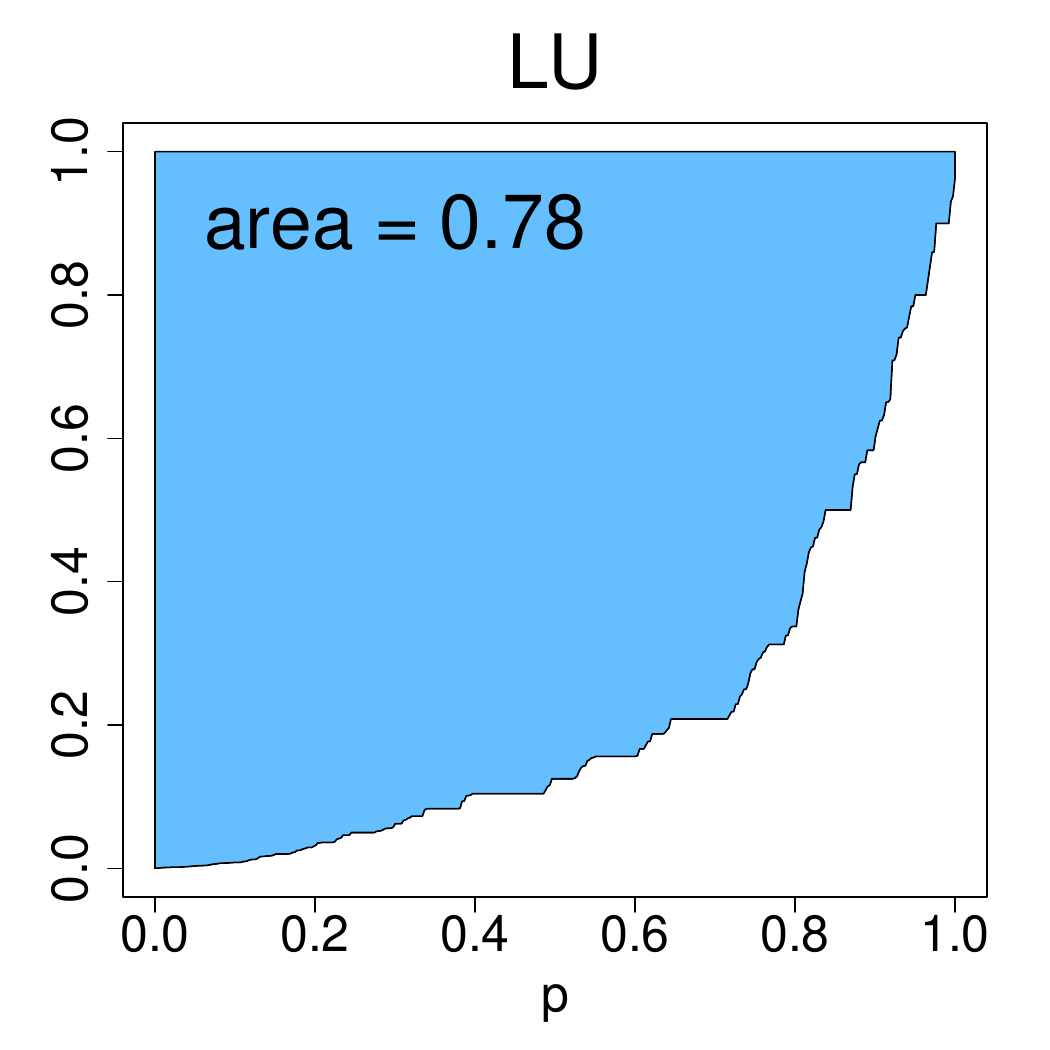}\qquad
\includegraphics[width=0.237\linewidth]{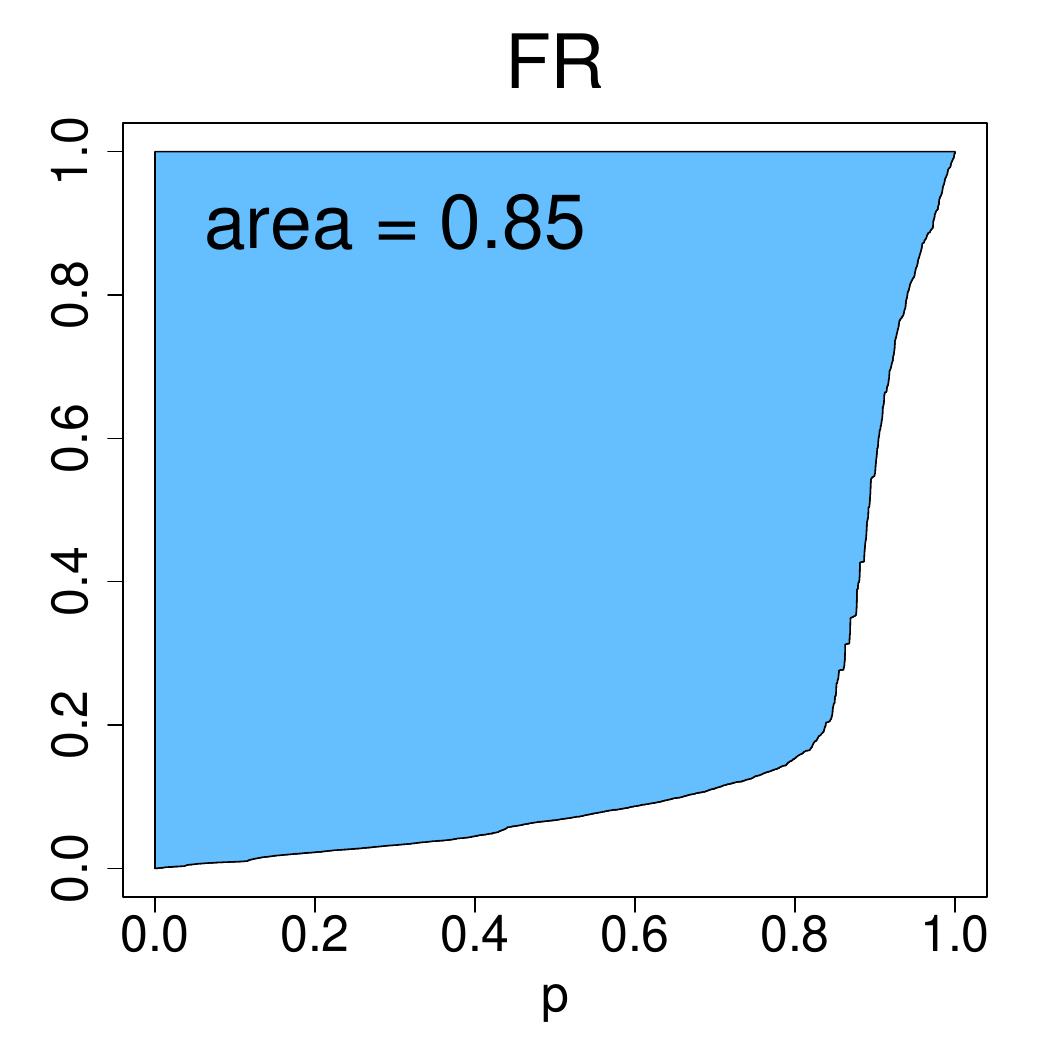}\qquad
\\
\includegraphics[width=0.237\linewidth]{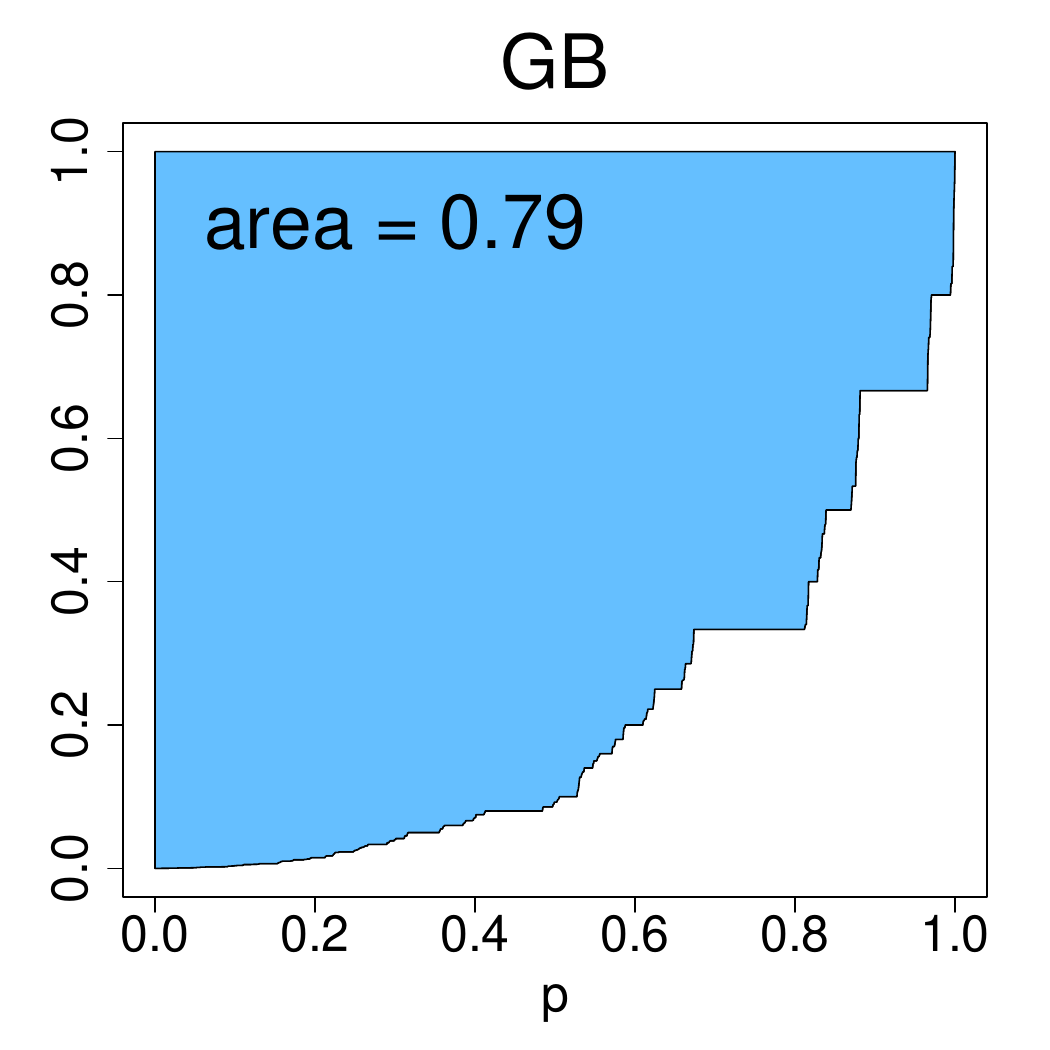}\qquad
\includegraphics[width=0.237\linewidth]{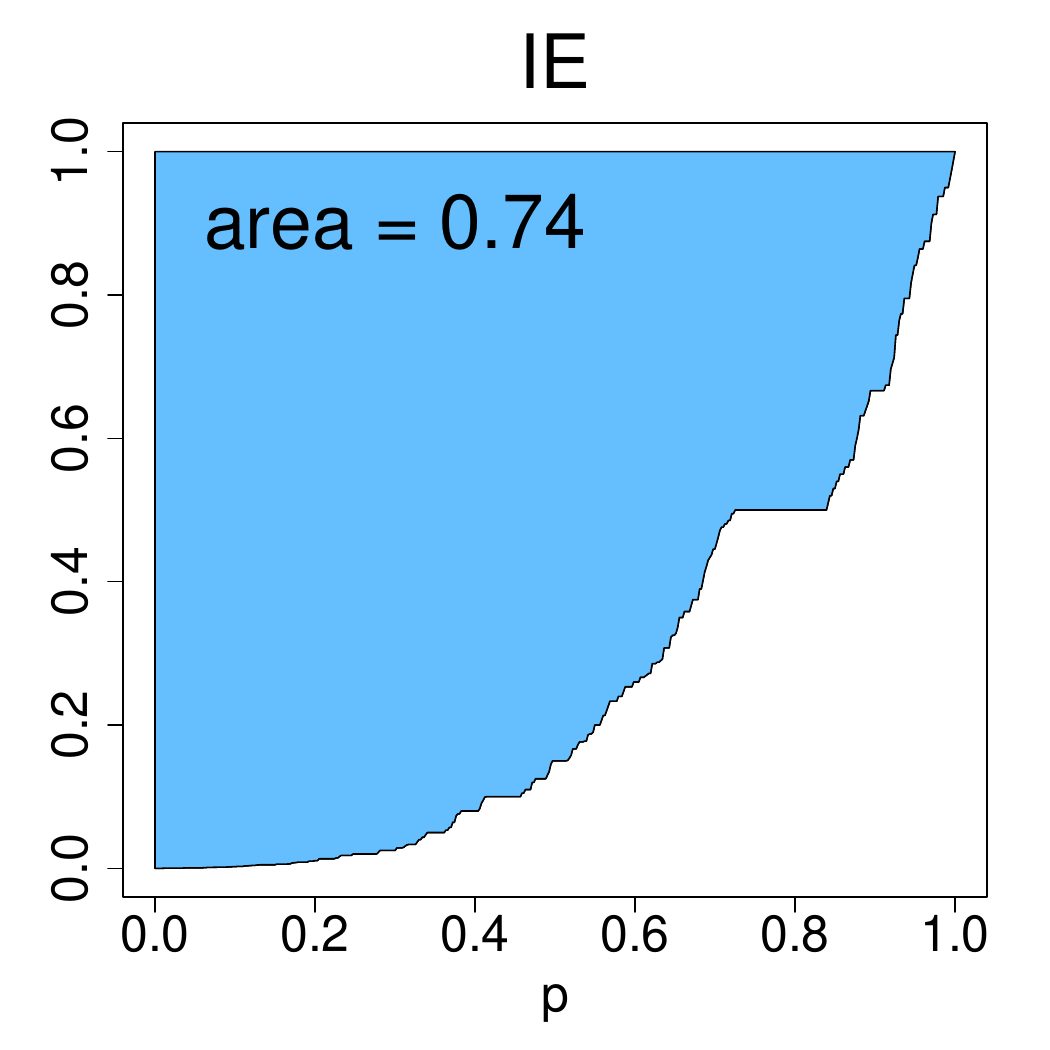}\qquad
\includegraphics[width=0.237\linewidth]{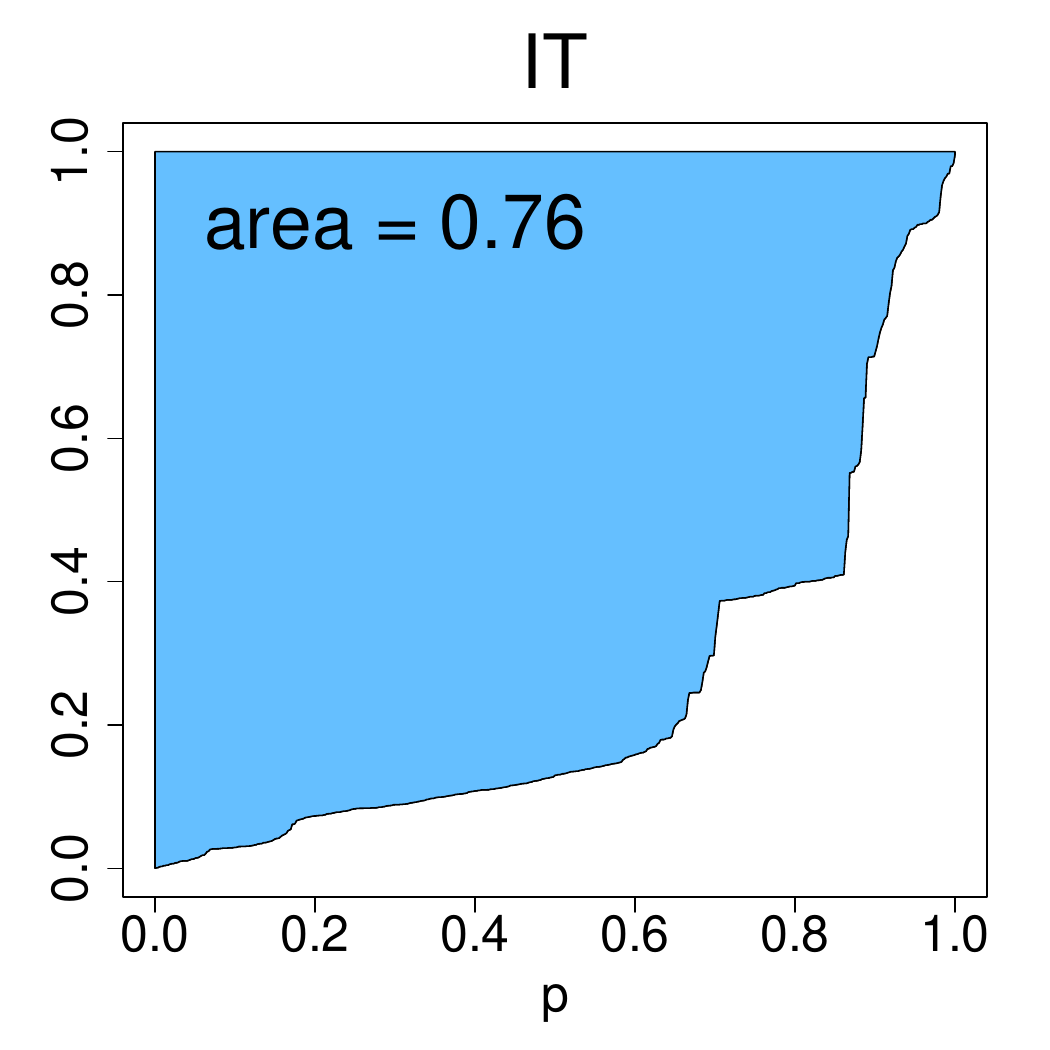}\qquad
\\
\includegraphics[width=0.237\linewidth]{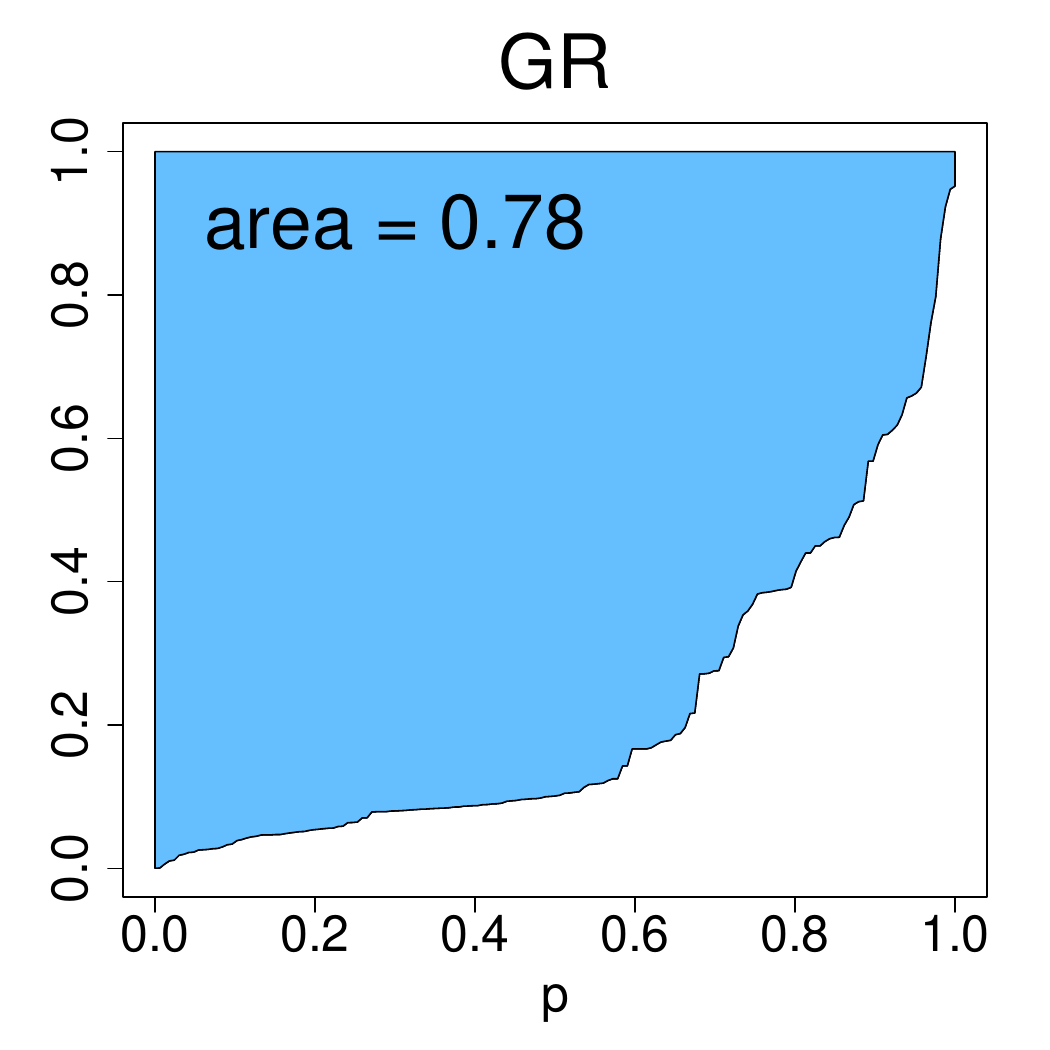}\qquad
\includegraphics[width=0.237\linewidth]{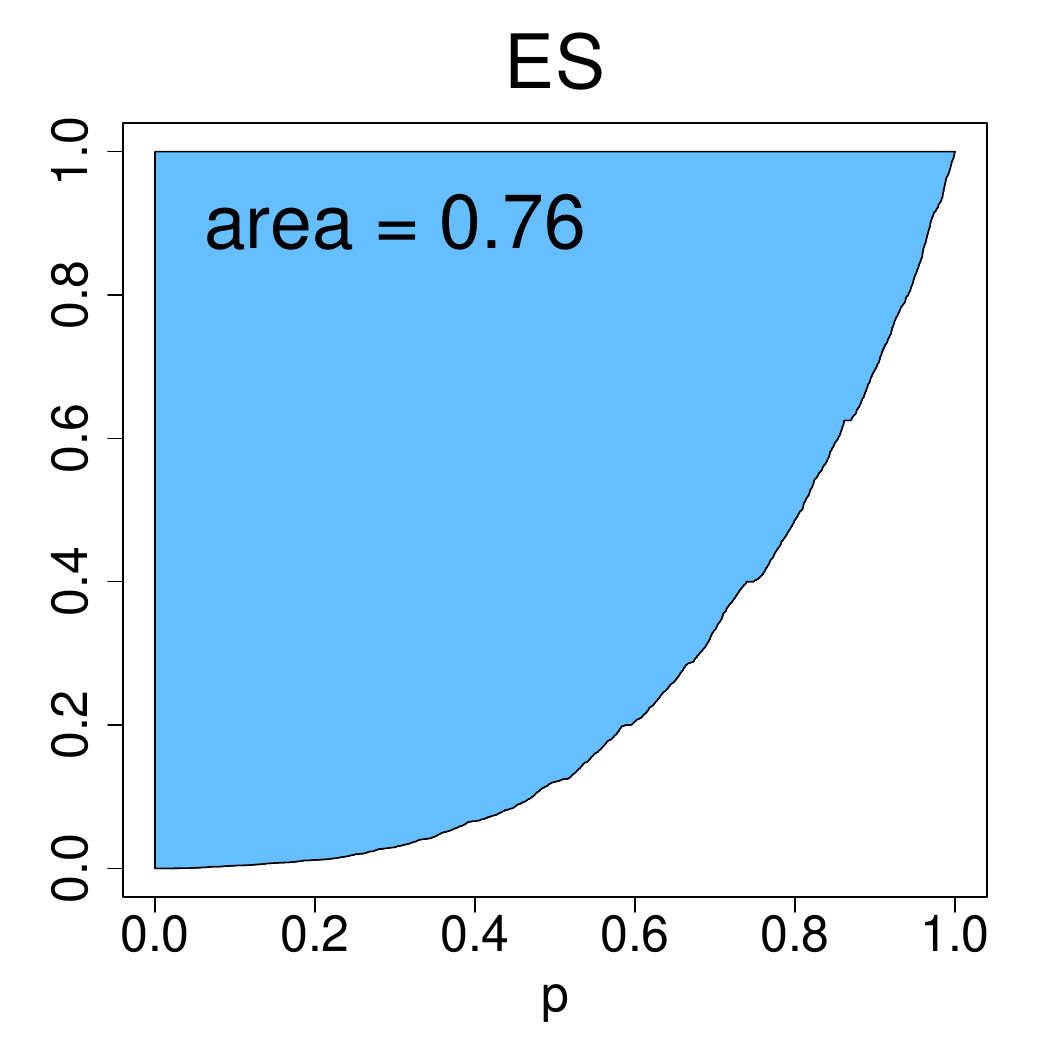}\qquad
\includegraphics[width=0.237\linewidth]{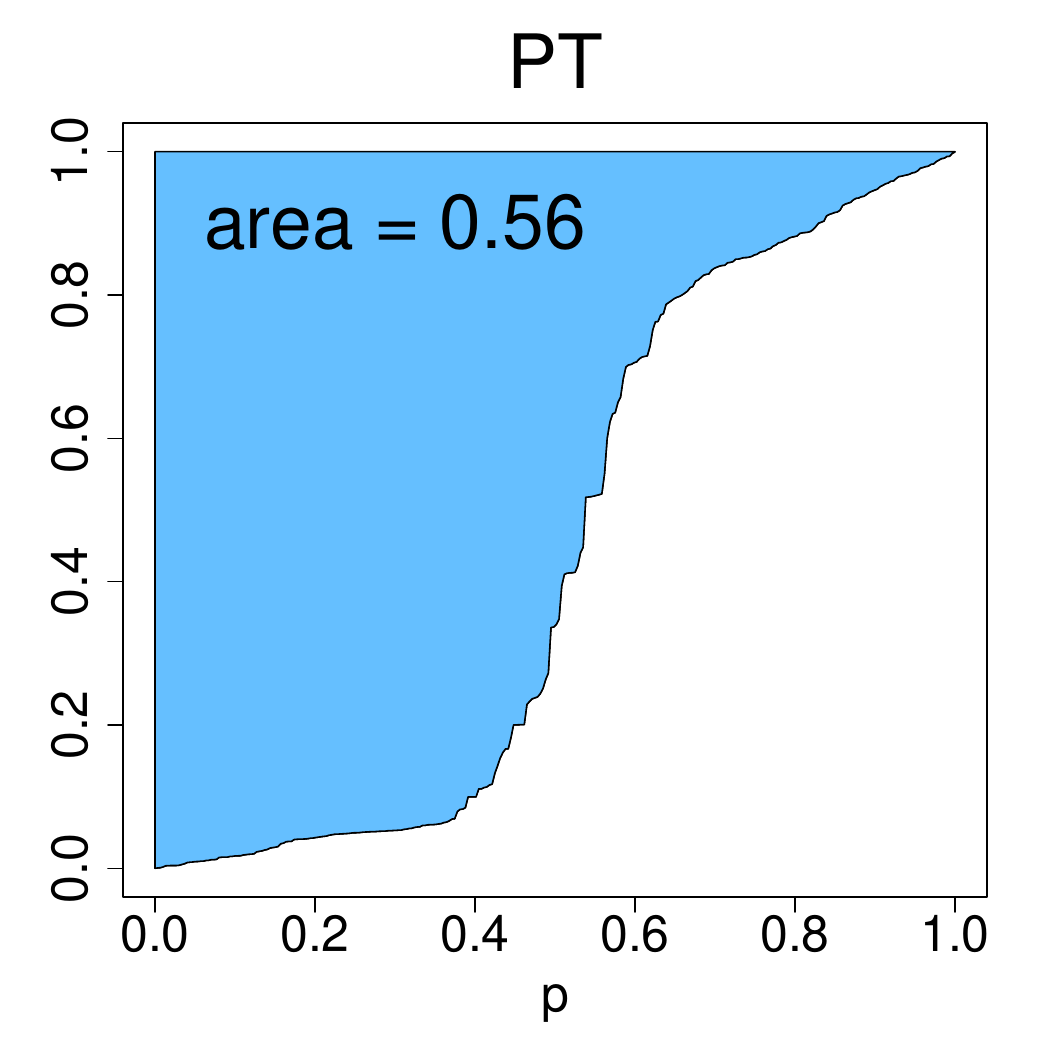}\qquad
\\
\includegraphics[width=0.237\linewidth]{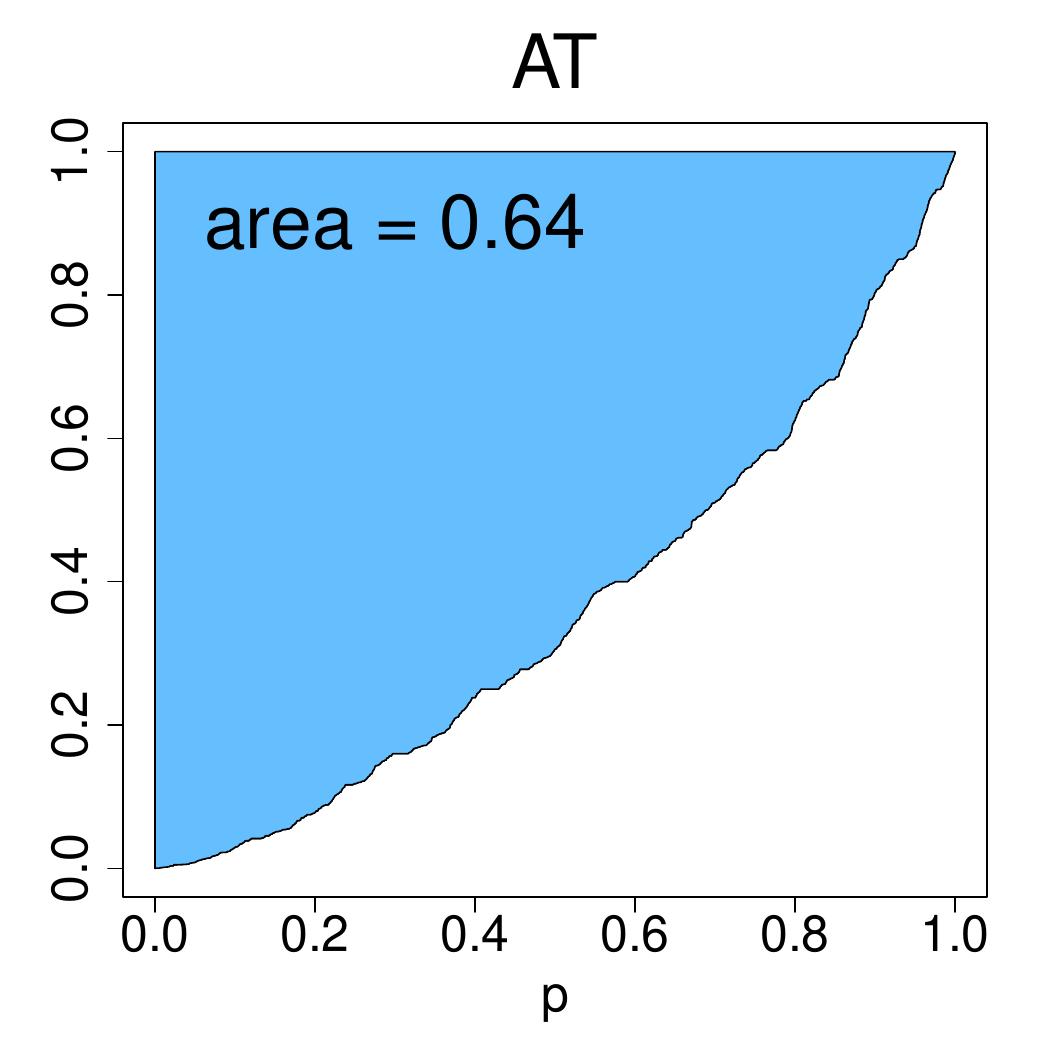}\qquad
\includegraphics[width=0.237\linewidth]{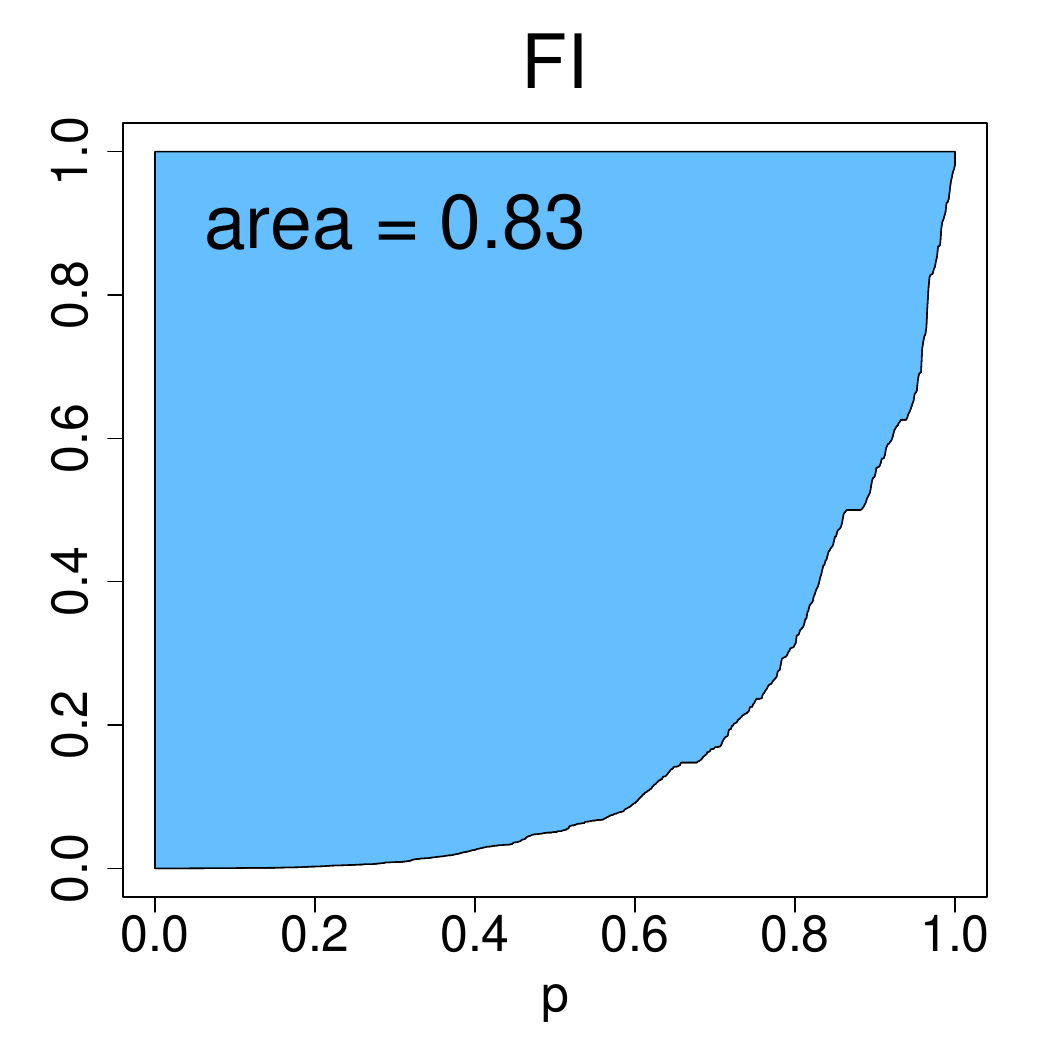}\qquad
\includegraphics[width=0.237\linewidth]{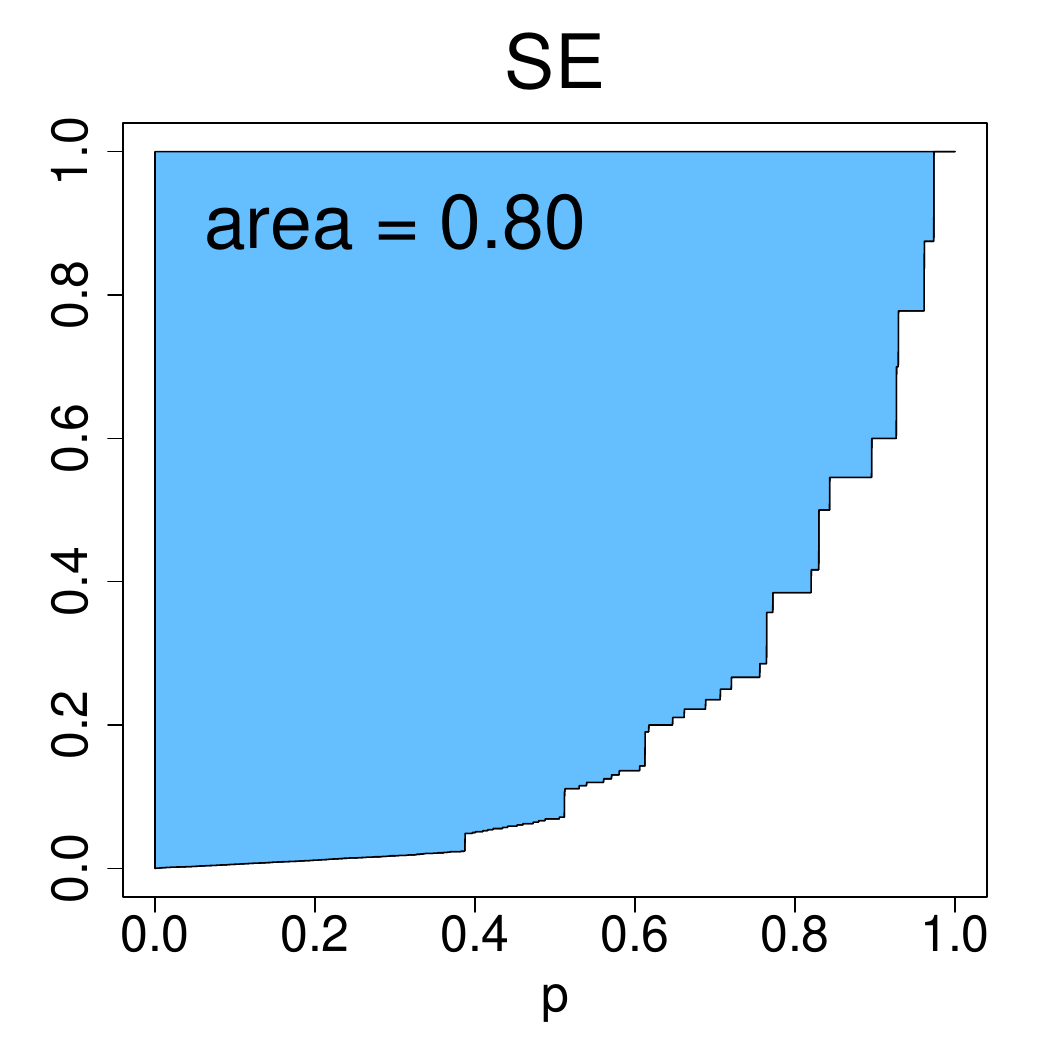}\qquad
\caption{The income-equality curve $\psi_{3,n}$ and the shaded-in area (i.e., $\Psi_{3,n}$) above it for the fifteen European countries, where $n=n_P$ is specified in Table~\ref{tab-real} \citep[based on][]{echp2001}.}
\label{fig:3}
\end{center}
\end{figure}

Figures~\ref{fig:1}--\ref{fig:3}
depict the three income-equality curves $\psi_{k,n}$  for the fifteen European countries specified in Table~\ref{tab-real}, with the shaded-in  areas above them depicting the values of the indices $\Psi_{k,n}$. The curves have been obtained via formulas~\eqref{curve-1}--\eqref{curve-3} by replacing $Q$ by $Q_n$ given by equation~\eqref{equantile function} with $n=n_P$, where $n_P$ is the number of people in the sample who possess capital incomes, and $n_T$ is the total sample size of the given country.

Comparing the plots of Figures~\ref{fig:1}--\ref{fig:3} derived from the actual data with the ones of Figures~\ref{fig-1}--\ref{fig-3} generated from the parametric distributions, for most of the countries we see that the distributions of capital incomes are similar to Pareto. The only exception is Portugal, where the three equality curves $\psi_{k,n}$ behave differently: having found no apparent correspondence with any of the parametric models of Section~\ref{parametrics}, the histogram of Portugal suggests a bimodal distribution. For all the other countries, the histograms are strongly skewed  with strictly decreasing bars when viewing from left to right, thus following the familiar $J$-shape mimicking the power law of the Pareto density.

\subsection{A comparison with capital incomes from the EU-SILC 2018 survey}

To get an insight into more recent European situation, we further analyse data coming from the EU Statistics on Income and Living Conditions survey \citep{EU-SILC2018}, which substituted the ECHP survey after its eighth wave in 2001.

We note at the outset that in the EU-SILC survey, the capital incomes are available only at the level of households, and sample sizes are approximately seven times larger if compared with the earlier ECHP survey. Hence, the EU-SILC data give rise to more accurate estimates. In our study we use the following variables: 
\begin{description}
\item[\rm HY040G:] income from rental of a property or land.
\item[\rm HY090G:] interests, dividends, profit from capital investments in unincorporated business.
\item[\rm PY080:] pensions received from individual private plans.
\end{description}
As the data refer to households, an equivalence scale needs to be employed to make meaningful comparisons of monetary incomes of social units with different numbers of inhabitants, and to also take into account the  economies of scale (within each household) with regard to the consumption of certain goods. An equivalence scale acts as a weight, giving rise to an \textit{equivalence income} that can be used for inequality, poverty and welfare analyses. We opt for the modified Organization for Economic Cooperation and Development (OECD) equivalence scale, which gives weight 1 to the household head, 0.5 to the other adult members of the household, and 0.3 to the members under 14 years of age.

We analyse the same  fifteen European countries as in previous Section~\ref{survey2001}, and consider the 340,540 households surveyed by the EU-SILC in 2018.  A summary is provided in Table~\ref{tab-real2018}. 
\begin{sidewaystable}
\centering
\begin{tabular}{l|rr|rr|cccrccc|rrr}
  \hline
   Countries & Means & Medians  &\multicolumn{2}{c|}{Sample sizes} & \multicolumn{7}{c|}{Inequality indices} & \multicolumn{3}{c}{Ranks based on}
   \\
  &&& $n_{T}~~$ & $n_{P}~~$ & $G_{n}$ & $Z_{n}$ & $D_{n}$ & $G_{2,n}$ & $\Psi_{1,n}$ & $\Psi_{2,n}$ & $\Psi_{3,n}$ & $\Psi_{1,n}$ & $\Psi_{2,n}$ & $\Psi_{3,n}$ \\
  \hline
 DE & 1,515.759 & 147.000 & 25,784 & 20,332 & 0.845 & 0.922 & 0.980 & 8.711 & 0.549 & 0.940 & 0.762 &    3 &    4 &    3 \\ 
DK & 691.532 & 83.539 & 16,812 & 5,118 & 0.846 & 0.923 & 0.983 & 7.003 & 0.746 & 0.978 & 0.858 &   12 &   10 &    9 \\ 
NL & 1,542.372 & 85.333 & 24,986 & 19,192 & 0.914 & 0.957 & 0.991 & 16.521 & 0.644 & 0.955 & 0.806 &    6 &    5 &    5 \\ 
BE & 1,833.003 & 54.286 & 11,892 & 6,568 & 0.873 & 0.936 & 0.990 & 29.860 & 0.699 & 0.962 & 0.881 &    9 &    6 &   13 \\ 
 LU & 3,193.057 & 124.500 & 7,666 & 4,400 & 0.853 & 0.927 & 0.988 & 21.883 & 0.627 & 0.972 & 0.822 &    5 &    9 &    7 \\ 
FR & 4,300.964 & 453.333 & 21,752 & 17,828 & 0.848 & 0.924 & 0.984 & 8.048 & 0.745 & 0.979 & 0.862 &   11 &   11 &   10 \\ 
GB & 2,811.430 & 442.439 & 34,226 & 15,090 & 0.788 & 0.894 & 0.971 & 5.010 & 0.695 & 0.963 & 0.842 &    8 &    8 &    8 \\ 
 IE & 4,653.139 & 1,080.000 & 8,764 & 1,678 & 0.754 & 0.877 & 0.968 & 3.245 & 0.823 & 0.983 & 0.891 &   15 &   13 &   14 \\ 
 IT & 2,004.340 & 266.667 & 42,346 & 22,188 & 0.808 & 0.904 & 0.976 & 6.075 & 0.663 & 0.963 & 0.822 &    7 &    7 &    6 \\ 
 GR & 3,216.821 & 1,966.815 & 48,610 & 7,512 & 0.579 & 0.781 & 0.892 & 0.947 & 0.598 & 0.866 & 0.710 &    4 &    1 &    2 \\ 
 ES & 2,132.438 & 264.200 & 26,736 & 13,246 & 0.806 & 0.903 & 0.978 & 6.504 & 0.739 & 0.980 & 0.865 &   10 &   12 &   11 \\ 
 PT & 2,266.447 & 694.447 & 27,434 & 5,516 & 0.703 & 0.849 & 0.941 & 2.292 & 0.528 & 0.910 & 0.705 &    1 &    2 &    1 \\ 
 AT & 1,699.386 & 103.740 & 12,206 & 8,598 & 0.877 & 0.938 & 0.987 & 14.358 & 0.543 & 0.934 & 0.765 &    2 &    3 &    4 \\ 
 FI & 3,525.164 & 203.167 & 19,664 & 16,008 & 0.854 & 0.927 & 0.988 & 14.831 & 0.809 & 0.993 & 0.900 &   14 &   15 &   15 \\ 
 SE & 312.698 & 33.170 & 11,662 & 9,138 & 0.836 & 0.918 & 0.983 & 7.880 & 0.761 & 0.985 & 0.868 &   13 &   14 &   12 \\ 
   \hline
\end{tabular}
    \caption{The income-inequality indices  $G_{n}$, $Z_{n}$,  $D_{n}$, $G_{2,n}$, and the new indices $\Psi_{1,n}$, $\Psi_{2,n}$, $\Psi_{3,n}$ for the fifteen European countries with $n=n_P$,  where $n_P$ is the number of people in the sample who possess capital incomes, and $n_T$ is the total sample size of the given country  \citep[based on][]{EU-SILC2018}.}
    \label{tab-real2018}
\end{sidewaystable}
For a useful comparison of means and medians, we apply the official average national currency exchange rates (year 2018) for the three countries that have not adopted the Euro: Denmark, Great Britain, and Sweden, whose currencies are the Danish Krone, the British Pound, and the Swedish Krona, respectively. Hence, all the analyzed data are in Euro. 
  
The  differences between the means and medians in Table~\ref{tab-real2018} facilitate the assessment of skewness of income distributions. The list of countries with lower inequality (having a two-digit rank in at least one of the new indices) is comprised of Denmark, Benelux, France, Ireland, Spain, Finland and Sweden. To compare with the 2001 data, Ireland has joined the list while Germany, Luxembourg, Great Britain and Greece left it. Portugal, that  was  the country with the highest inequality in 2001, in 2018 was joined by Greece in the list for the primacy of the highest inequality, as seen from the rankings produced by the three new indices.
Figure~\ref{fig-real} (with $G_{2,n}$ excluded due to its large values)
\begin{figure}[h!]
    \centering
    \includegraphics[width=\textwidth]{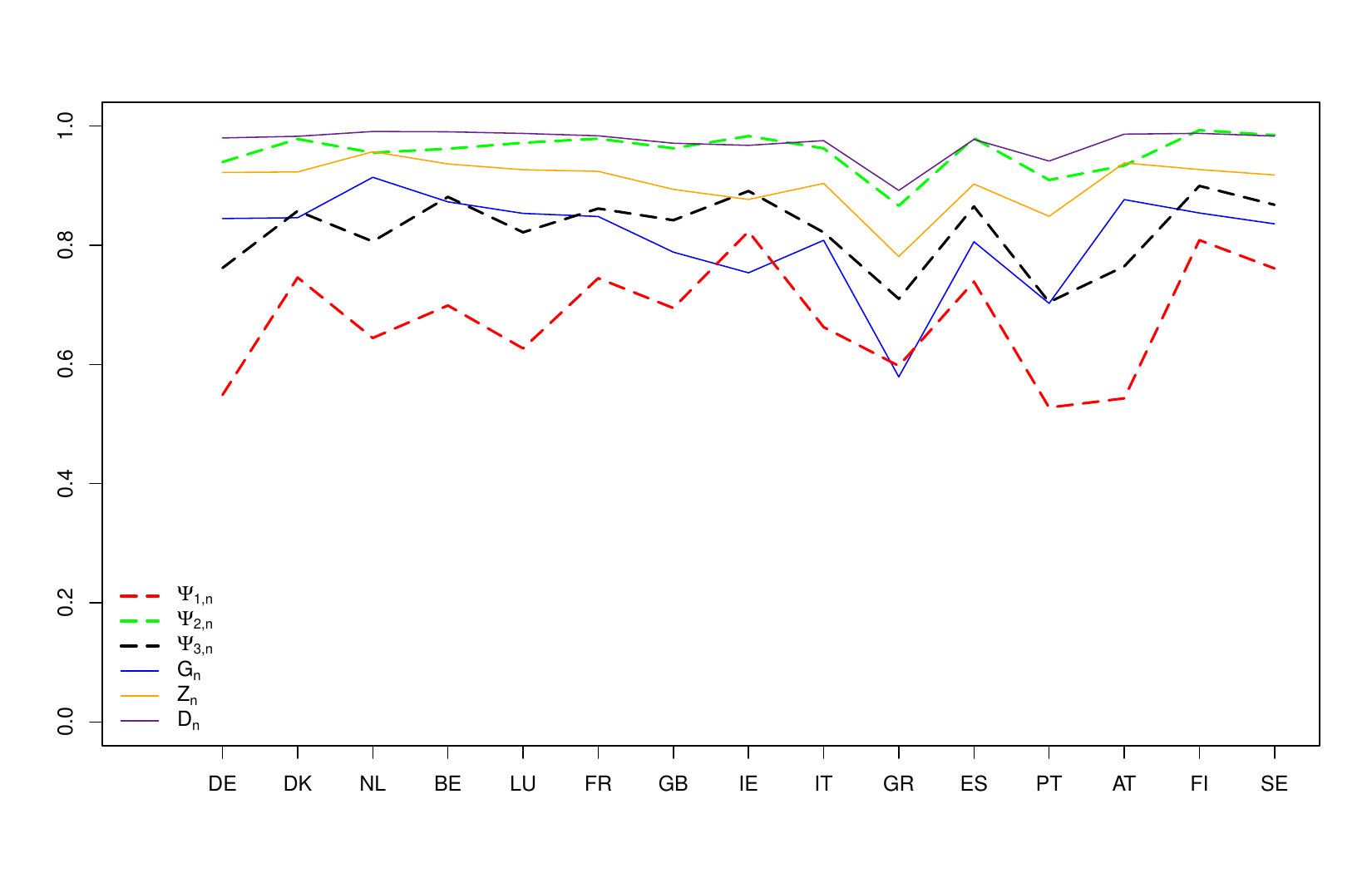}
    \caption{The income-inequality indices  $G_{n}$, $Z_{n}$,  $D_{n}$, and the new indices $\Psi_{k,n}$ for the fifteen European countries with $n=n_P$ specified in Table~\ref{tab-real2018} \citep[based on][]{EU-SILC2018}.}
    \label{fig-real2018}
\end{figure}
visualizes the index values calculated using formulas~\eqref{def-psi1}--\eqref{def-psi3} and reported in Table~\ref{tab-real2018}.

\begin{figure}[h!]
\begin{center}
\includegraphics[width=0.237\linewidth]{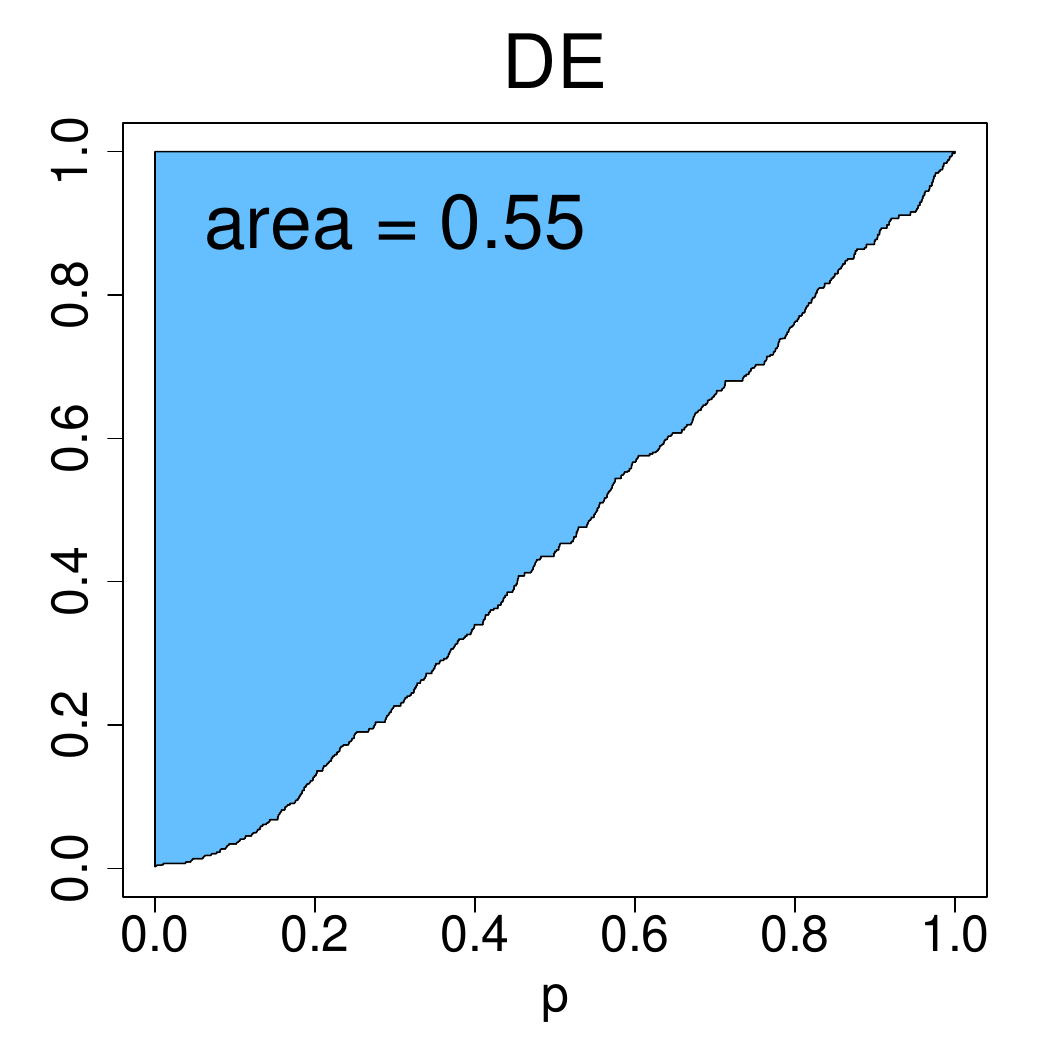}\qquad
\includegraphics[width=0.237\linewidth]{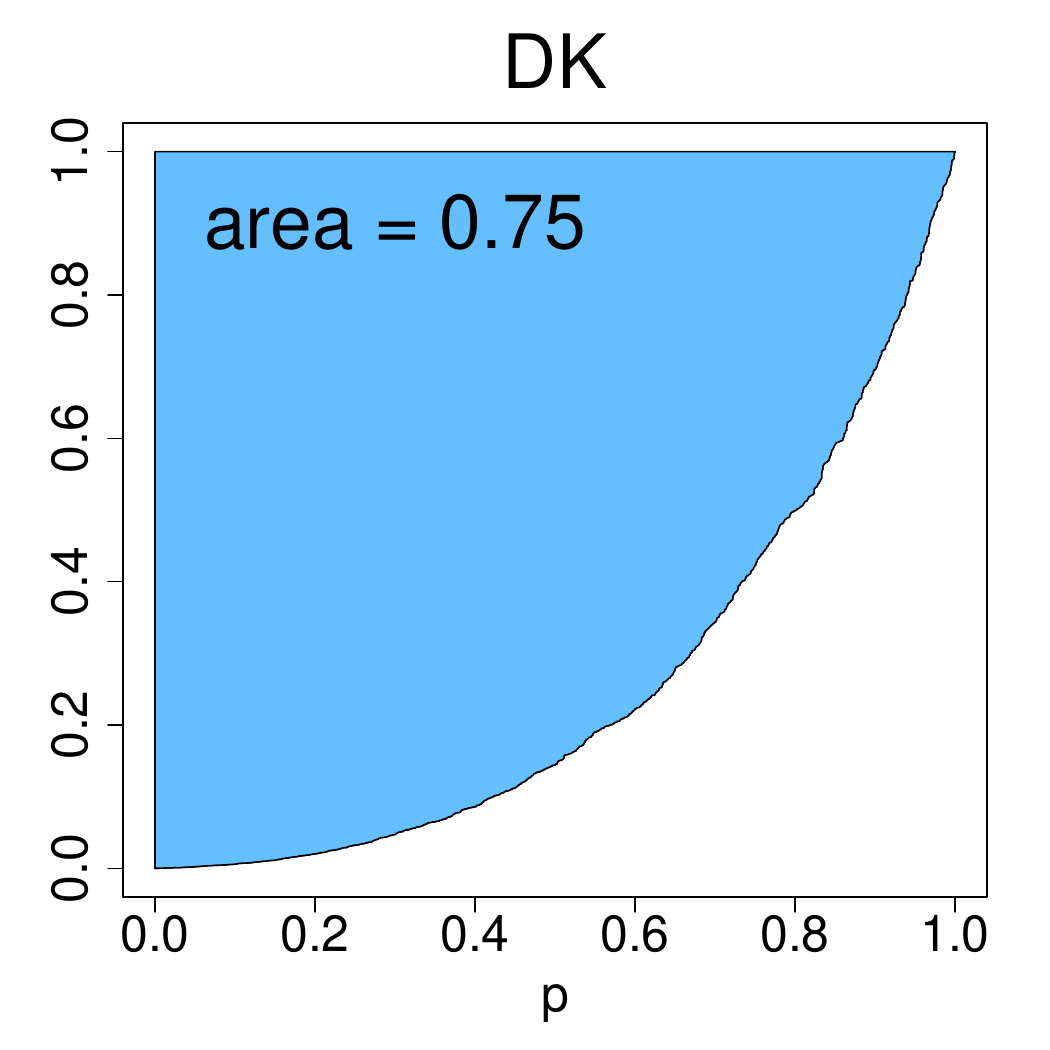}\qquad
\includegraphics[width=0.237\linewidth]{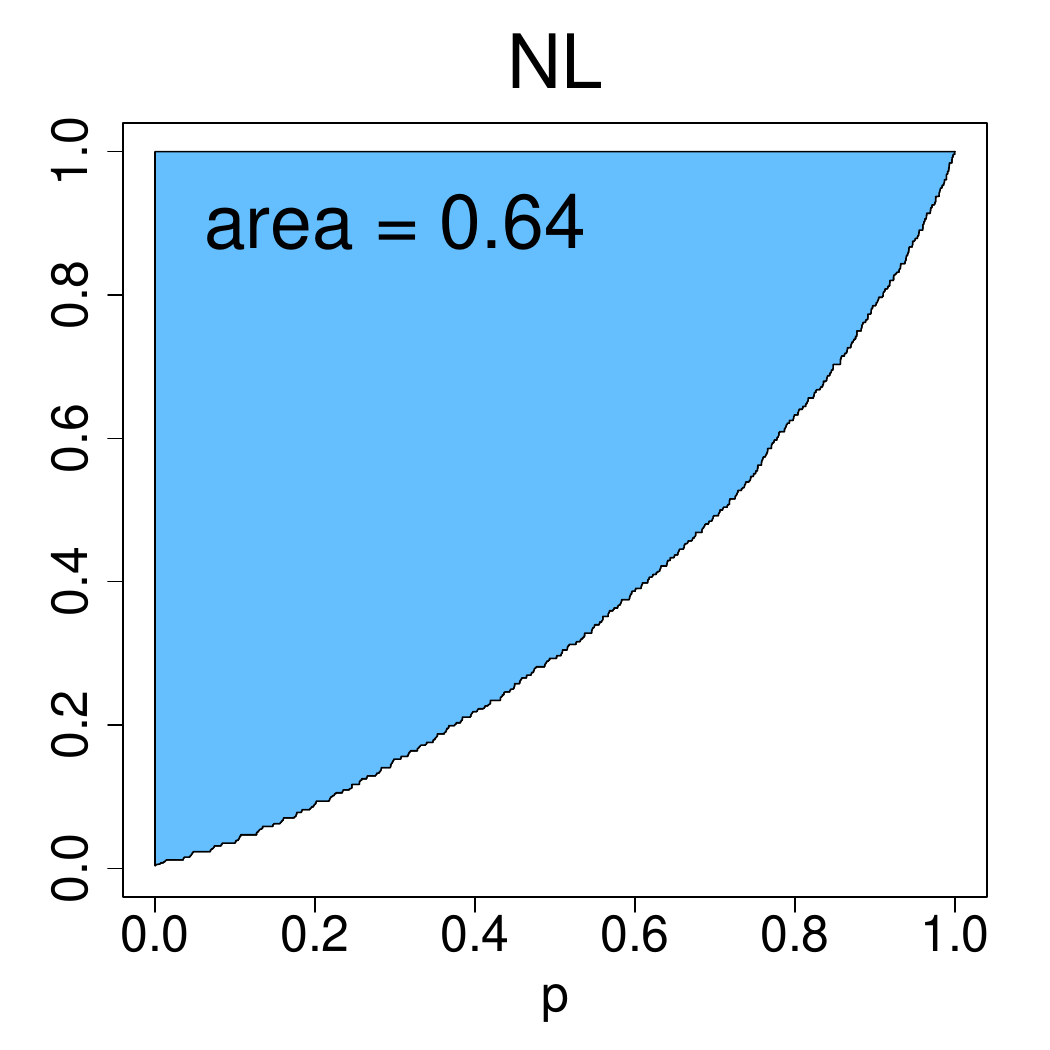}\qquad
\\
\includegraphics[width=0.237\linewidth]{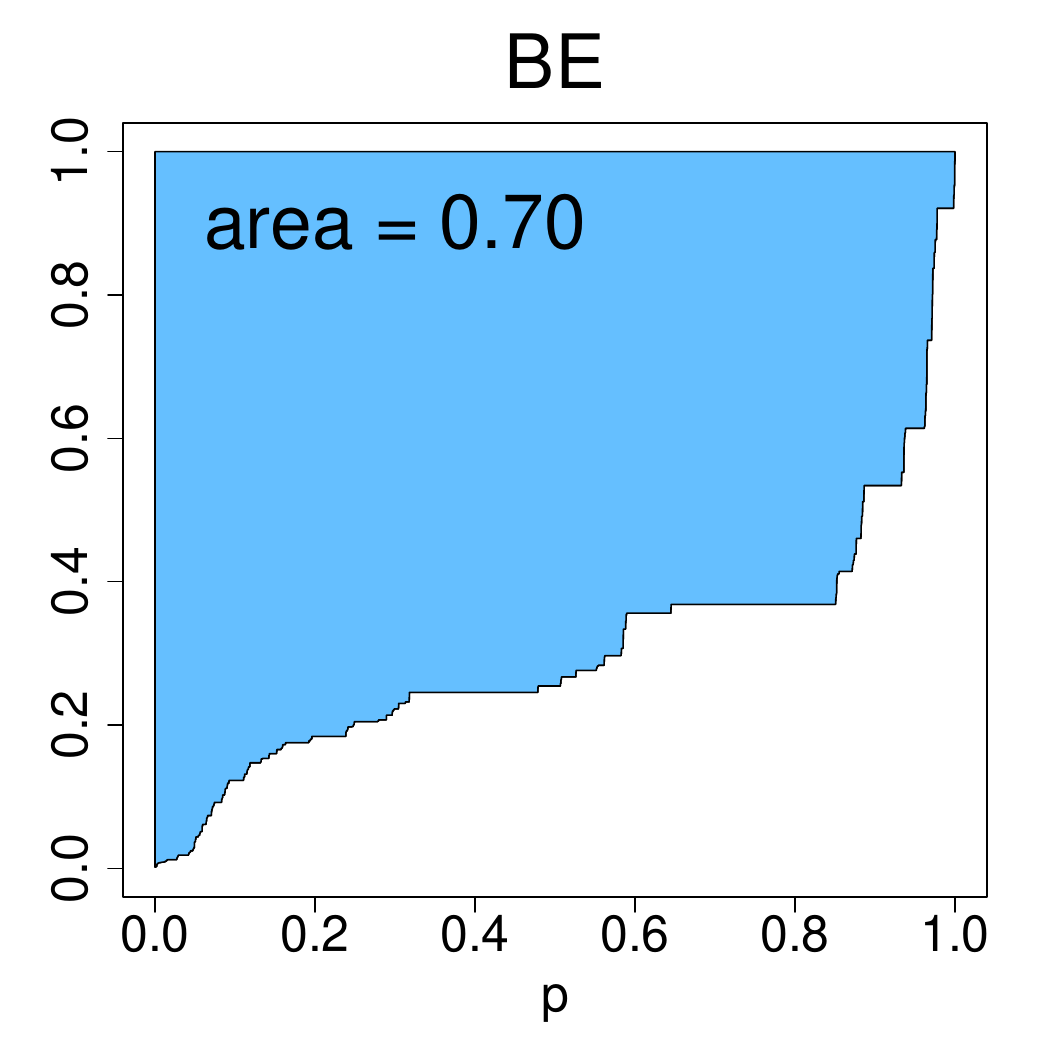}\qquad
\includegraphics[width=0.237\linewidth]{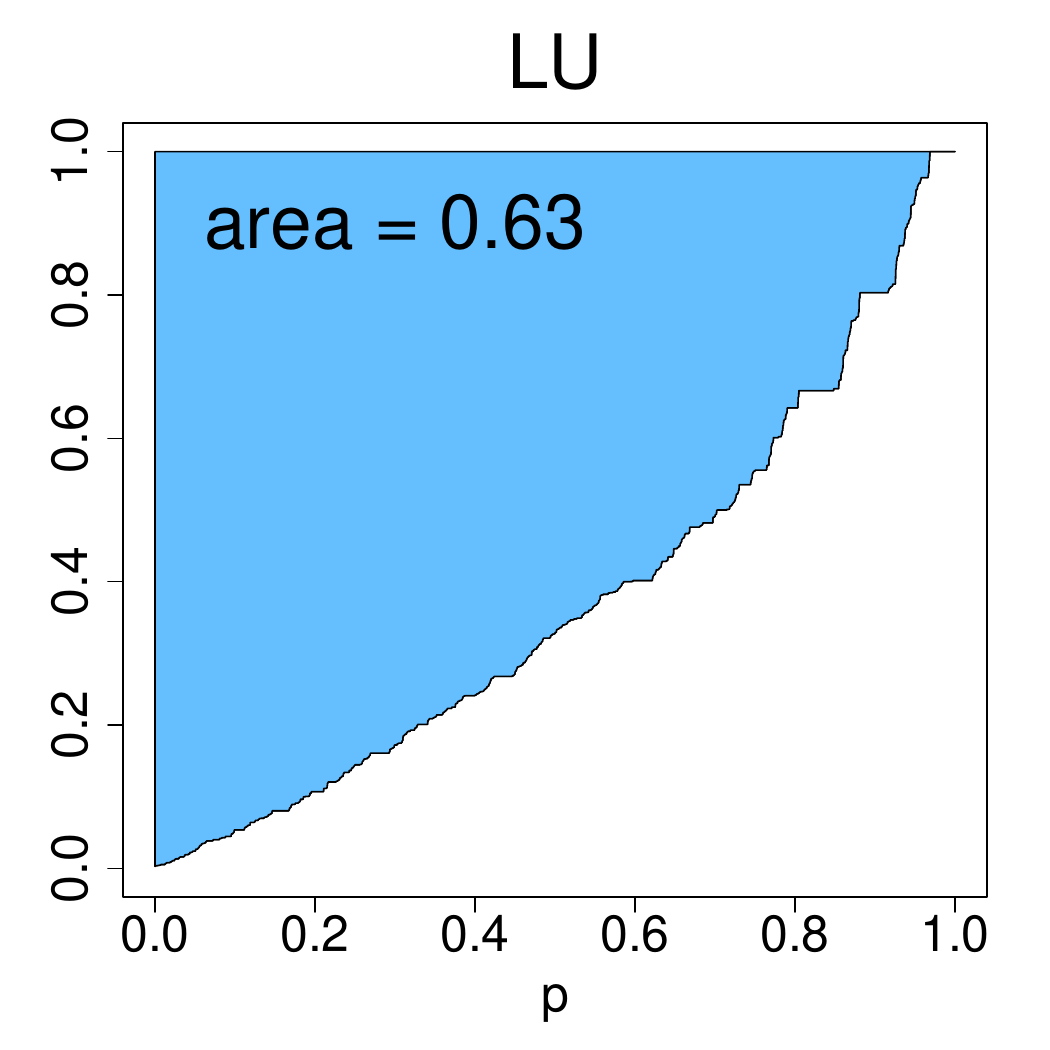}\qquad
\includegraphics[width=0.237\linewidth]{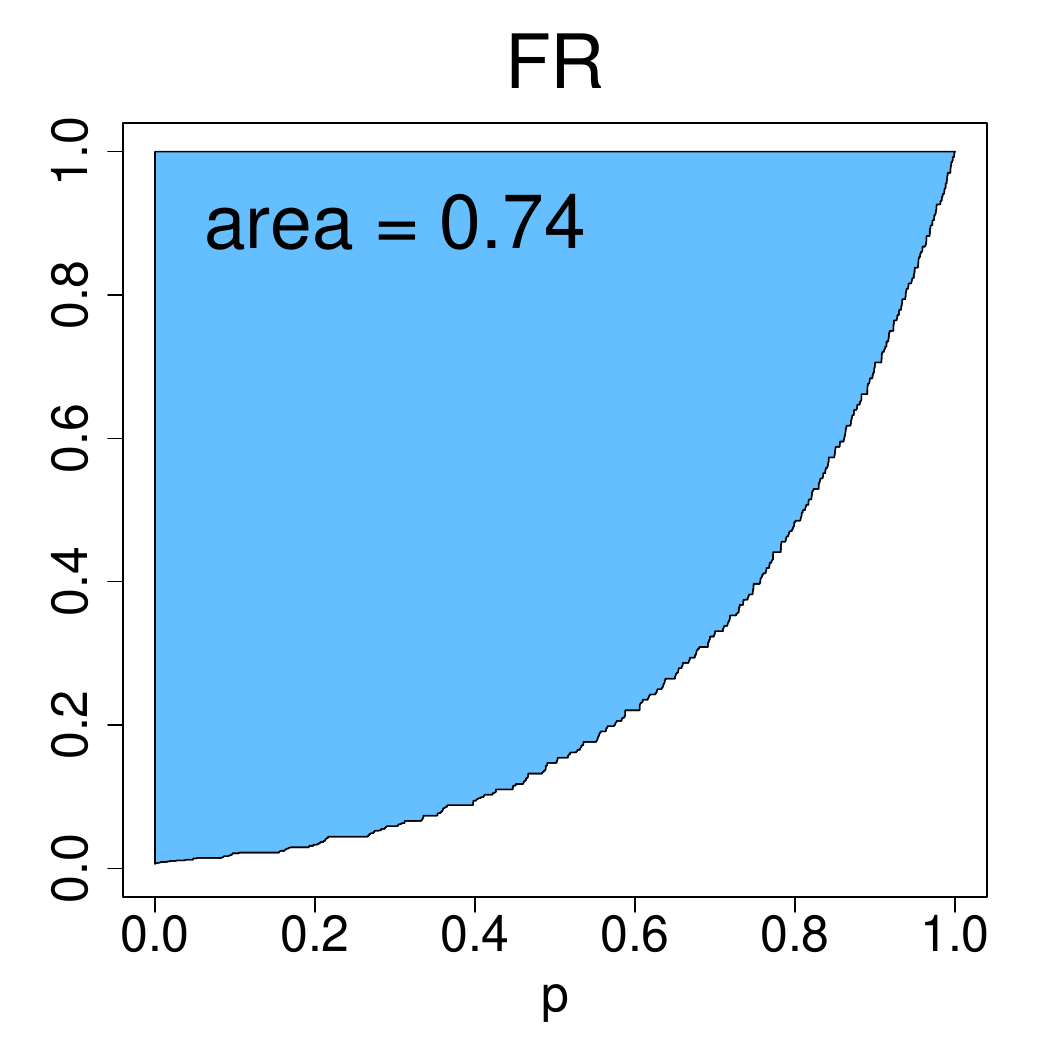}\qquad
\\
\includegraphics[width=0.237\linewidth]{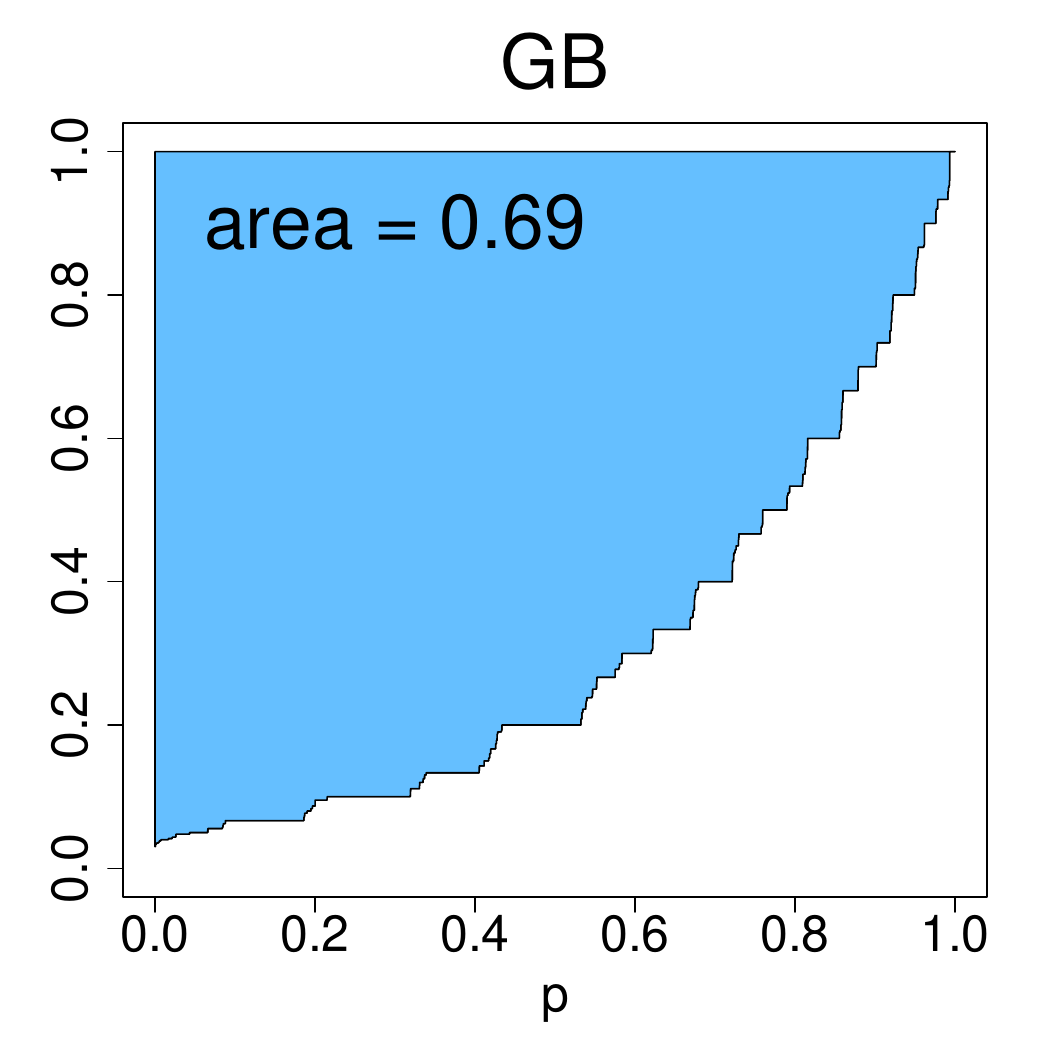}\qquad
\includegraphics[width=0.237\linewidth]{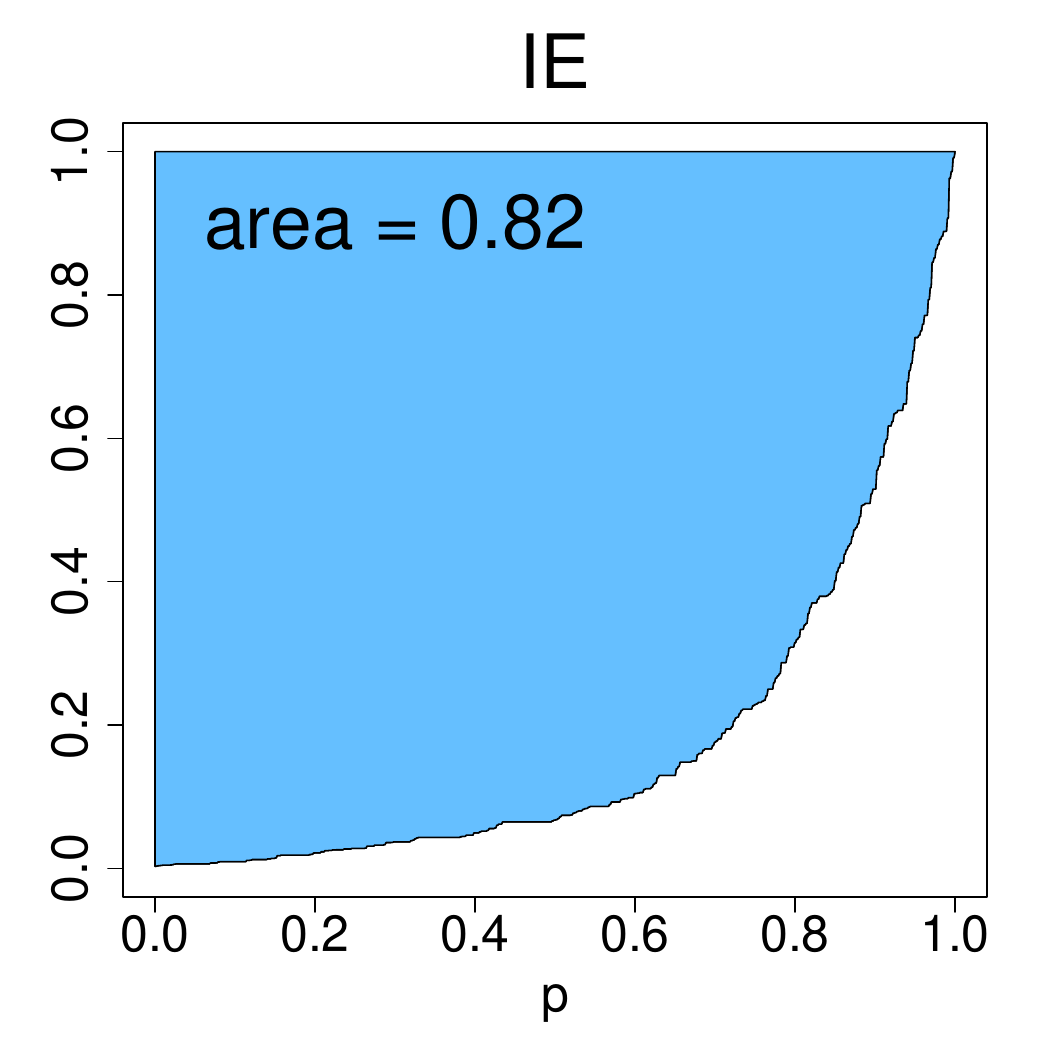}\qquad
\includegraphics[width=0.237\linewidth]{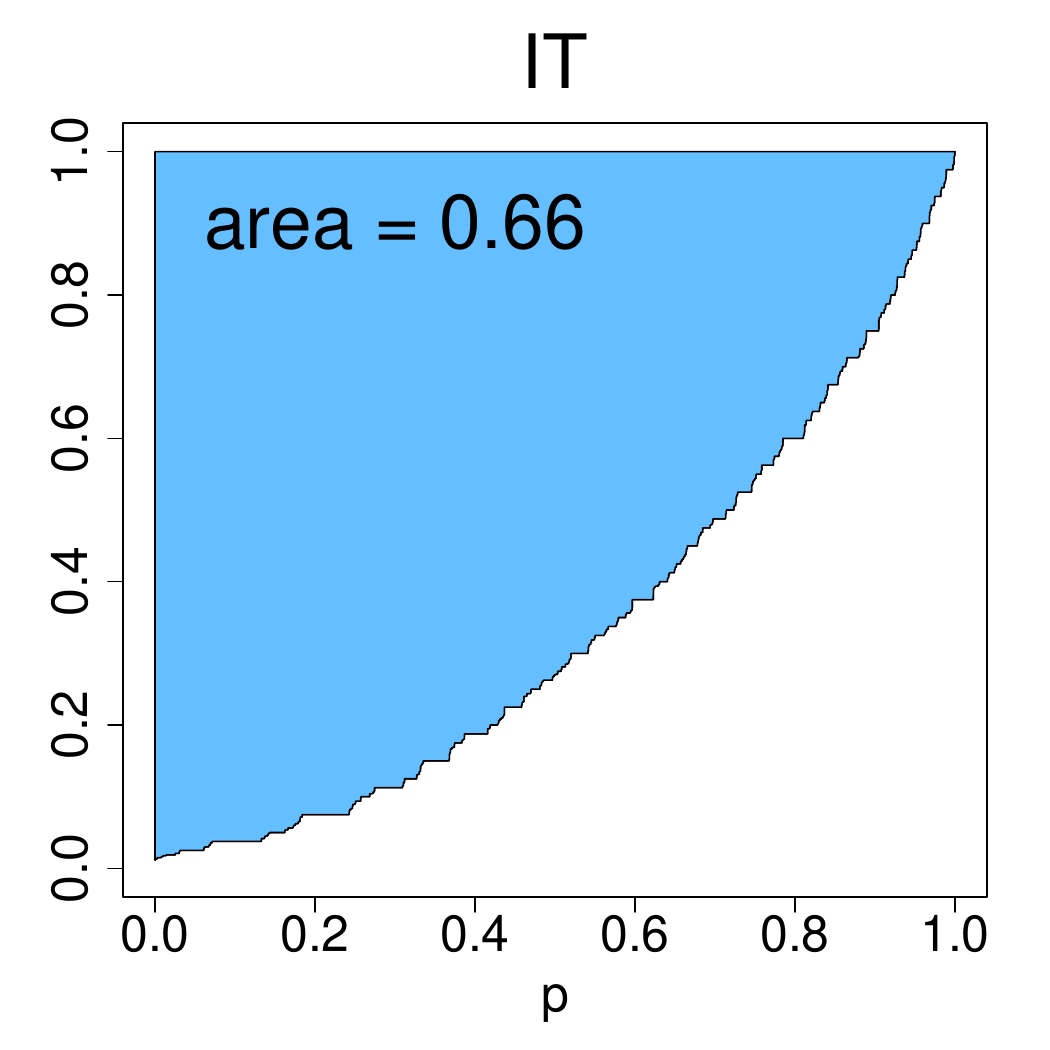}\qquad
\\
\includegraphics[width=0.237\linewidth]{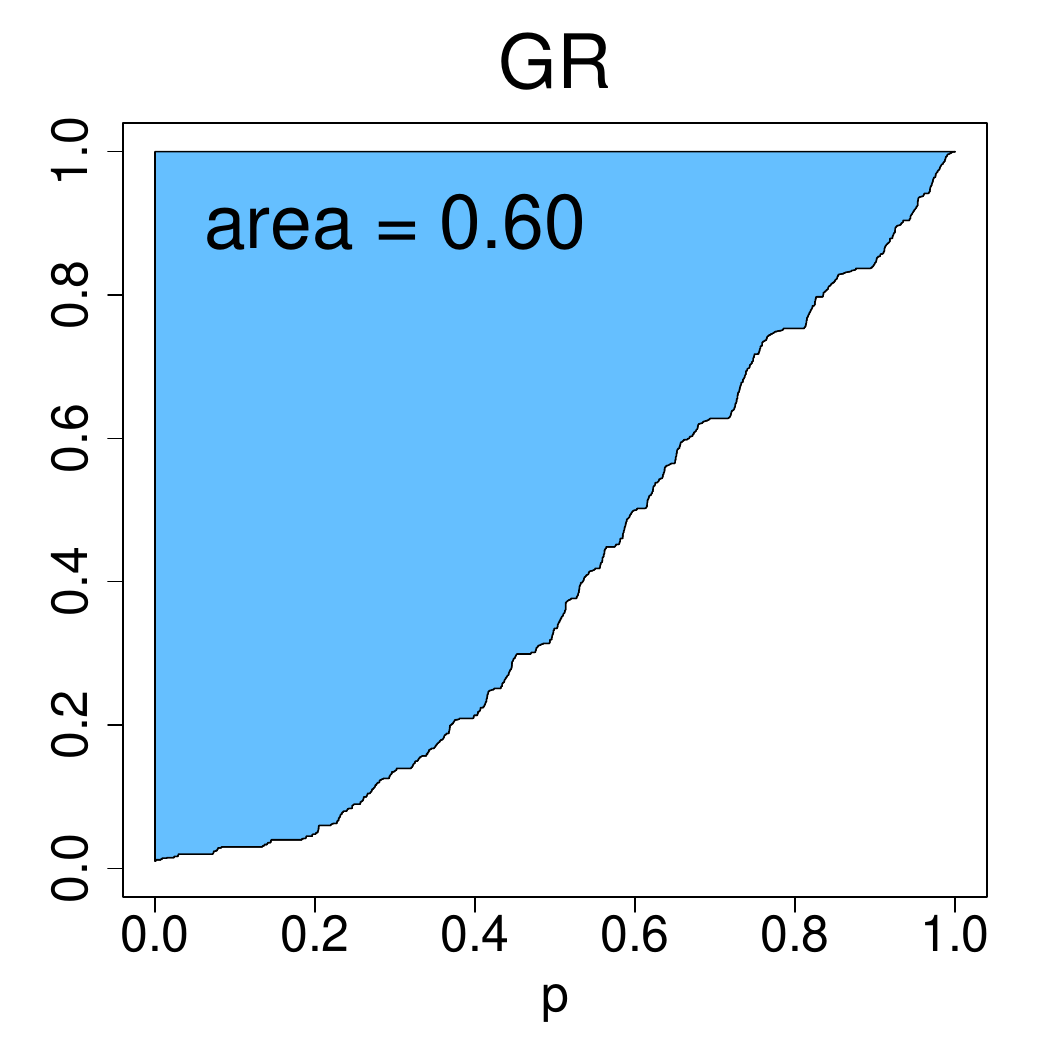}\qquad
\includegraphics[width=0.237\linewidth]{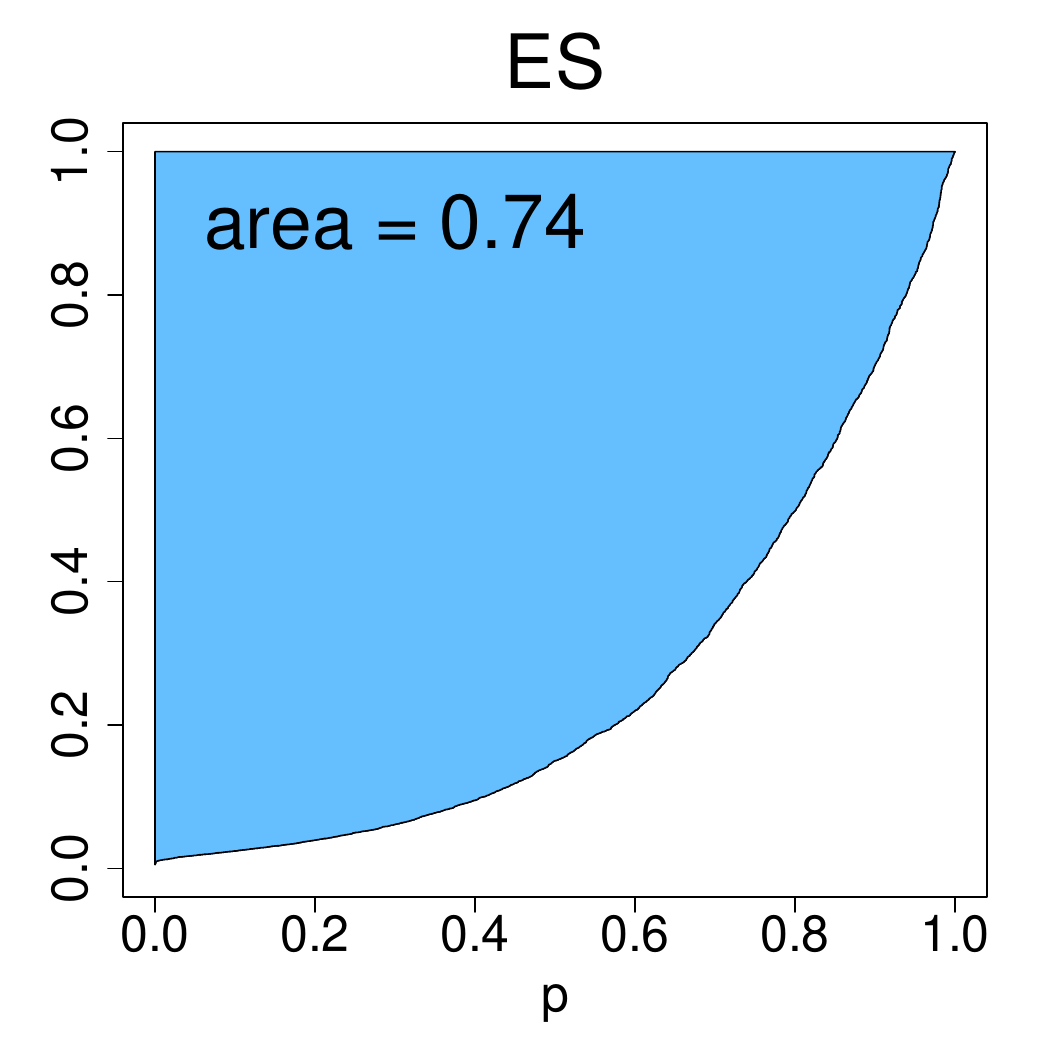}\qquad
\includegraphics[width=0.237\linewidth]{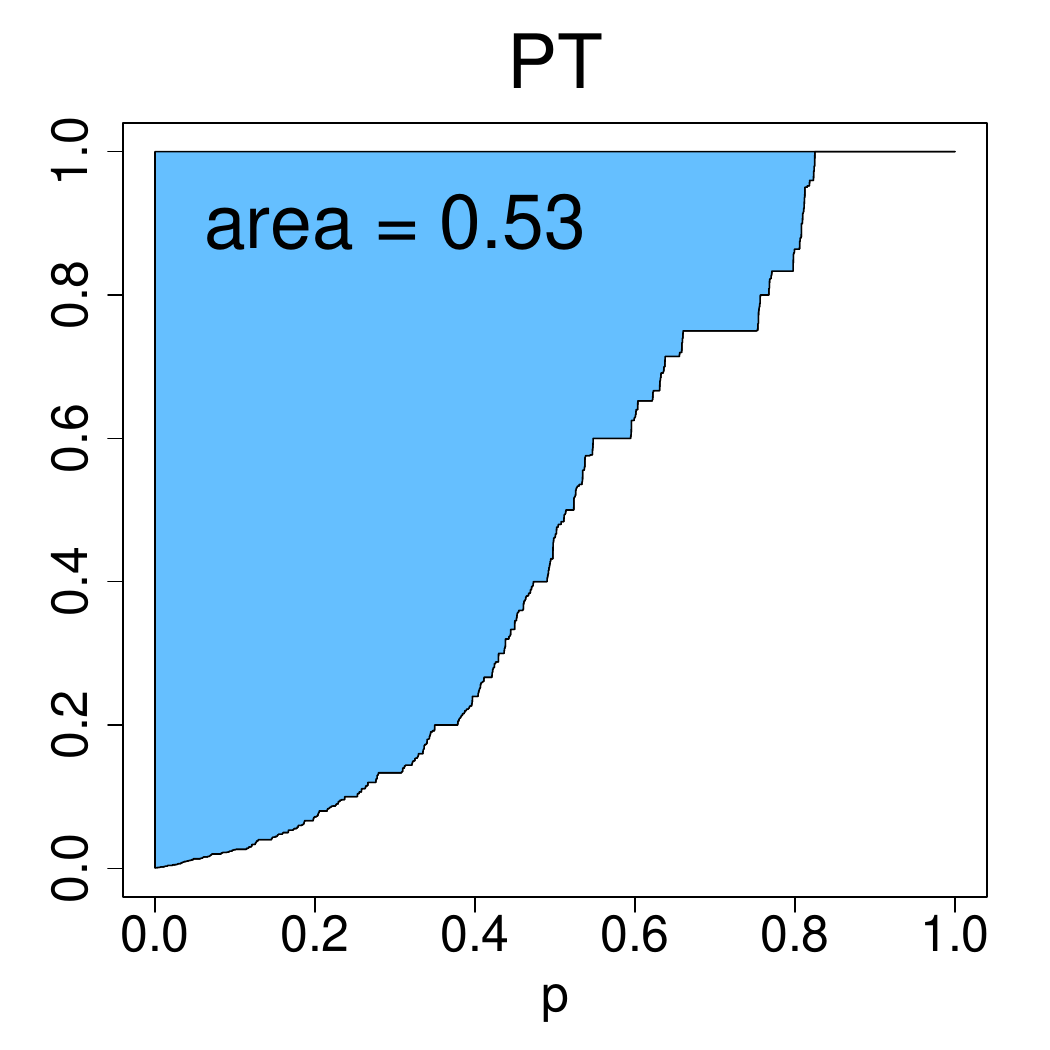}\qquad
\\
\includegraphics[width=0.237\linewidth]{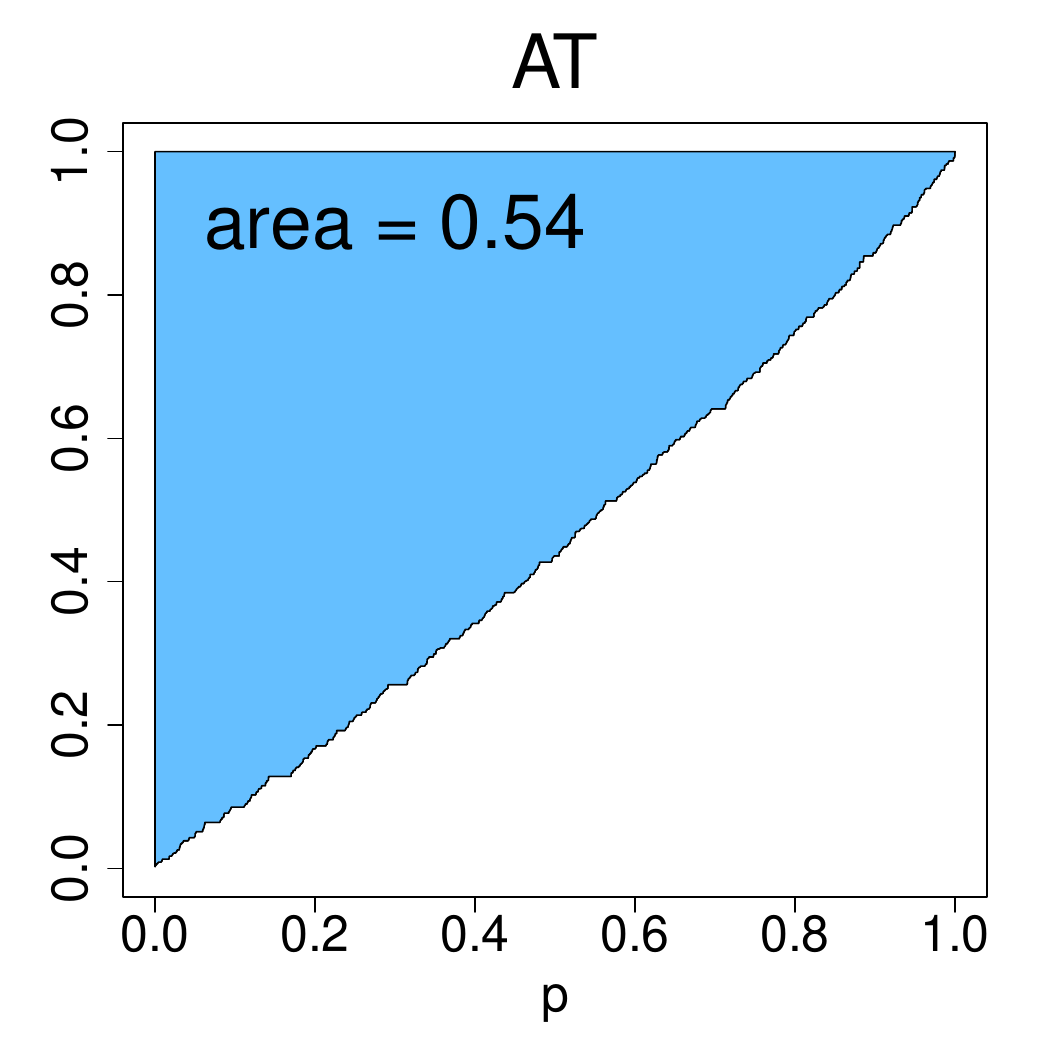}\qquad
\includegraphics[width=0.237\linewidth]{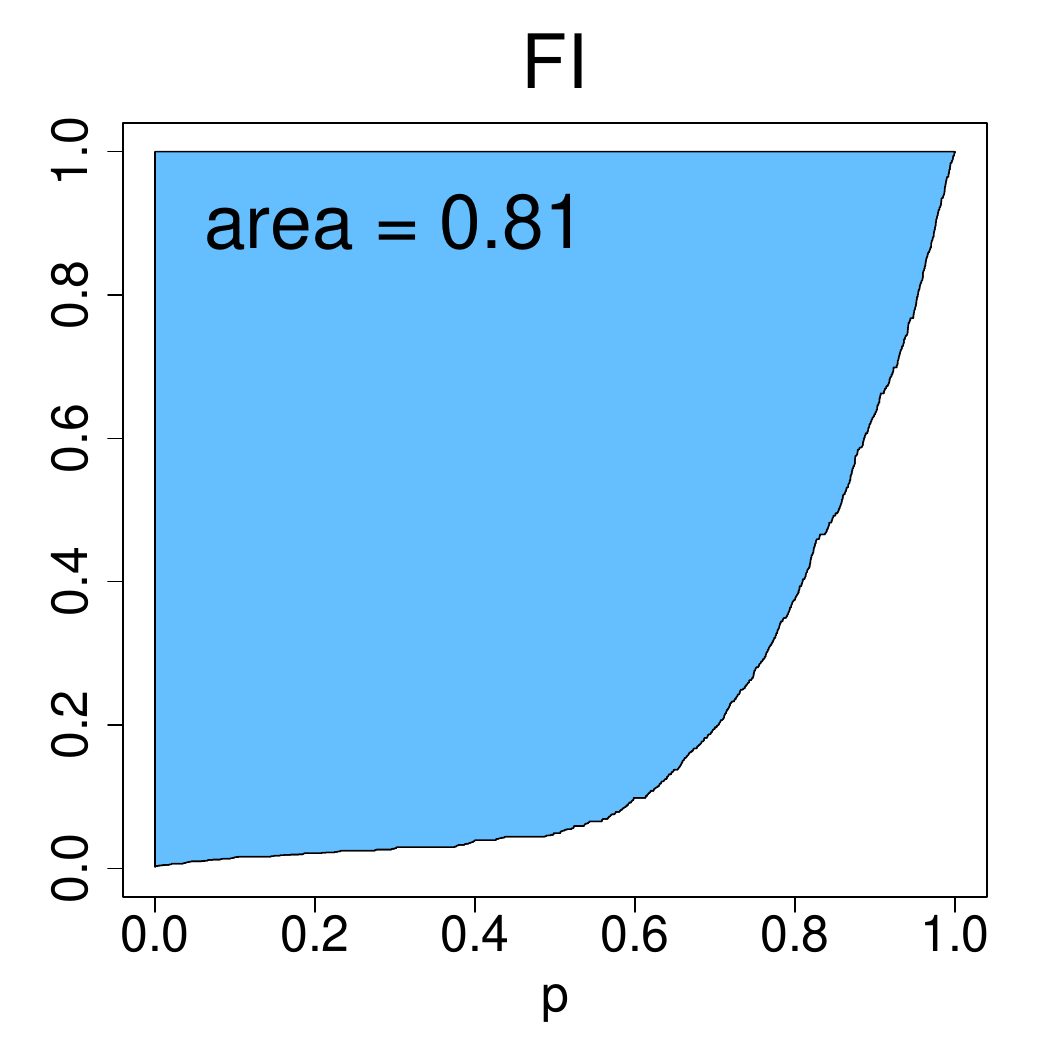}\qquad
\includegraphics[width=0.237\linewidth]{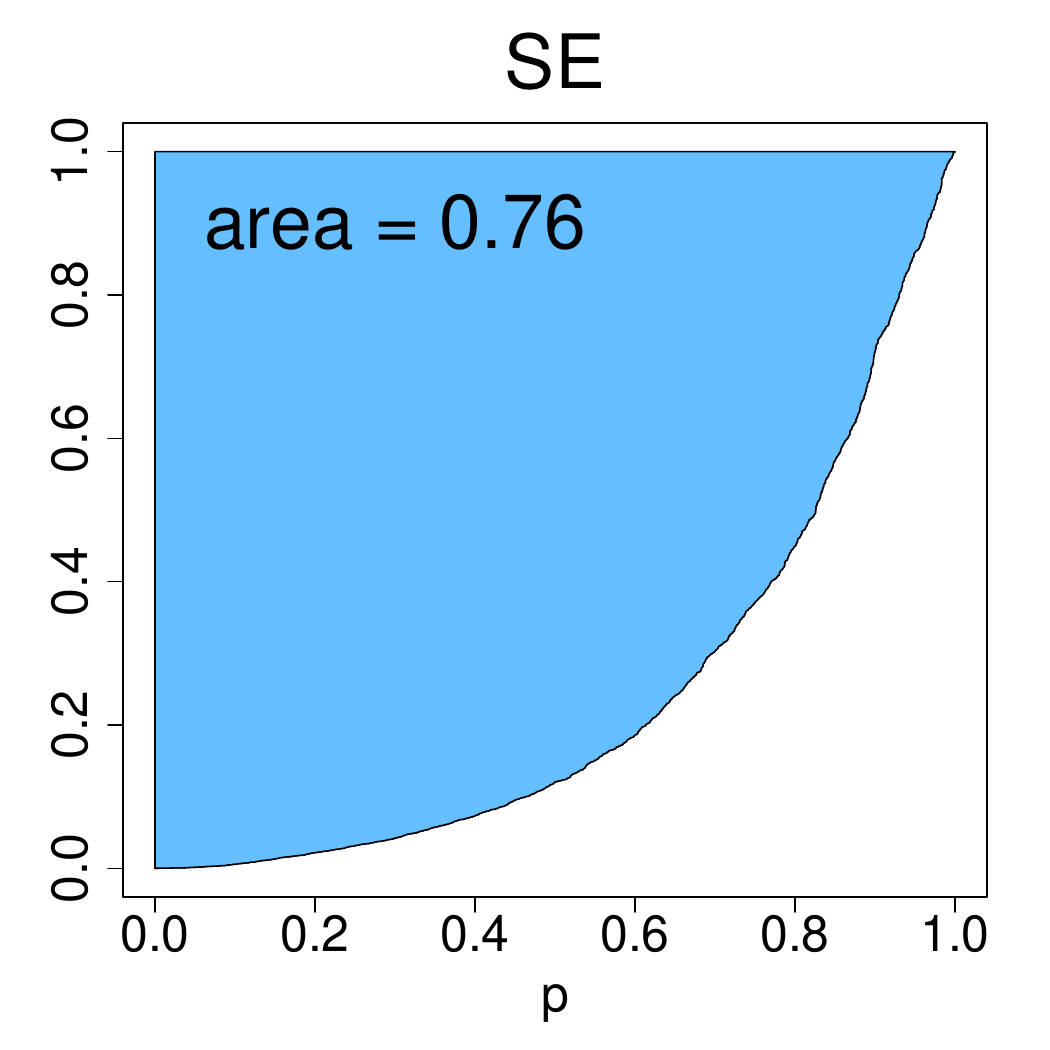}\qquad
\caption{The income-equality curve $\psi_{1,n}$ and the shaded-in area (i.e., $\Psi_{1,n}$) above it for the fifteen European countries, where $n=n_P$ is specified in Table~\ref{tab-real2018}  \citep[based on][]{EU-SILC2018}.}
\label{fig:4}
\end{center}
\end{figure}

\begin{figure}[h!]
\begin{center}
\includegraphics[width=0.237\linewidth]{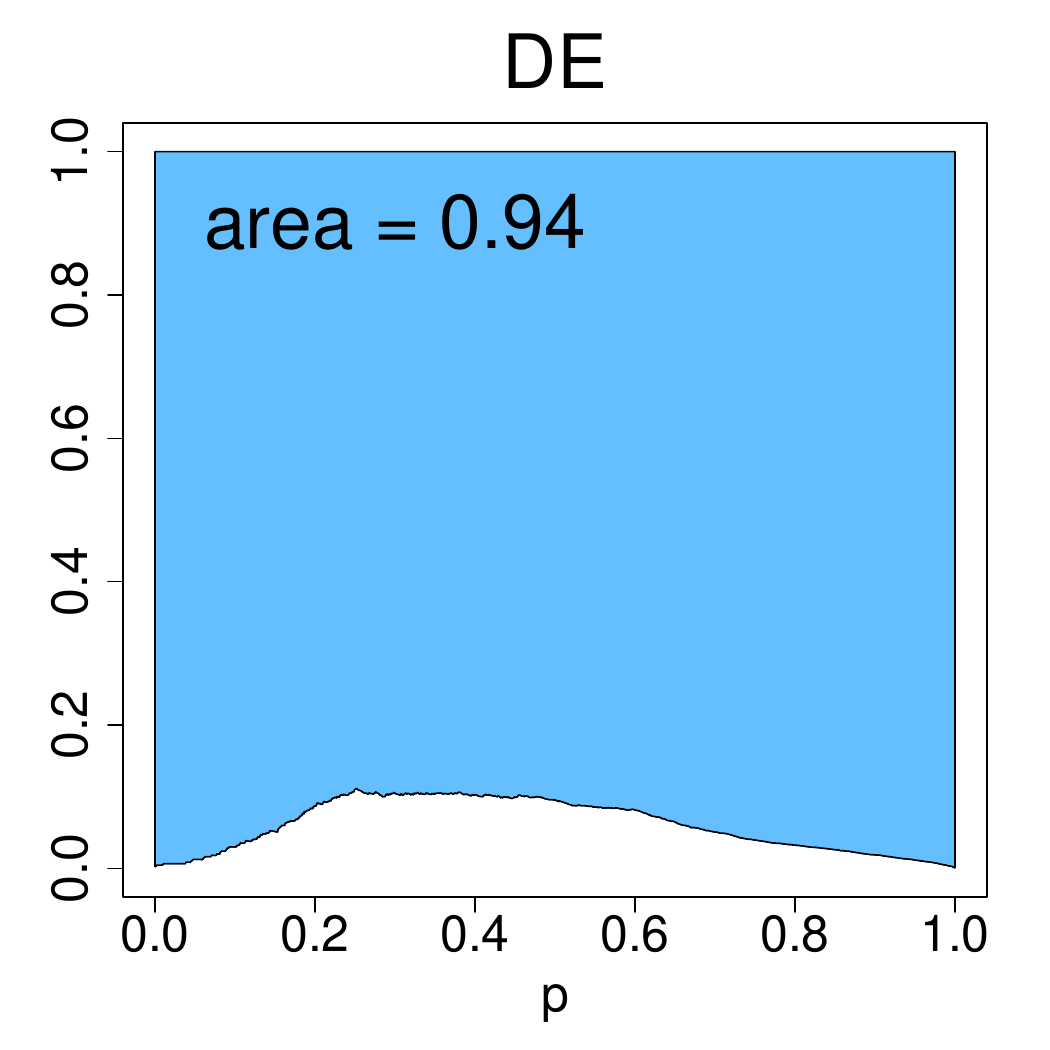}\qquad
\includegraphics[width=0.237\linewidth]{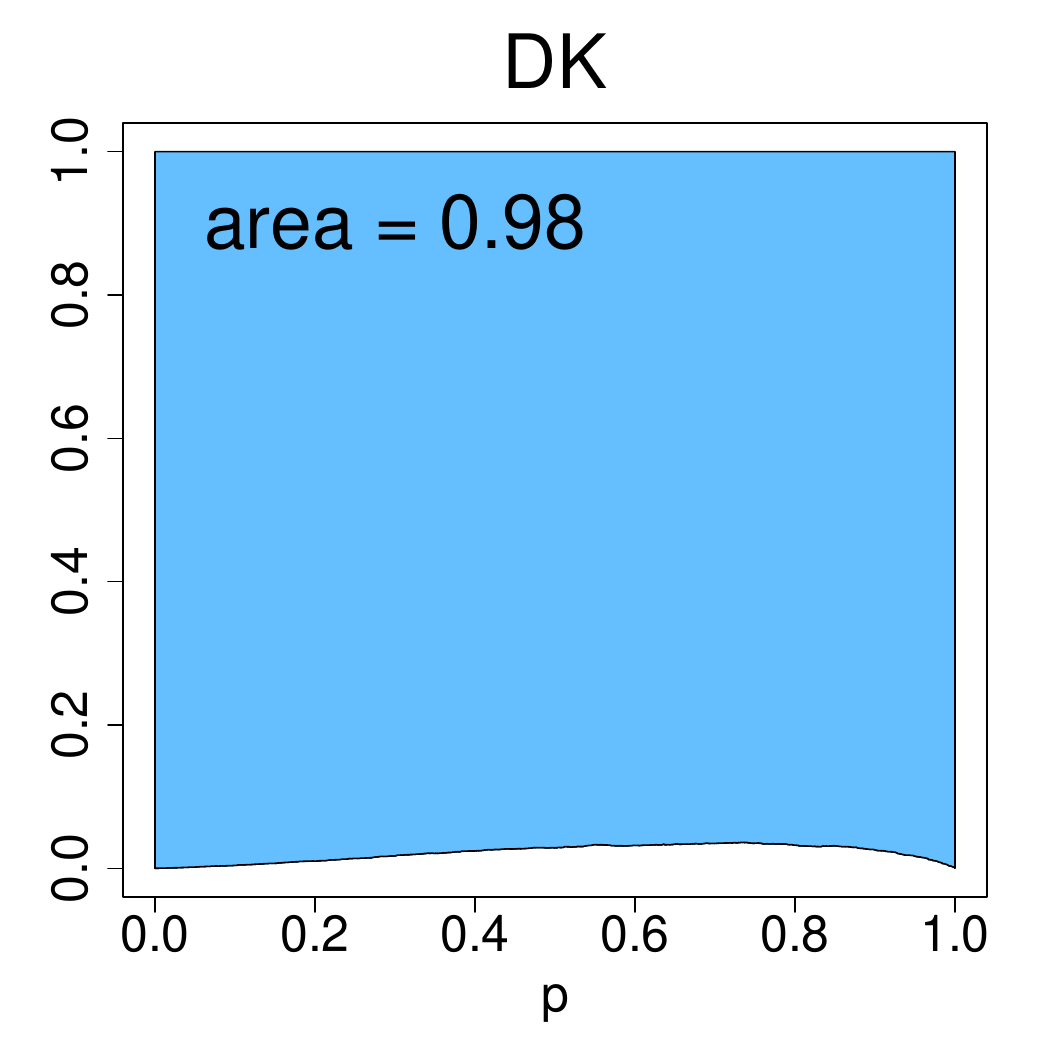}\qquad
\includegraphics[width=0.237\linewidth]{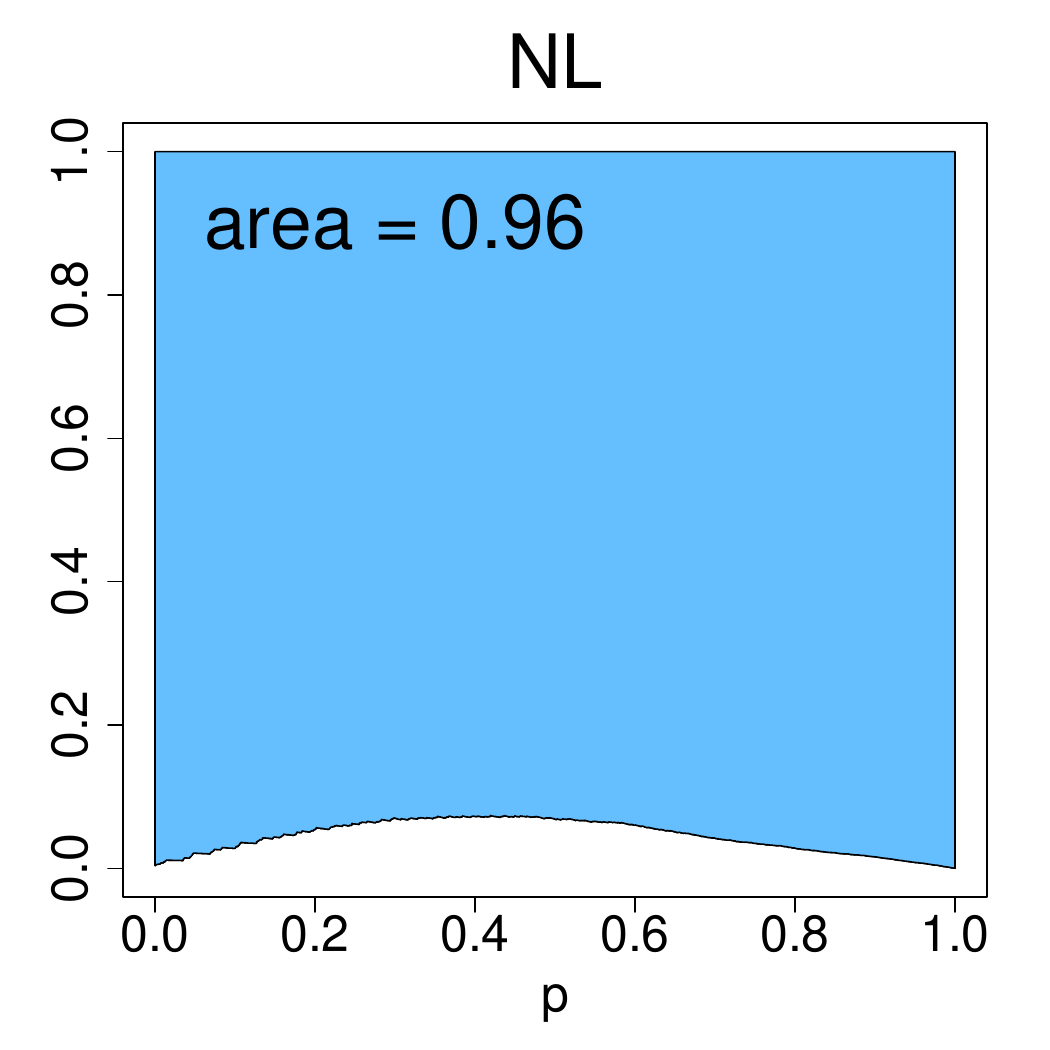}\qquad
\\
\includegraphics[width=0.237\linewidth]{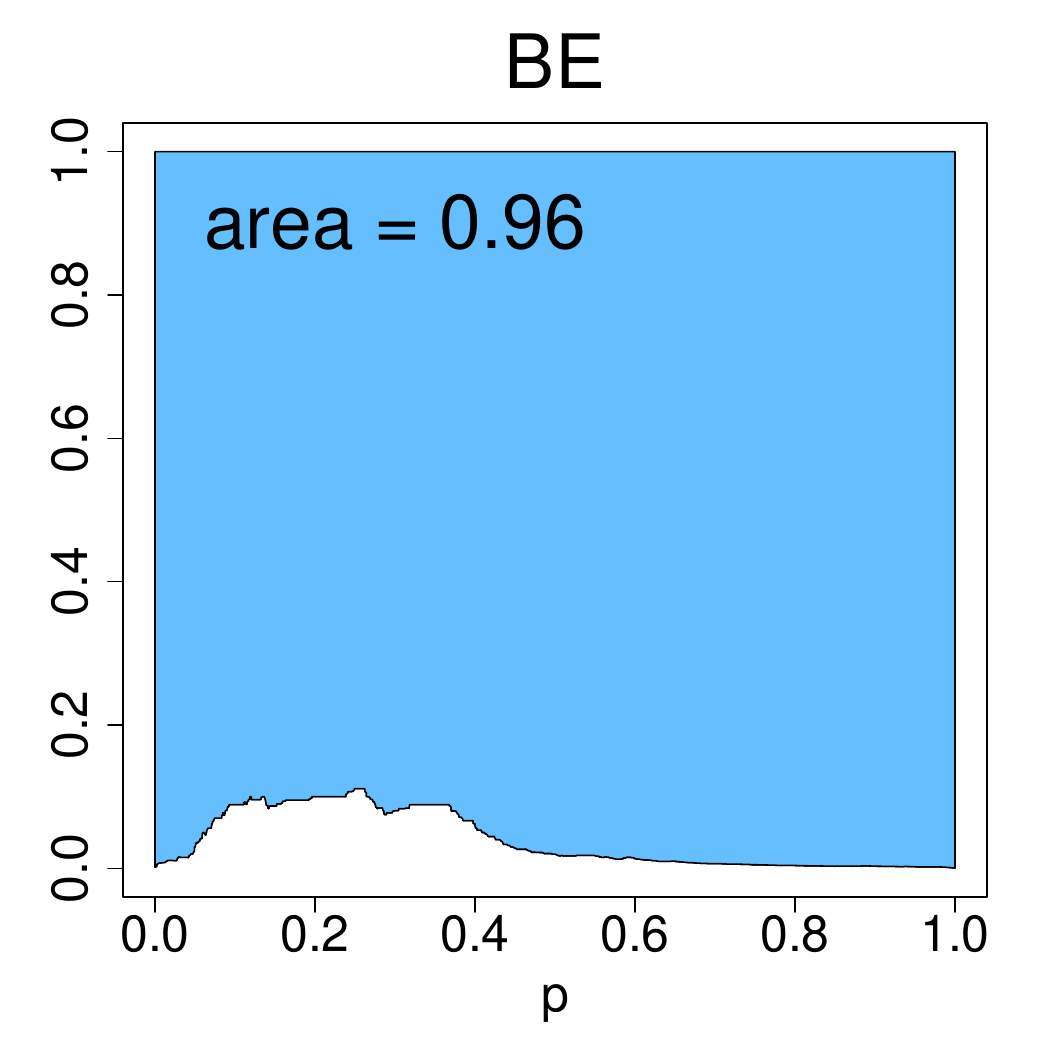}\qquad
\includegraphics[width=0.237\linewidth]{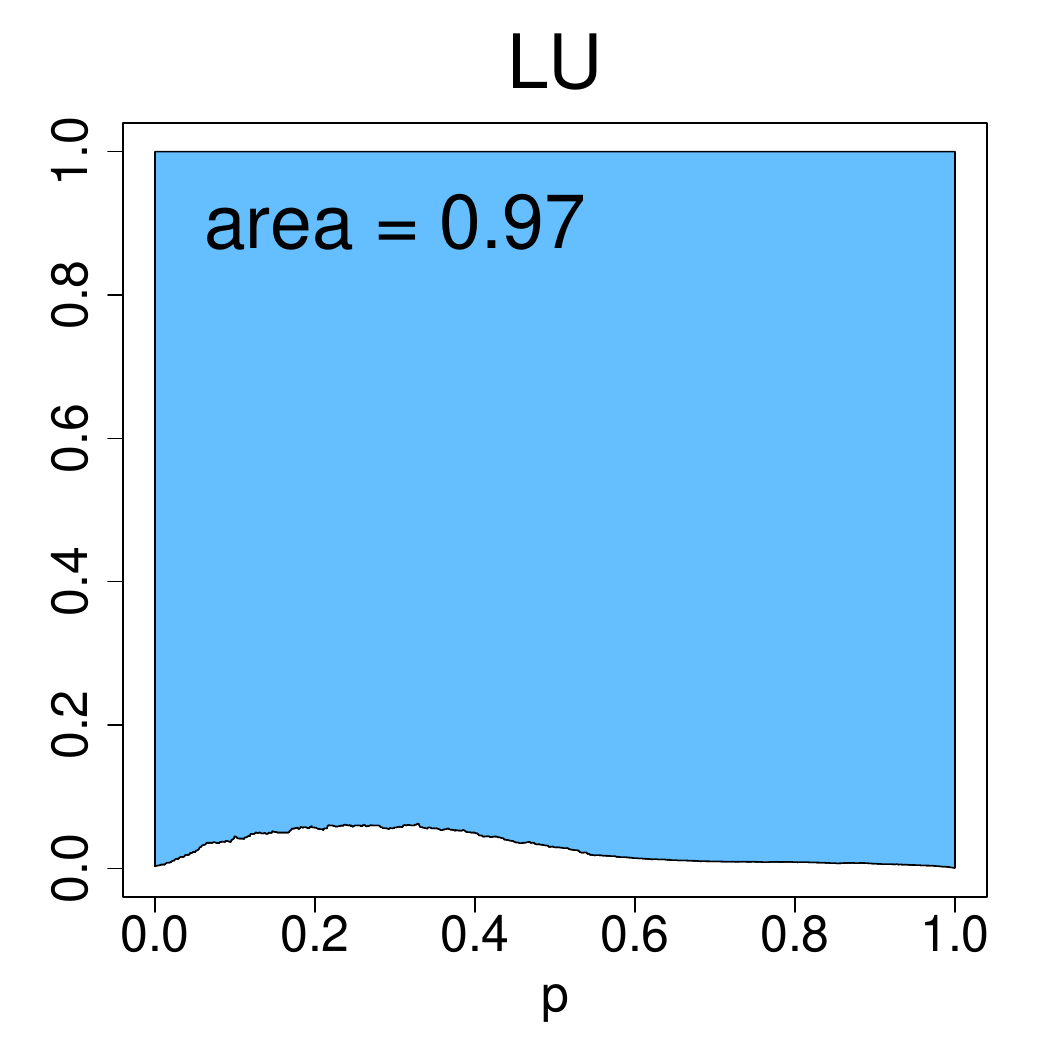}\qquad
\includegraphics[width=0.237\linewidth]{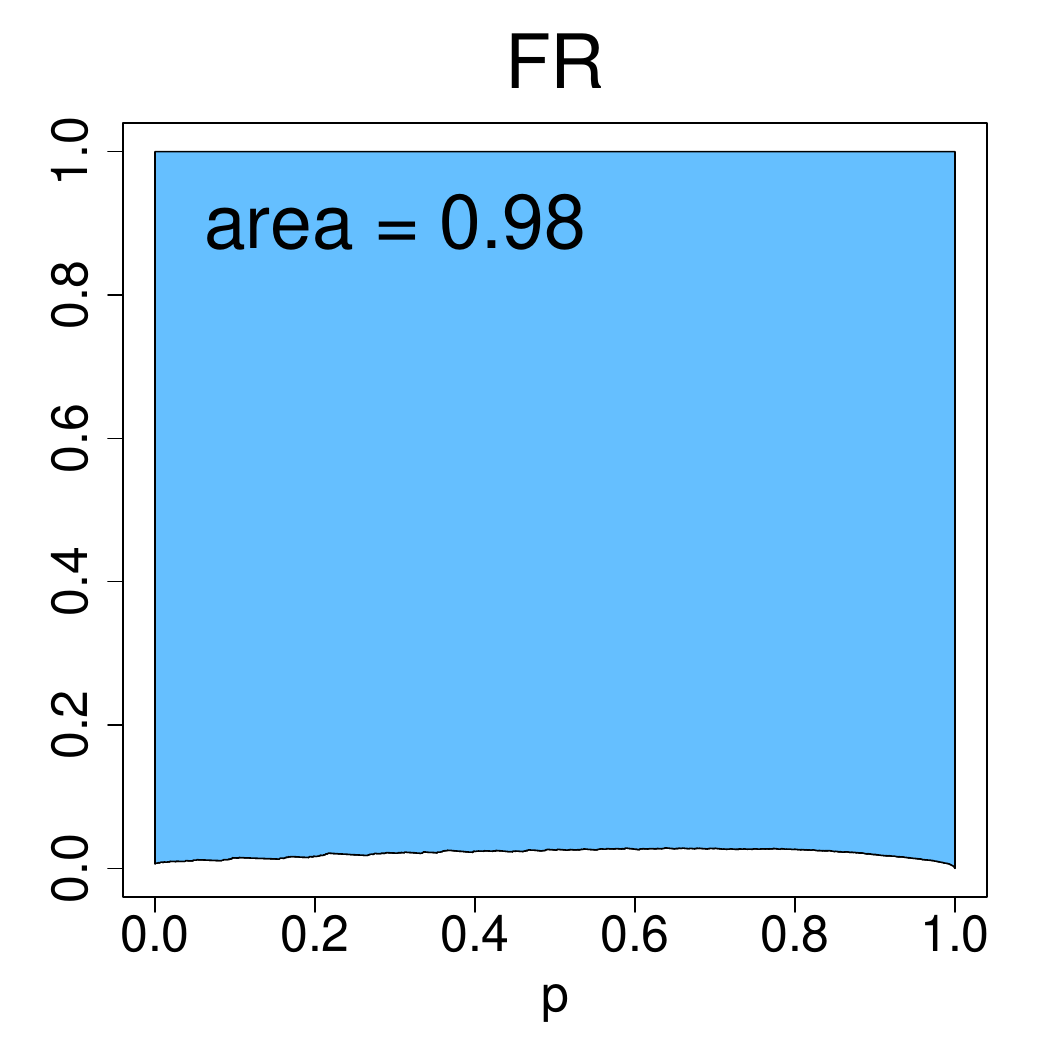}\qquad
\\
\includegraphics[width=0.237\linewidth]{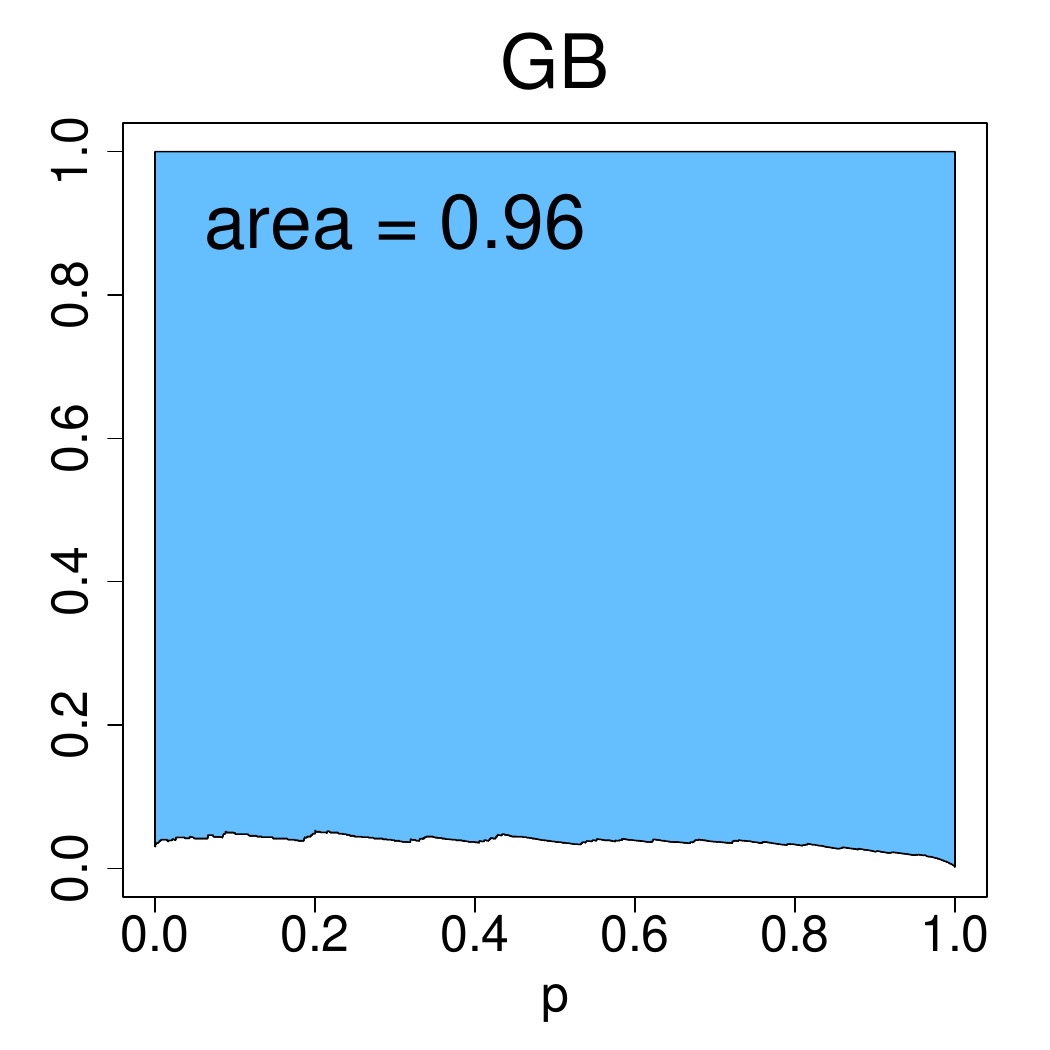}\qquad
\includegraphics[width=0.237\linewidth]{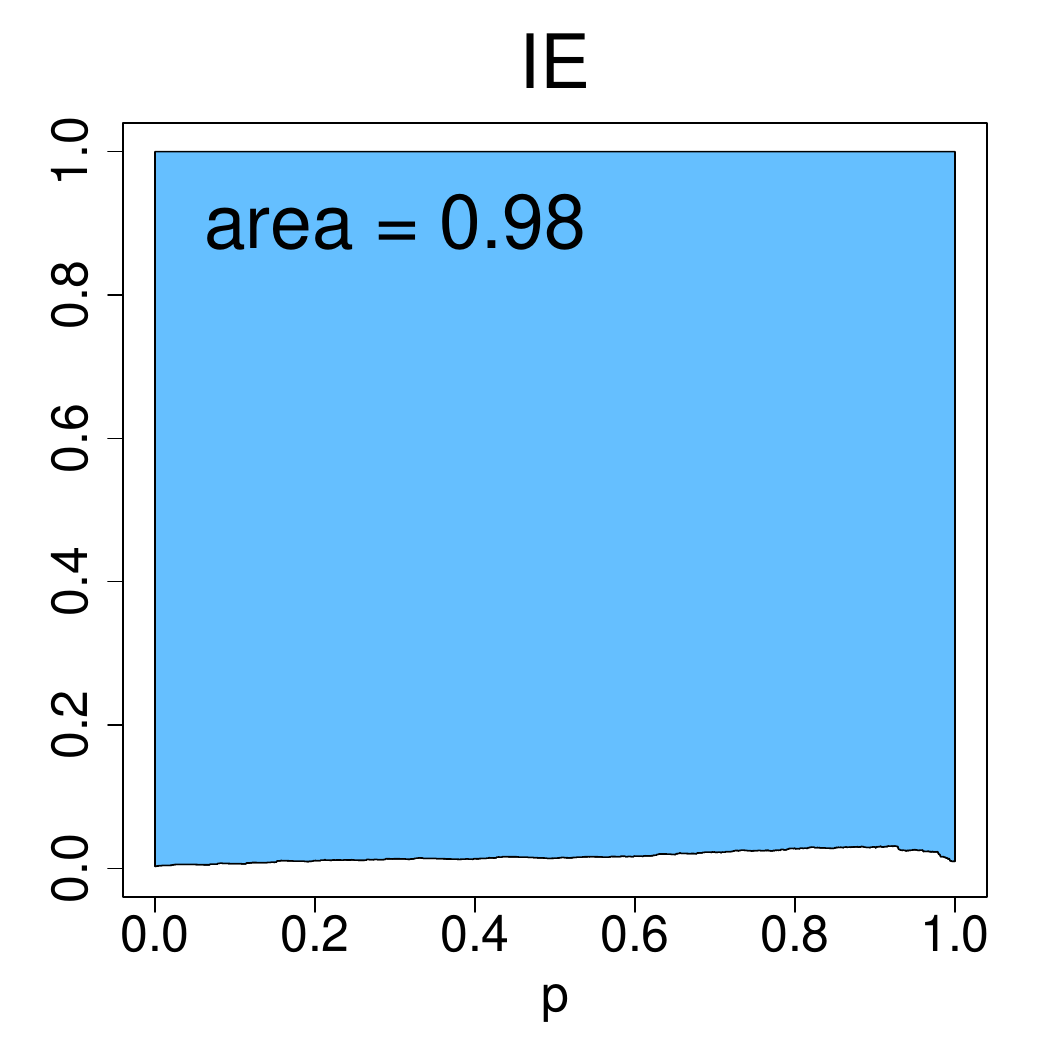}\qquad
\includegraphics[width=0.237\linewidth]{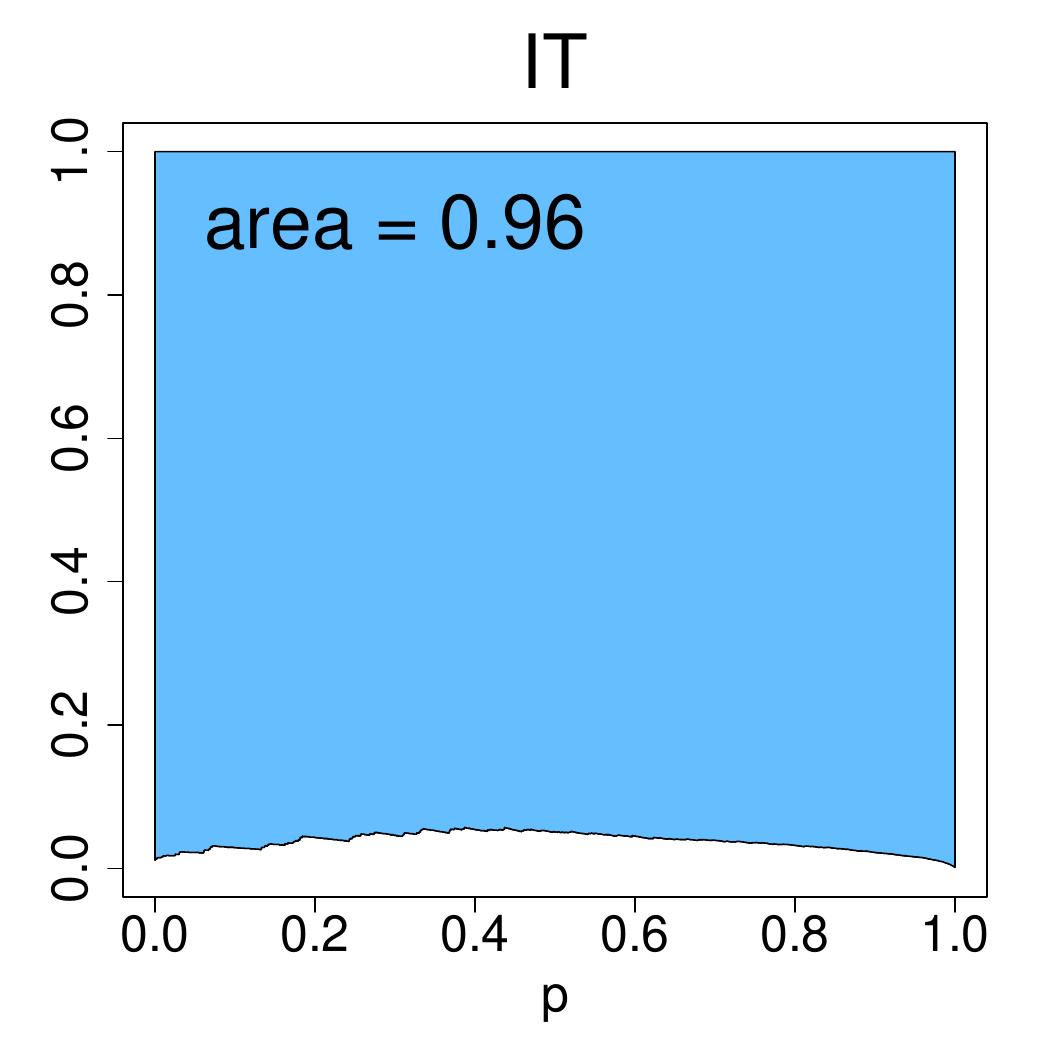}\qquad
\\
\includegraphics[width=0.237\linewidth]{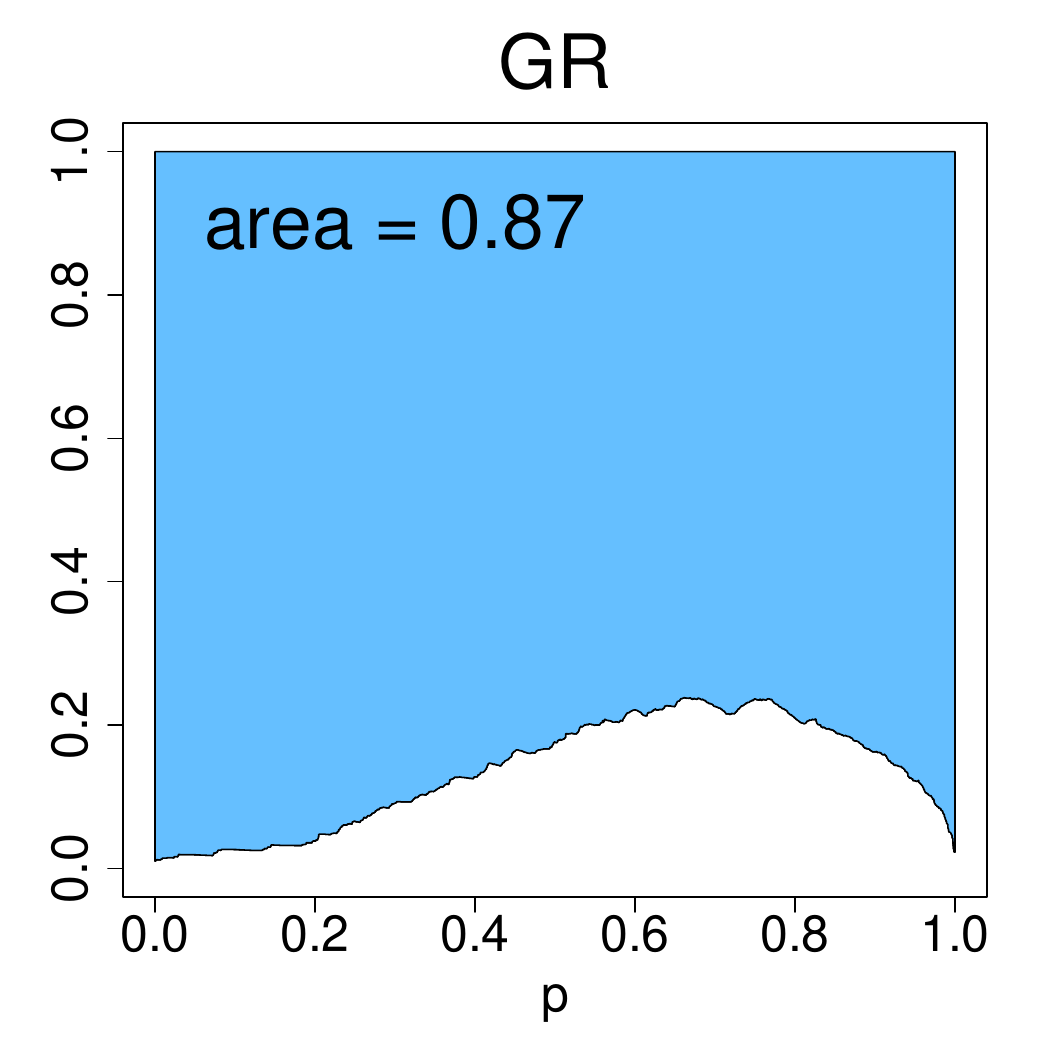}\qquad
\includegraphics[width=0.237\linewidth]{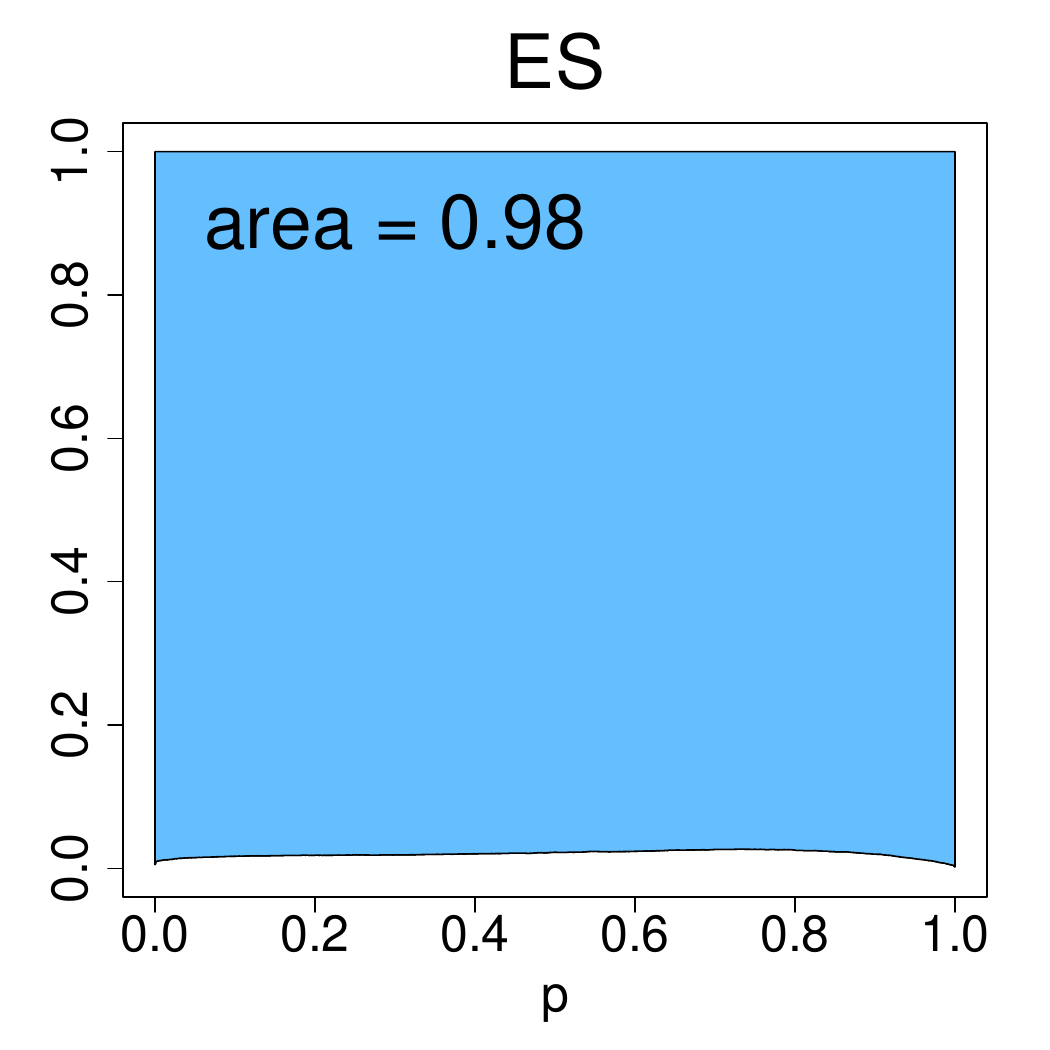}\qquad
\includegraphics[width=0.237\linewidth]{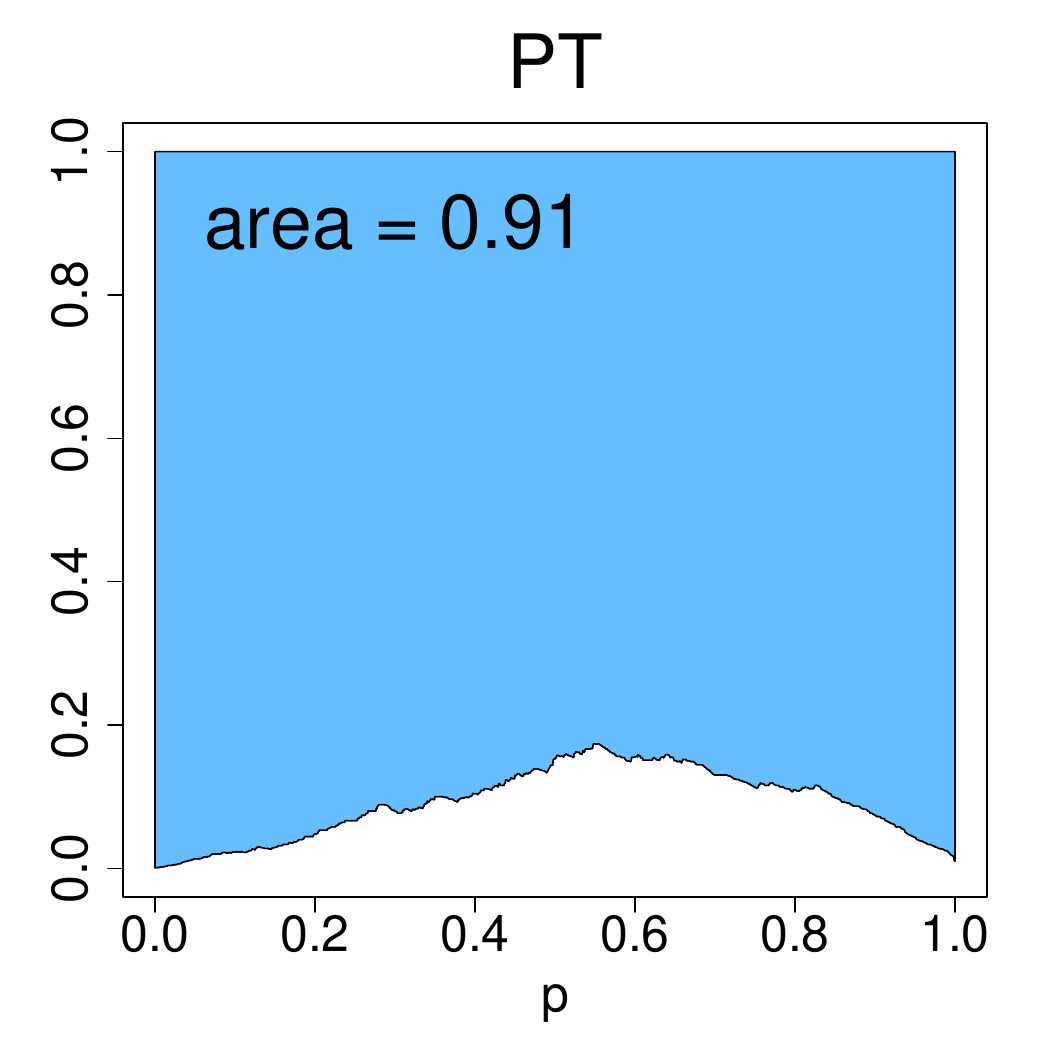}\qquad
\\
\includegraphics[width=0.237\linewidth]{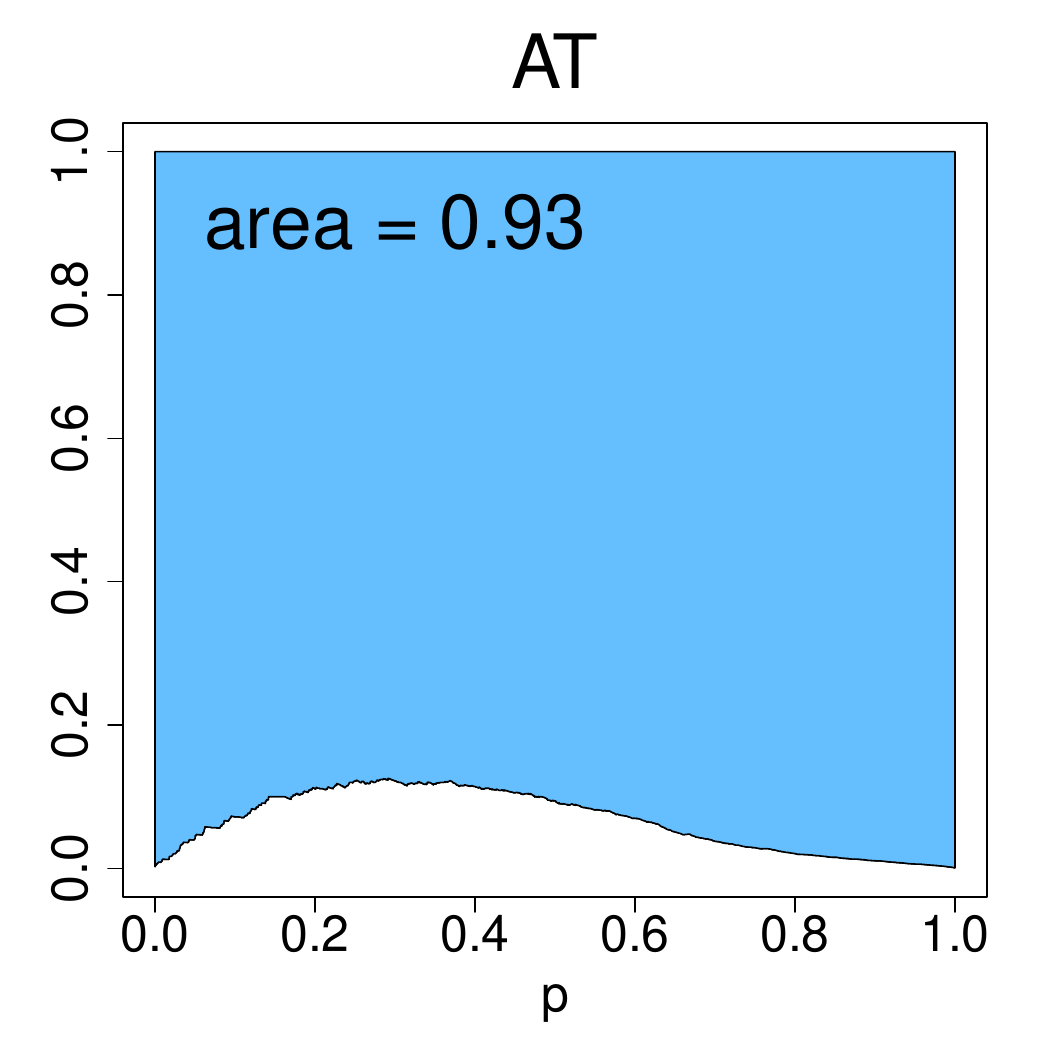}\qquad
\includegraphics[width=0.237\linewidth]{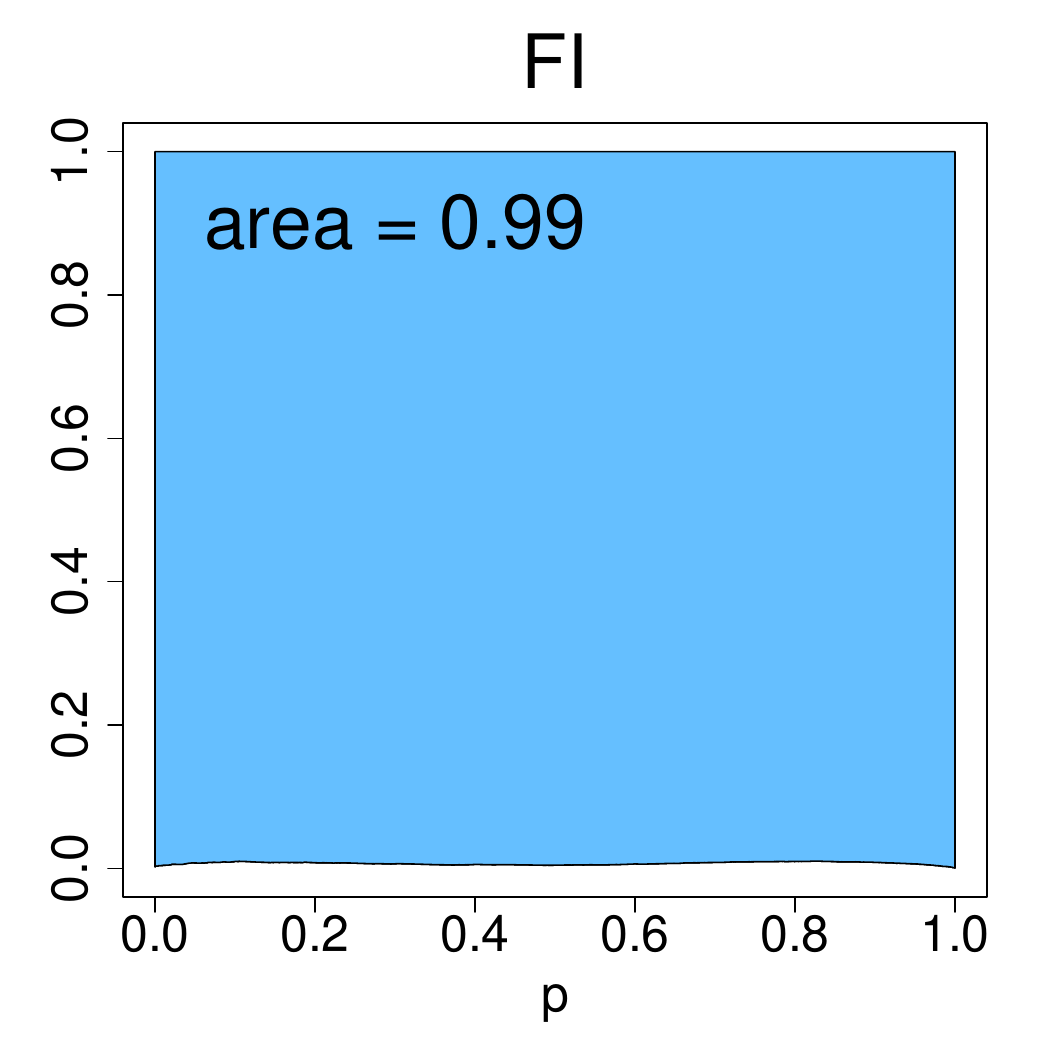}\qquad
\includegraphics[width=0.237\linewidth]{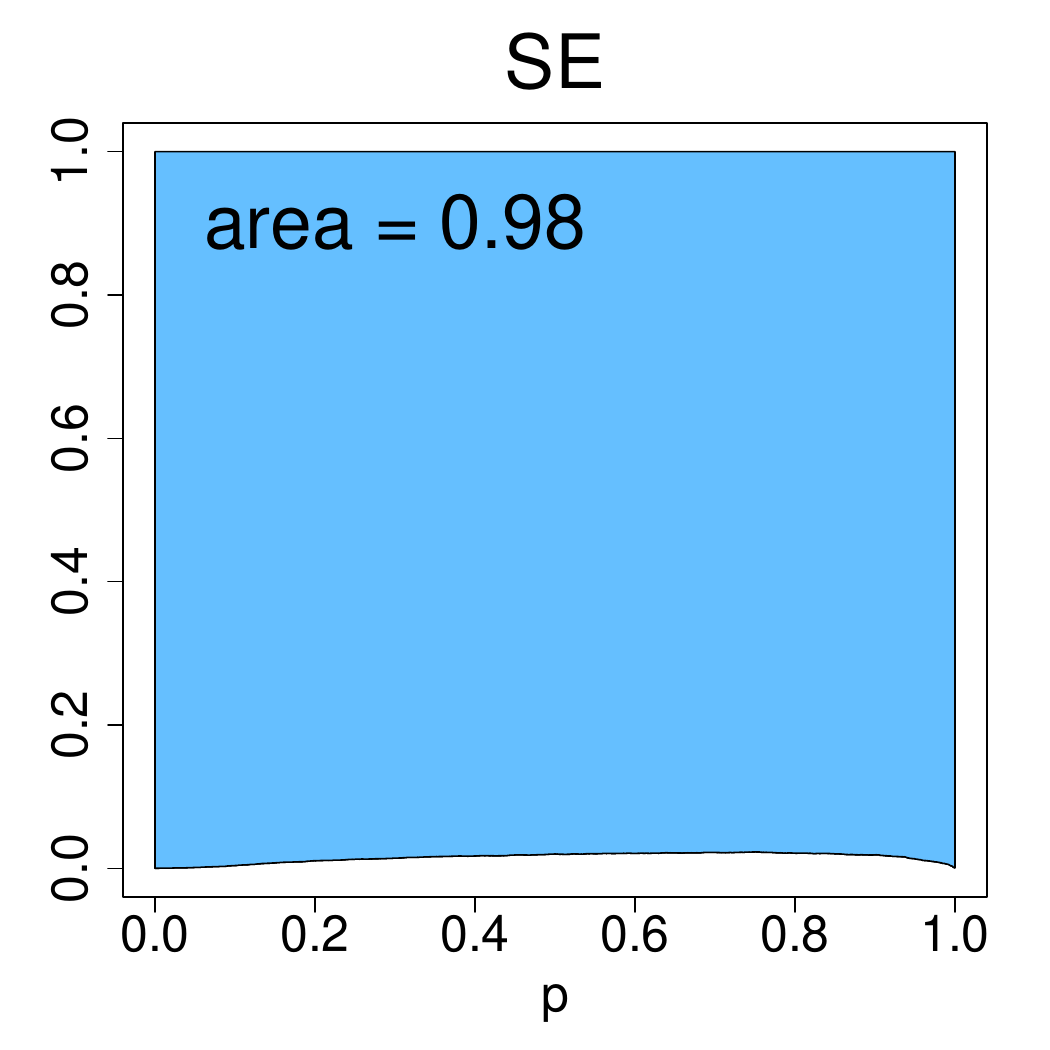}\qquad
\caption{The income-equality curve $\psi_{2,n}$ and the shaded-in area (i.e., $\Psi_{2,n}$) above it for the fifteen European countries, where $n=n_P$ is specified in Table~\ref{tab-real2018}  \citep[based on][]{EU-SILC2018}.}
\label{fig:5}
\end{center}
\end{figure}

\begin{figure}[h!]
\begin{center}
\includegraphics[width=0.237\linewidth]{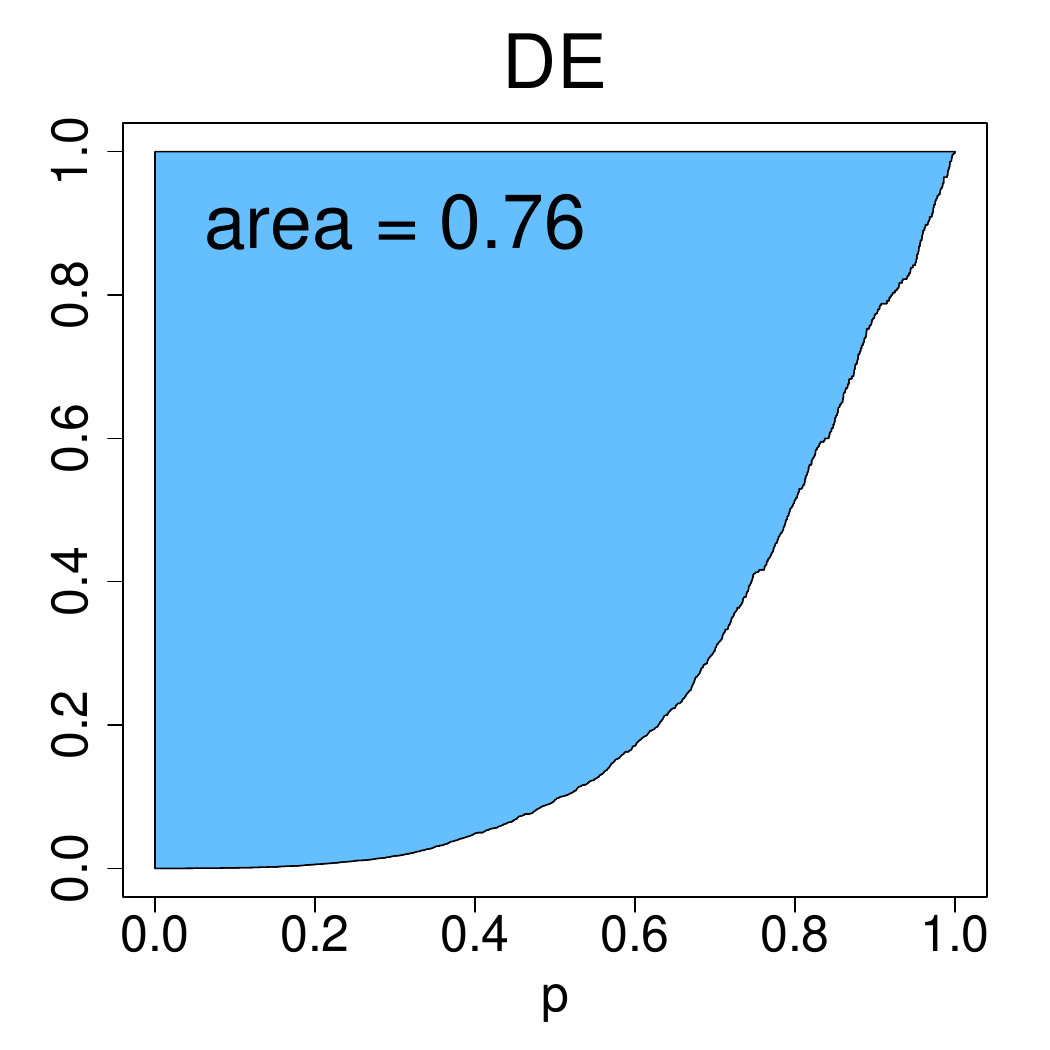}\qquad
\includegraphics[width=0.237\linewidth]{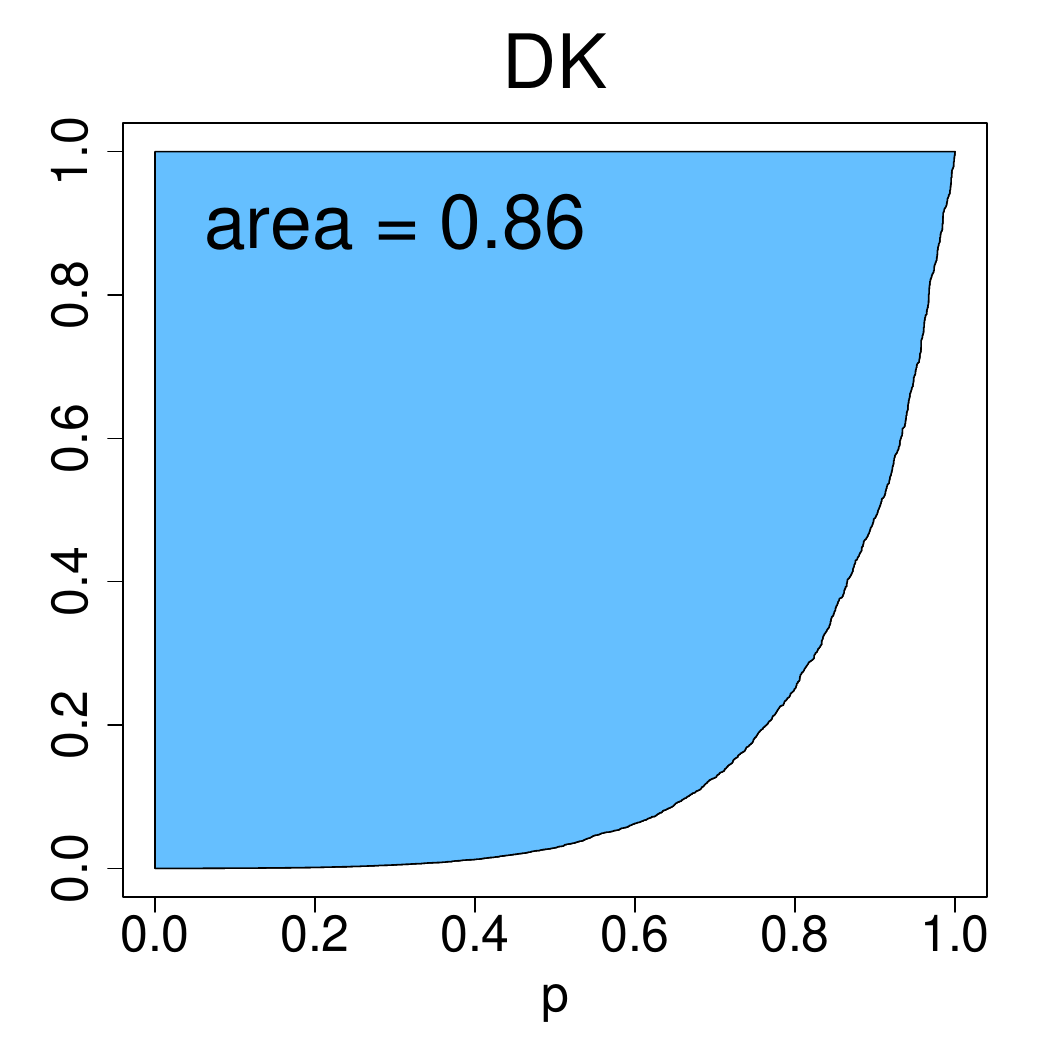}\qquad
\includegraphics[width=0.237\linewidth]{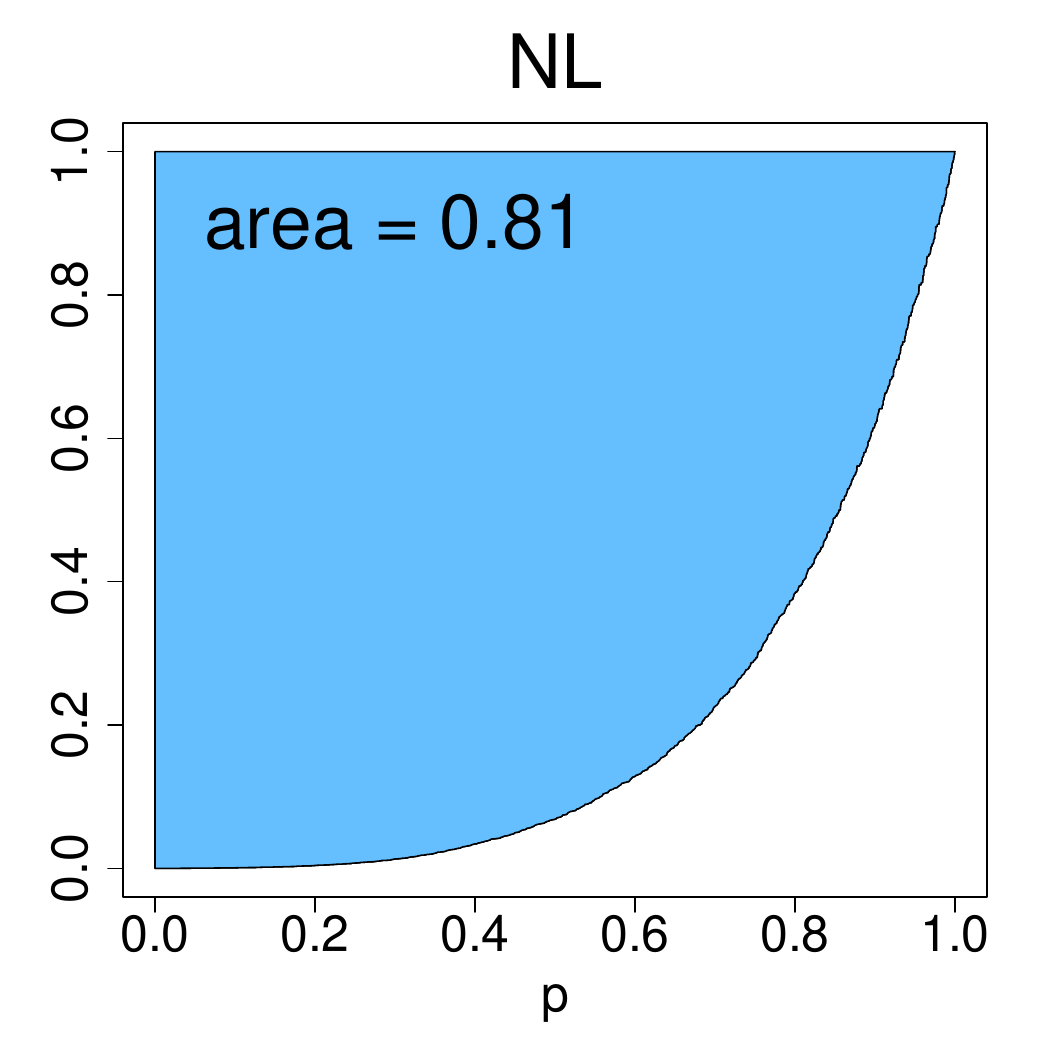}\qquad
\\
\includegraphics[width=0.237\linewidth]{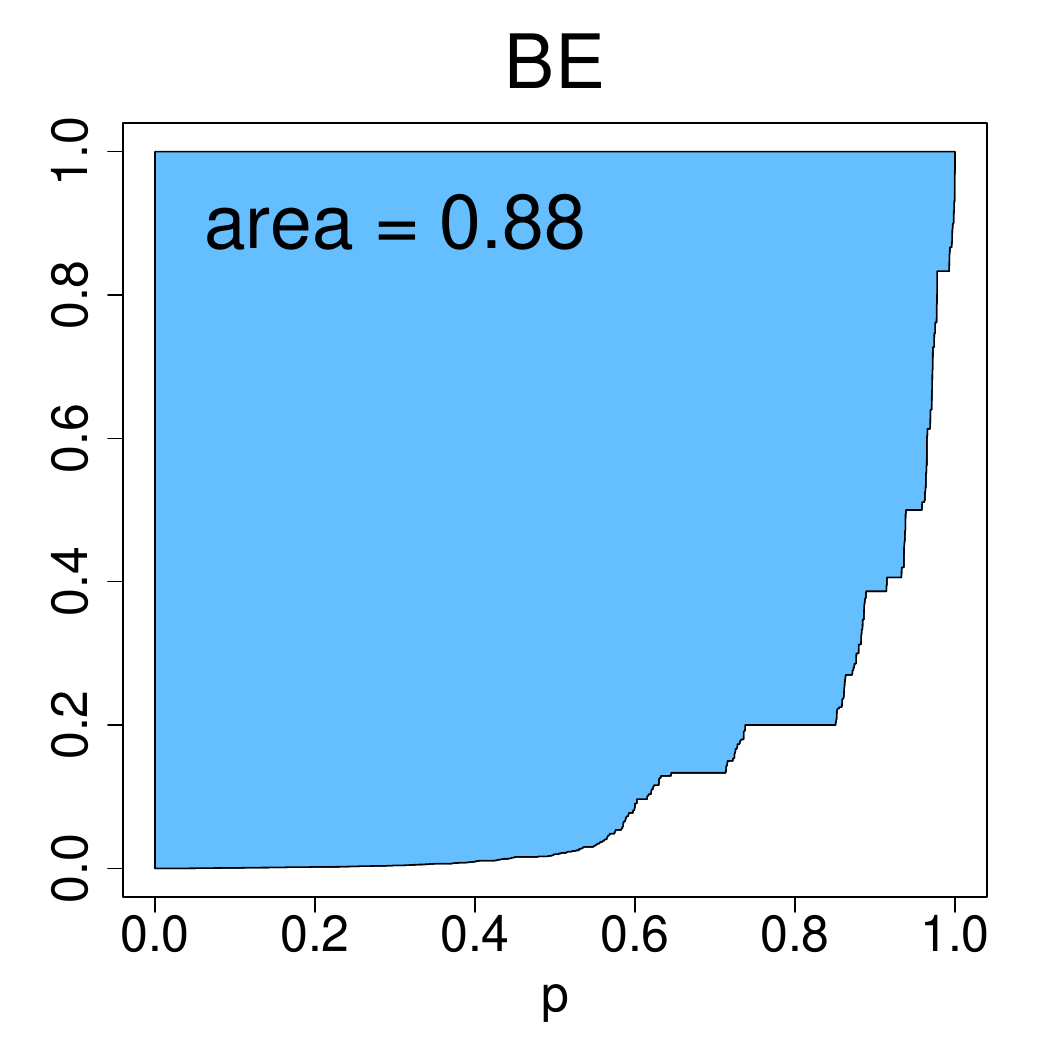}\qquad
\includegraphics[width=0.237\linewidth]{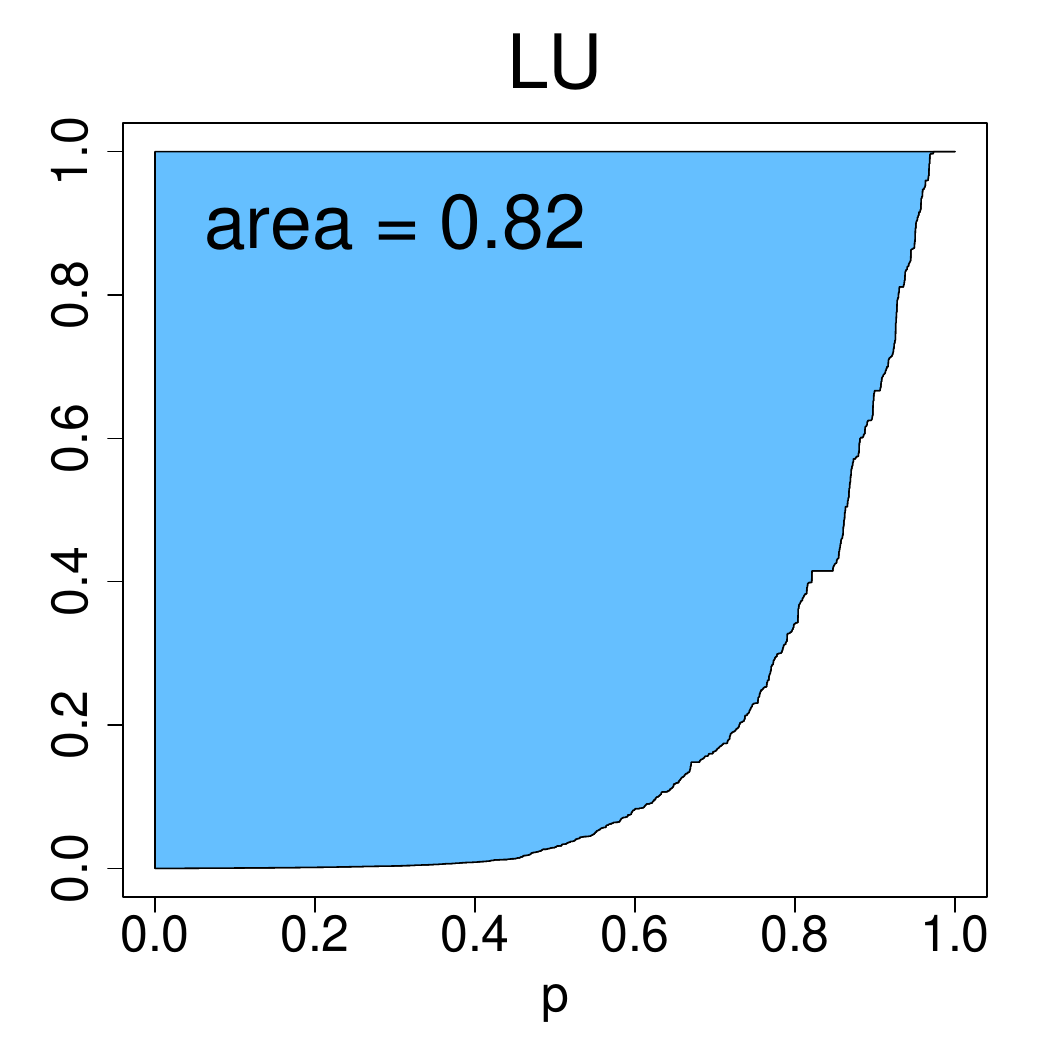}\qquad
\includegraphics[width=0.237\linewidth]{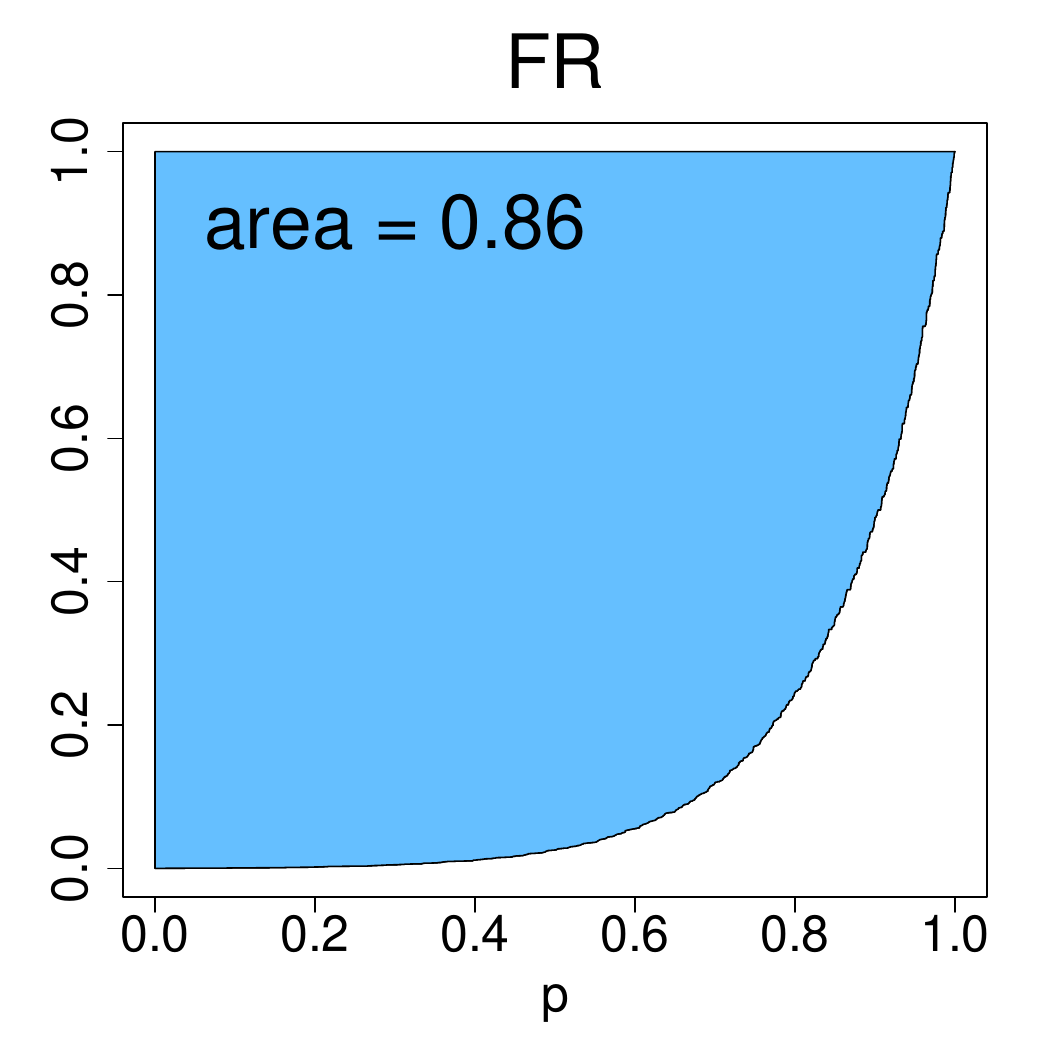}\qquad
\\
\includegraphics[width=0.237\linewidth]{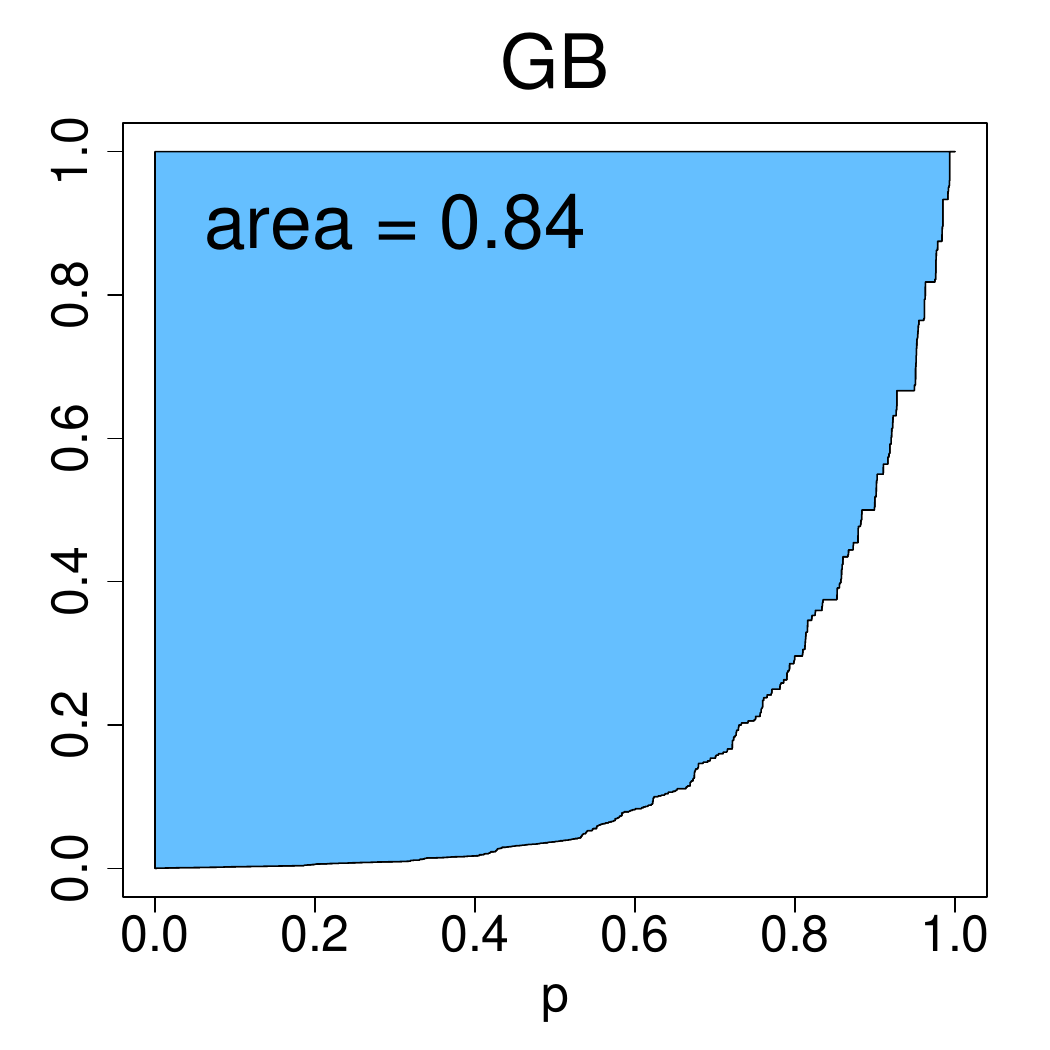}\qquad
\includegraphics[width=0.237\linewidth]{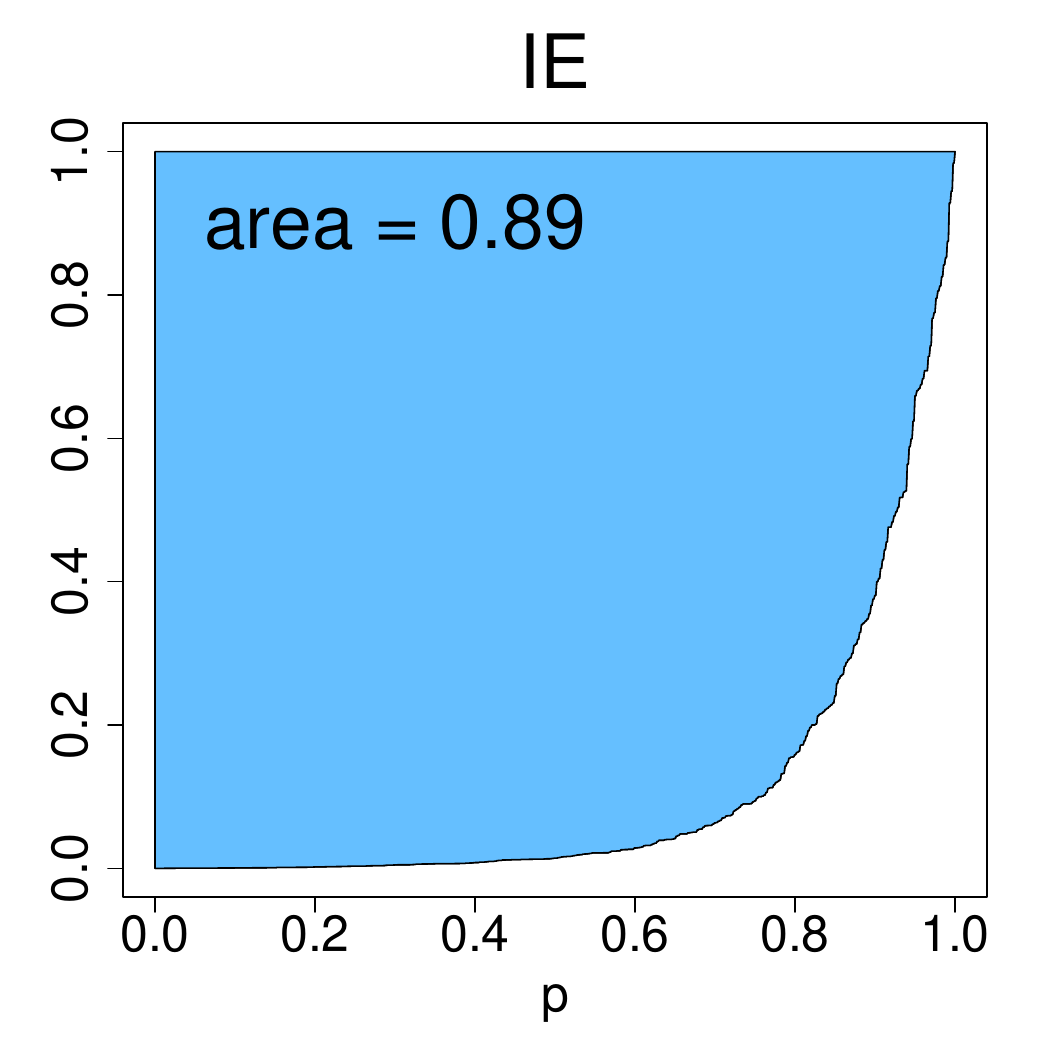}\qquad
\includegraphics[width=0.237\linewidth]{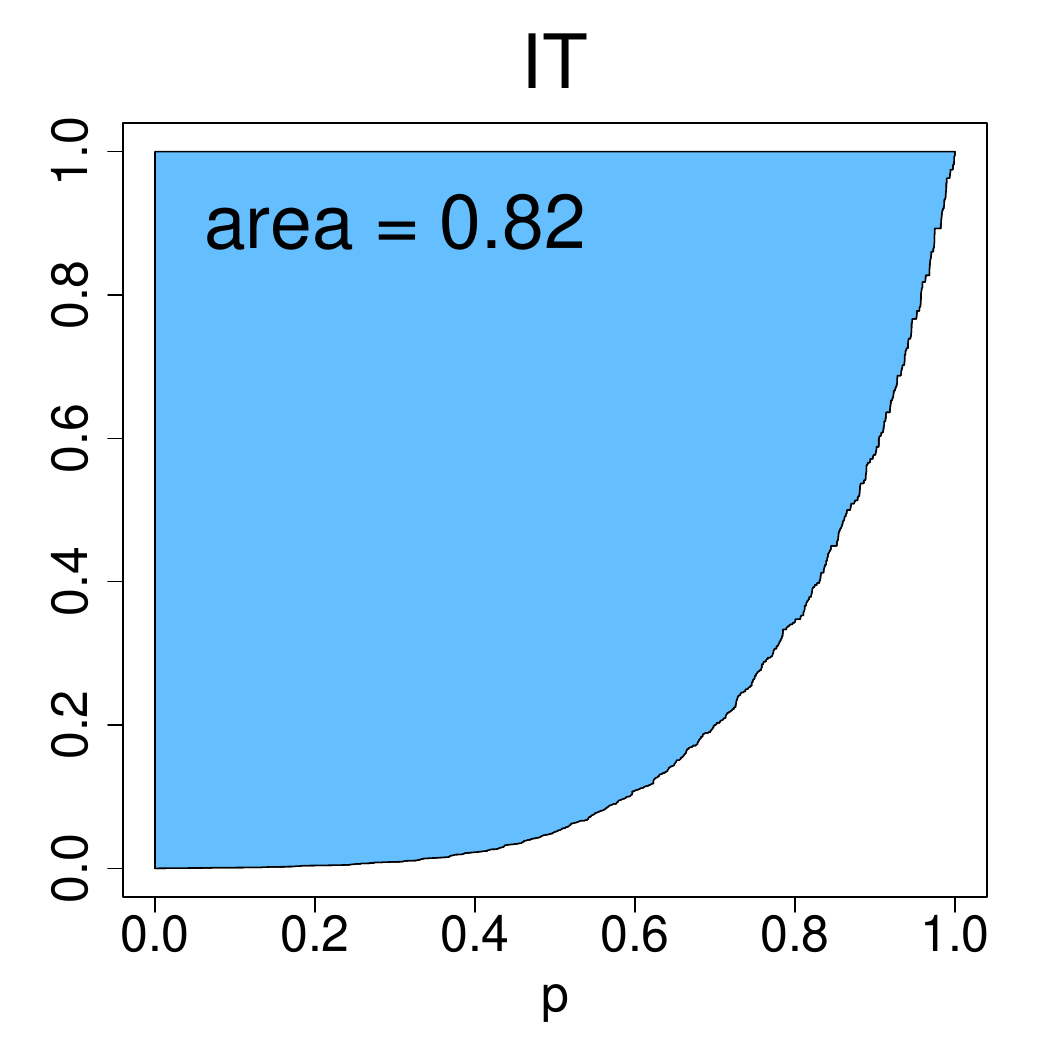}\qquad
\\
\includegraphics[width=0.237\linewidth]{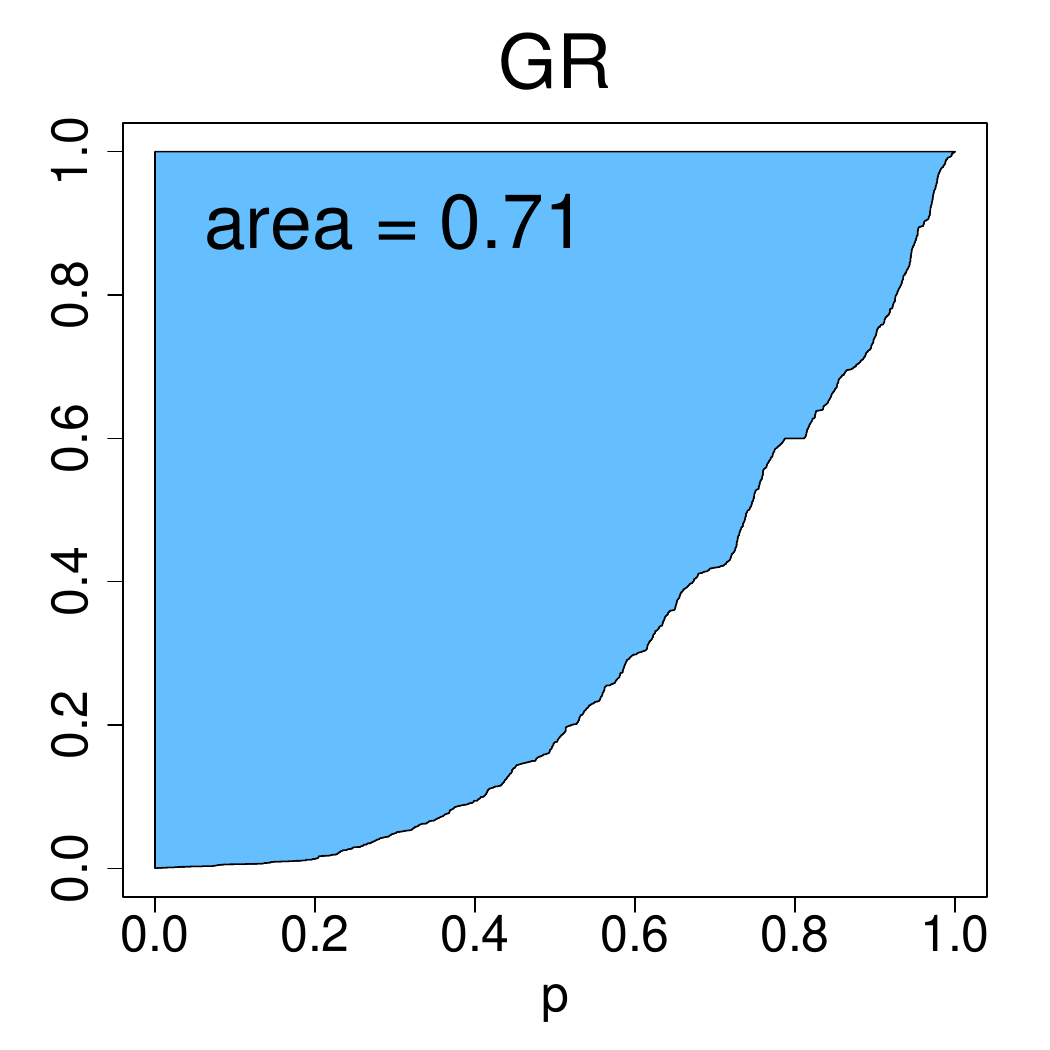}\qquad
\includegraphics[width=0.237\linewidth]{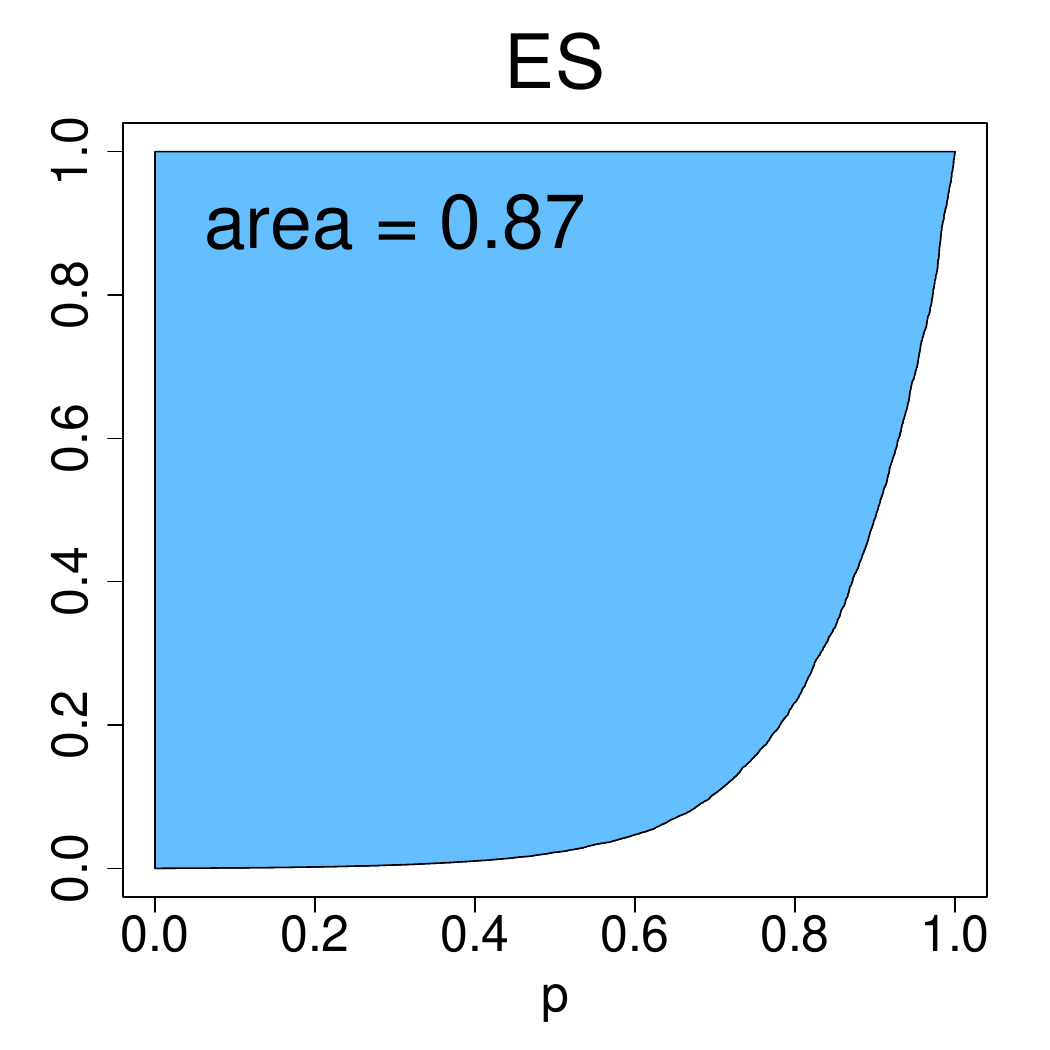}\qquad
\includegraphics[width=0.237\linewidth]{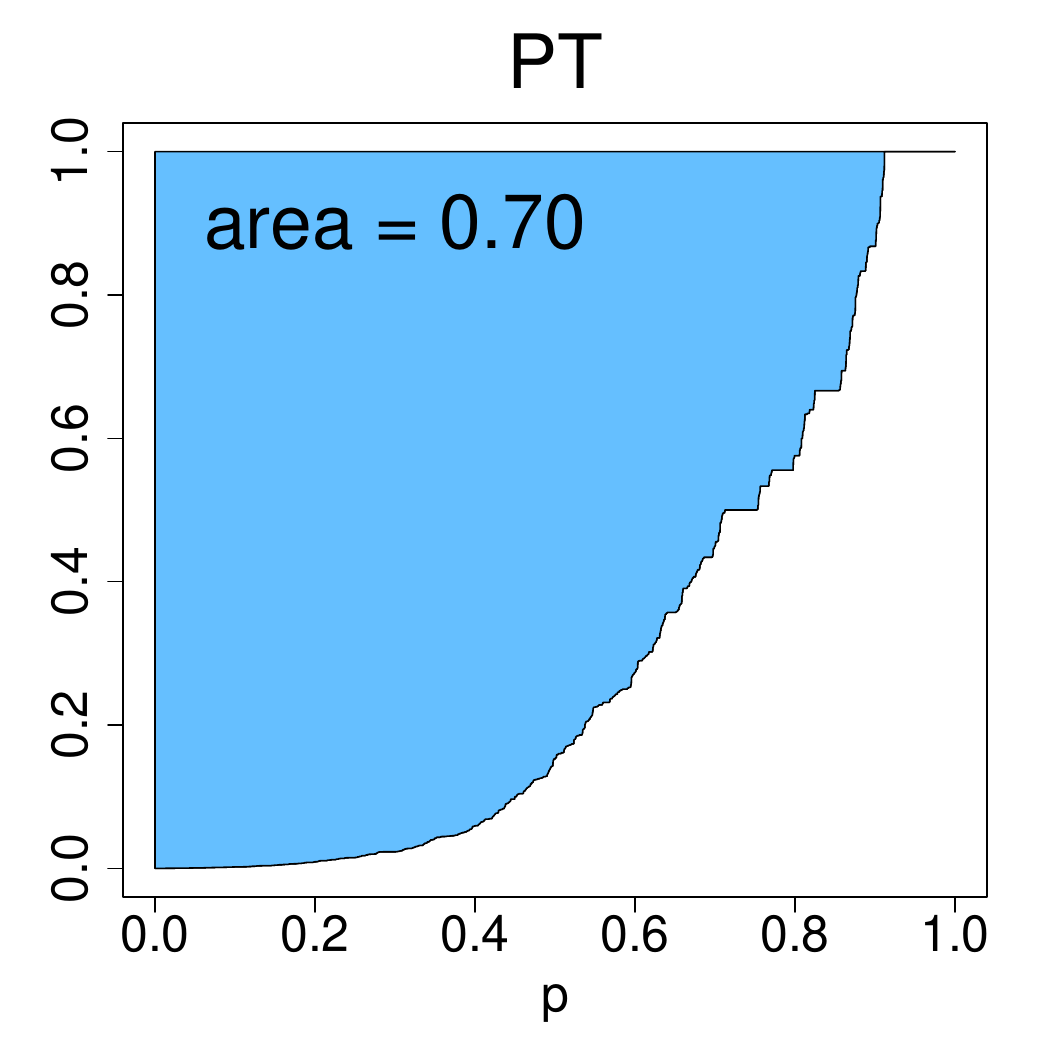}\qquad
\\
\includegraphics[width=0.237\linewidth]{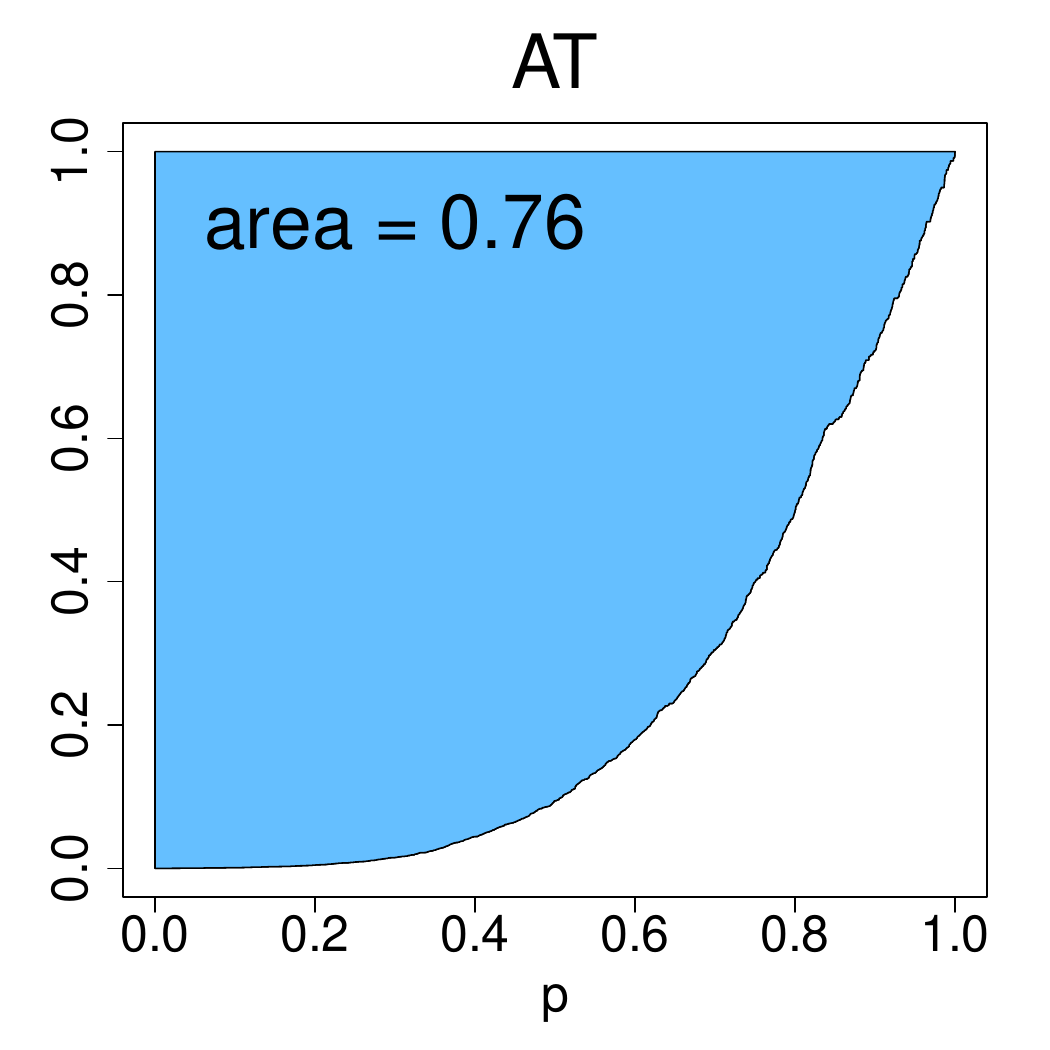}\qquad
\includegraphics[width=0.237\linewidth]{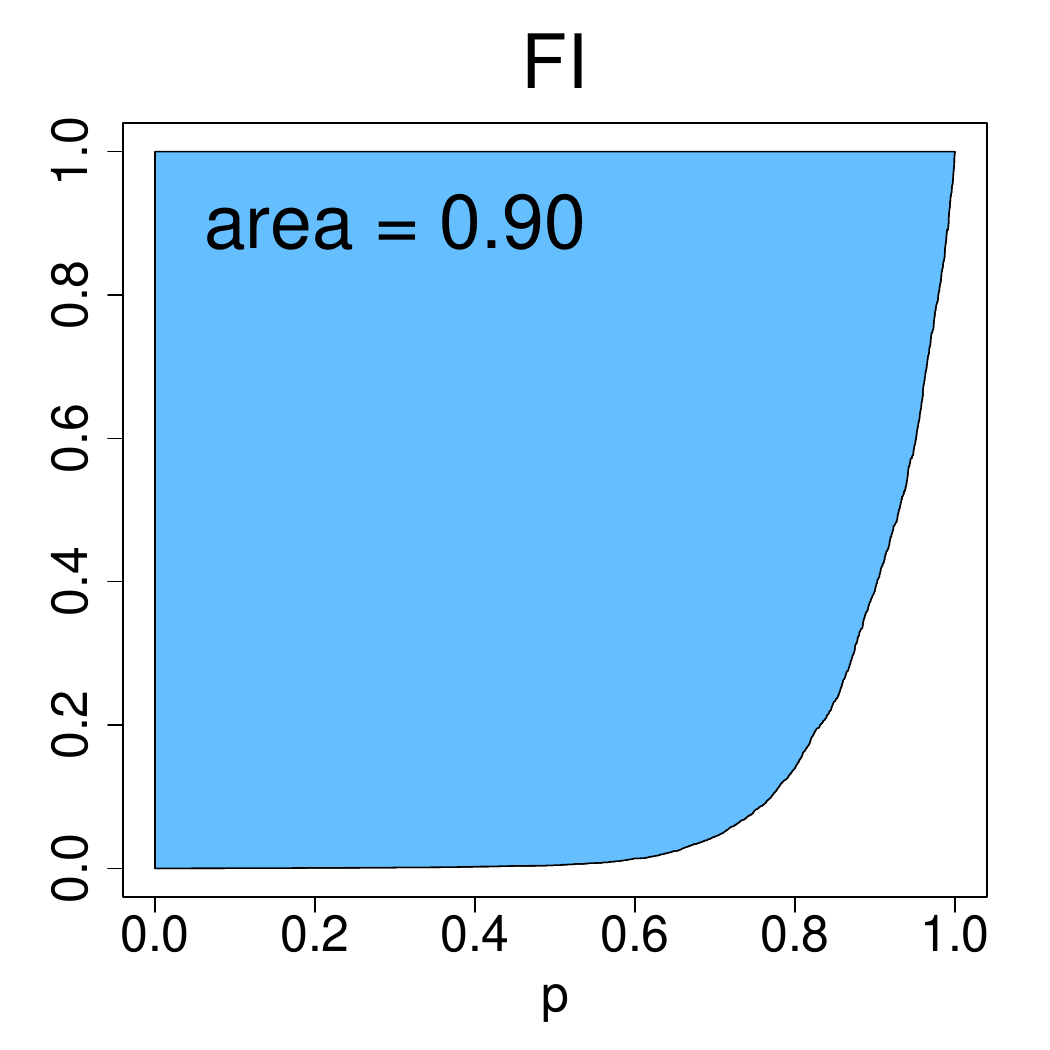}\qquad
\includegraphics[width=0.237\linewidth]{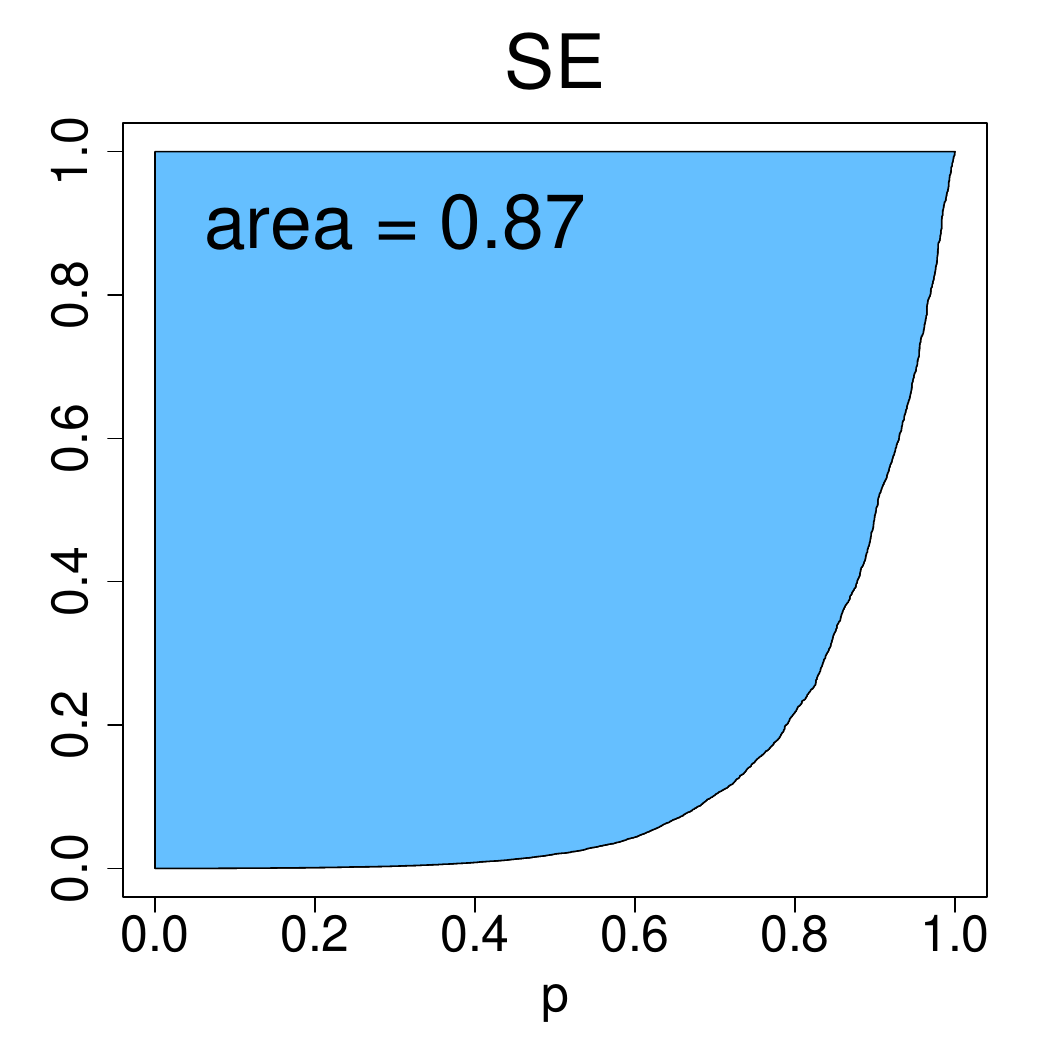}\qquad
\caption{The income-equality curve $\psi_{3,n}$ and the shaded-in area (i.e., $\Psi_{3,n}$) above it for the fifteen European countries, where $n=n_P$ is specified in Table~\ref{tab-real2018}  \citep[based on][]{EU-SILC2018}.}
\label{fig:6}
\end{center}
\end{figure}

Figures~\ref{fig:4}--\ref{fig:6}
depict the three income-equality curves $\psi_{k,n}$  for the fifteen European countries specified in Table~\ref{tab-real2018}, with the shaded-in  areas above them depicting the values of the indices $\Psi_{k,n}$. We observe from the simultaneous inspection of the plots that, in general, the Pareto models fit well the data, and  that the Gamma distribution ($\theta,\alpha=0.5$) can be a good model for the capital incomes in Denmark, France, Ireland, Spain and Sweden.

\section{The effects of income transfers on the new indices}
\label{est-pd}

Consider $n$ persons whose ordered incomes we denote by $X_{1:n}< \cdots < X_{n:n}$. Choose any pair from these persons and call them $L$ and $H$. The person $L \in \{1,\dots , n-1\}$ possesses income $X_{L:n}$ and the person $H \in \{2,\dots, n\}$ possesses income $X_{H:n}$. We assume $L<H$.  Hence, $L$ has less income than $H$, that is, $X_{L:n}< X_{H:n}$. Denote $\mathbf{X}=(X_{1:n}, \dots , X_{n:n})$.

Assume now that $H$ transfers a positive amount $c>0$ to $L$ without changing the income ordering among the $n$ persons. The transfer produces $\mathbf{X}'=(X_{1:n}', \dots , X_{n:n}')$ with the same ordering $X_{1:n}'< \cdots < X_{n:n}'$ of the coordinates as in the case of $\mathbf{X}$. (See Appendix~\ref{technicalities} for additional technical details.) Succinctly, we denote the transfer by
\begin{equation}\label{hlc}
L \stackrel{c}{\longleftarrow} H
\end{equation}
and read it, e.g., ``$L$ receives amount $c$ from $H$'' or ``$H$ transfers amount $c$ to $L$.''
We are interested in how the three indices $\Psi_{k,n}=\Psi_{k,n}(\mathbf{X})$ react to such transfers, that is, when $\mathbf{X}$ turns into $\mathbf{X}'$.

\begin{figure}[h!]
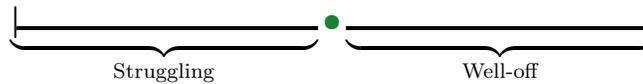

\begin{center}
\[
\underbrace{\boldsymbol{|}\hspace{-0.5mm}\rule{40mm}{0.5mm}}_{\textrm{Struggling}}
 {\boldsymbol{\gr \bullet}}
\underbrace{\rule{40mm}{0.5mm}\hspace{-0.5mm}\boldsymbol{|}}_{\textrm{Well-off}}
\]
\end{center}
  \caption{The median (green) delineates the struggling group from the well-off.}\label{fig-sw51}
\end{figure}

In addition to $L$ and $H$, we also involve the ``median'' person
\[
M:=\lceil n/2 \rceil
\]
whose income is $X_{M:n}=Q_n(1/2)$ as per equation~\eqref{equantile function} with $p=1/2$. Any person $P$ with income above the median (i.e., when $P>M$) is called \textit{well-off}, and any person $P$ with income below the median (i.e., when $P<M$) is called  \textit{struggling} (see Figure~\ref{fig-sw51}).
In what follows, we shall be interested in the effects of transfer~\eqref{hlc} on the new three indices when both $L$ and $H$ are well-off, both are struggling, and when one of them (i.e., $L$) is struggling and the other one (i.e., $H$) is well-off.

Before going into details, we note that the classical Pigou-Dalton principle (PDP) -- when it holds -- says that $\Psi_{k,n}(\mathbf{X})\ge \Psi_{k,n}(\mathbf{X}')$ in its weak form and $\Psi_{k,n}(\mathbf{X})> \Psi_{k,n}(\mathbf{X}')$ in its strong form. As we shall soon see, the three new indices will tell us a richer story. Based on it, we shall be able to choose a preferred index, or at least be prompted to think outside the box, which is necessary as \citet{ac1999} have convincingly argued.

\subsection{Index $\Psi_{1,n}$}

\begin{property}\label{prop53}
In the case of struggling $L$ and well-off $H$ (i.e., $L <M<H$), the transfer $L \stackrel{c}{\longleftarrow} H $  diminishes the value of the index $\Psi_{1,n}$, that is, we have $\Psi_{1,n}(\mathbf{X})> \Psi_{1,n}(\mathbf{X}')$.
\end{property}

\begin{property}\label{prop51}
When both $L$ and $H$ are well-off (i.e., $M <L<H$), or when both are struggling (i.e., $L<H< M $), the transfer $L \stackrel{c}{\longleftarrow} H $ does not change the value of the index $\Psi_{1,n}$, that is, we have $\Psi_{1,n}(\mathbf{X})= \Psi_{1,n}(\mathbf{X}')$.
\end{property}

These two properties say that in order to decrease income inequality based on the index $\Psi_{1,n}$, a well-off person needs to transfer some amount to a struggling person, whereas any transfer between two well-off persons or between two struggling ones does not make any difference.

\subsection{Index $\Psi_{2,n}$}

The index $\Psi_{2,n}$ is more sensitive to transfers than the previous index. Specifically, we shall see from the following properties that $\Psi_{2,n}$ decreases when  $L \stackrel{c}{\longleftarrow} H $, unless both  $H$ and $L$ are well-off and  $H$ transfers to $L$ only a small amount $c>0$.

\begin{property}\label{prop56}
In the case of struggling $L$ and well-off $H$  (i.e., $L< M < H$), or when both $L$ and $H$ are struggling  (i.e., $L<H< M $), the transfer $L \stackrel{c}{\longleftarrow} H $ diminishes the value of the index $\Psi_{2,n}$, that is, $\Psi_{2,n}(\mathbf{X})> \Psi_{2,n}(\mathbf{X}')$.
\end{property}

\begin{property}\label{prop55}
When both $L$ and $H$ are well-off (i.e., $M <L<H$), the transfer $L \stackrel{c}{\longleftarrow} H $ implies $\Psi_{2,n}(\mathbf{X})> \Psi_{2,n}(\mathbf{X}')$ when
\begin{equation}\label{delt-1}
c> c_2:={X_{L-M :n}X^2_{H:n} - X_{H-M :n}X^2_{L:n}
\over X_{L-M :n}X_{H:n} + X_{H-M :n}X_{L:n} } .
\end{equation}
Furthermore, we have $\Psi_{2,n}(\mathbf{X})= \Psi_{2,n}(\mathbf{X}')$ when $c=c_2$, and
$\Psi_{2,n}(\mathbf{X})< \Psi_{2,n}(\mathbf{X}')$ when $c<c_2$.
\end{property}

Hence, the index $\Psi_{2,n}$ avoids giving the impression of inequality reduction when only a small amount is transferred among well-off persons. In other words, for the index to decrease in the case of two well-off persons, the richer one needs to transfer a sufficiently large amount in order to qualify for inequality reduction.
Next is an example illustrating Properties~\ref{prop56} and~\ref{prop55}.

\begin{example}\label{example-51}
Consider a group of seven persons, among whom there are three struggling ones (denoted by $S$'s) and three well-off persons (denoted by $W$'s). The person $M$ has the median income $X_{M:7}$ among these seven persons, and thus a ``$7$'' in its notation. Let their incomes be
\begin{align}
\mathbf{X}
&=(X_{1:7},X_{2:7},X_{3:7},X_{4:7},X_{5:7},X_{6:7},X_{7:7})
\notag
\\
&=(X_{S_1:7},X_{S_2:7},X_{S_3:7},X_{M:7},X_{W_1:7},X_{W_2:7},X_{W_3:7})
\notag
\\
&= (\underbrace{1, 3, 5,}_{\textrm{Incomes of $S$'s}} \overbrace{7,}^{\textrm{Income of $M$}} \underbrace{10, 20, 24}_{\textrm{Incomes of $W$'s}}) .
\label{incomes-0}
\end{align}
The index of inequality for this vector is $\Psi_{2,n}=0.8472$. Hence, $n=7$ and thus $M=\lceil 3.5 \rceil =4$, which gives the median income $X_{4:7}=7$. There are three struggling persons $S_1=1$, $S_2=2$, and $S_3=3$ with incomes $1$, $3$, and $5$, respectively, and three well-off persons $W_1=5$, $W_2=6$, and $W_3=7$ with incomes $10$, $20$, and $24$, respectively (see the top-left panel in Figure~\ref{transfers2}
\begin{figure}[h!]
\centering
\includegraphics[width=1.0\linewidth]{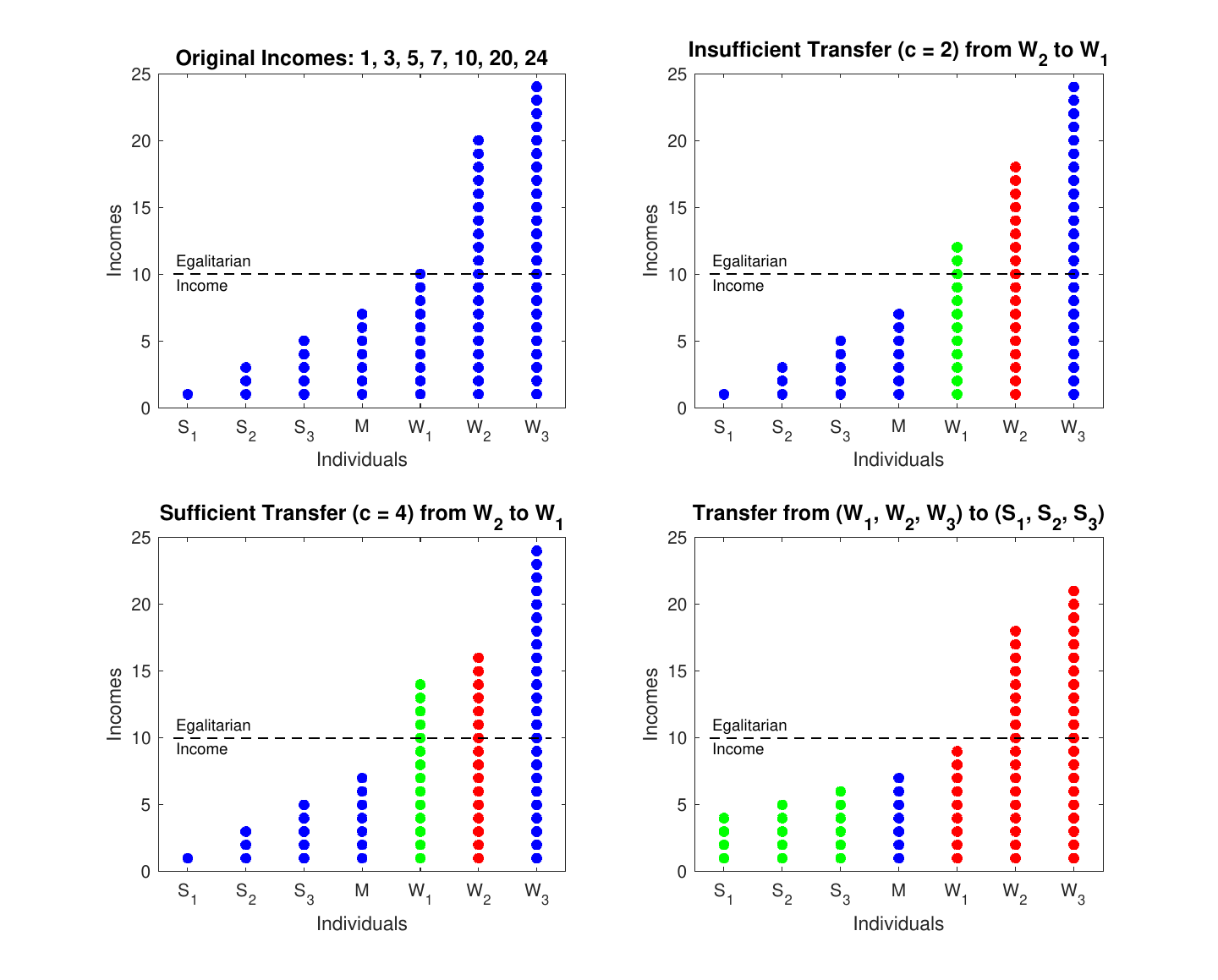}
\caption{Distributions of incomes with dots representing units, or amounts, of income: the blue dots correspond to the original distribution of incomes, the red ones correspond to reduced incomes due to transfers, and the green dots correspond to increased incomes.}
\label{transfers2}
\end{figure}
for a visualization). The horizontal dashed line in each panel of Figure~\ref{transfers2}, noted as ``egalitarian income'' and plotted at the height $10$, refers to the egalitarian redistribution of the above specified incomes (whose sum is equal to $70$) among the seven participating persons.

Choose now two well-off persons, say $W_1(=L)=5$ and $W_2(=H)=6$. We have $L-M=1$ and $H-M=2$. Condition~\eqref{delt-1} is equivalent to
\[
c>c_2= {1\times 20^2 - 3\times 10^2
\over 1\times 20 + 3\times 10 } = 2.
\]
For the ordering of incomes to remain the same after the transfer $L \stackrel{c}{\longleftarrow} H $, we need the restriction
\[
c< {X_{H:7}-X_{L:7} \over 2}=5.
\]
Hence, to decrease income inequality according to the index $\Psi_{2,n}$, the person $H$ needs to transfer to $L$ more than $2$, but less than $5$ to avoid swapping the position with $L$.

The top-right panel of Figure~\ref{transfers2} depicts the transfer from $W_2=6$ to $W_1=5$ of  the insufficient for inequality decrease amount $c=2$, in which case we have the distribution
\begin{equation}\label{incomes-1}
(1, 3, 5, 7, 12, 18, 24)
\end{equation}
with the value of the index remaining the same, that is, $\Psi_{2,n}=0.8472$.

The bottom-left panel of Figure~\ref{transfers2} depicts the transfer from $W_2=6$ to $W_1=5$ of the sufficient for inequality decrease amount $c=4$, in which case we have
\begin{equation}\label{incomes-2}
(1, 3, 5, 7, 14, 16, 24)
\end{equation}
with the value of the index $\Psi_{2,n}=0.8442$.

We now consider a more complex situation, depicted in the bottom-right panel of Figure~\ref{transfers2}, when every well-off person commits to improving the incomes of the three struggling persons, with the final distribution of incomes becoming
$(4, 5, 6, 7, 9, 18, 21) $.
We achieve this distribution in several steps, each reducing income inequality and maintaining the original ordering of the seven persons. Recall that we start from the vector $(1, 3, 5, 7, 10, 20, 24) $, whose inequality index is $\Psi_{2,n}=0.8472$, and the steps could be these:
The transfer $ S_3 \stackrel{1}{\longleftarrow} W_1 $ results in the distribution
\begin{equation}\label{incomes-3}
(1, 3, 6, 7, 9, 20, 24)
\end{equation}
 with the index $\Psi_{2,n}=0.8296$.
The transfer $ S_2\stackrel{2}{\longleftarrow} W_2 $ results in
\begin{equation}\label{incomes-4}
 (1, 5, 6, 7, 9, 18, 24)
\end{equation}
with the index $\Psi_{2,n}=0.7870$.
Finally, the transfer $ S_1\stackrel{3}{\longleftarrow} W_3 $ results in the distribution
\begin{equation}\label{incomes-5}
 (4, 5, 6, 7, 9, 18, 21)
\end{equation}
depicted in the bottom right panel of Figure~\ref{transfers2} and having the index $\Psi_{2,n}=0.6640$.
All these are inequality-reducing transfers from well-off persons to struggling ones.

Alternatively, without delving into the psychology of people and thus plausibility of transfers, we can have the following steps, some of which involve two well-off persons and some involve both well-off and struggling persons, leading to the same end-result $(4, 5, 6, 7, 9, 18, 21) $, the same as above:
\begin{enumerate}[label={\rm$\arabic*$)}]
\item\label{step1} $W_1 \stackrel{3}{\longleftarrow} W_2 $ results in $ (1, 3, 5, 7, 13, 17, 24)$ with  $\Psi_{2,n}=0.8461$
\item\label{step2} $W_2 \stackrel{3}{\longleftarrow} W_3 $ results in $ (1, 3, 5, 7, 13, 20, 21)$ with  $\Psi_{2,n}=0.8450$
\item\label{step3} $ S_3\stackrel{1}{\longleftarrow} W_2 $ results in $ (1, 3, 6, 7, 13, 19, 21)$ with  $\Psi_{2,n}=0.8265$
\item\label{step4} $ S_2\stackrel{1}{\longleftarrow} W_2 $ results in $ (1, 4, 6, 7, 13, 18, 21)$ with  $\Psi_{2,n}=0.8050$
\item\label{step5} $ S_2 \stackrel{1}{\longleftarrow} W_1 $ results in $ (1, 5, 6, 7, 12, 18, 21)$ with  $\Psi_{2,n}=0.7844$
\item\label{step6} $ S_1\stackrel{3}{\longleftarrow} W_1 $ results in $ (4, 5, 6, 7, 9, 18, 21)$ with  $\Psi_{2,n}=0.6640$
\end{enumerate}
Step~\ref{step1} is justified by our earlier argument at the beginning of this example saying that any transfer higher than $2$ but less than $5$ from $W_2$ to $W_1$ is legitimate, and we transfer $c=3$. To justify Step~\ref{step2}, we note that we can only transfer less than $(24-17)/2=3.5$ but more than
$(3\times 24^2 - 5\times 17^2) /(3\times 24+ 5\times 17)= 1.8025$, and so we transfer $c=3$. All Steps~\ref{step3}--\ref{step6} are from well-off persons to struggling ones, and so the only requirement on the transfers is that they should maintain the original ordering of incomes.
This concludes Example~\ref{example-51}.
\end{example}

\subsection{Index $\Psi_{3,n}$}

\begin{property}\label{prop59}
In the case of struggling $L$ and well-off $H$   (i.e., $L <M<H$), the transfer $L \stackrel{c}{\longleftarrow} H $ diminishes the value of the index $\Psi_{3,n}$, that is, $\Psi_{3,n}(\mathbf{X})> \Psi_{3,n}(\mathbf{X}')$.
\end{property}

\begin{property}\label{prop57}
When both $L$ and $H$ are well-off (i.e., $M <L<H$), or when both are struggling (i.e., $L<H< M $),  the transfer $L \stackrel{c}{\longleftarrow} H $ increases the value of the index $\Psi_{3,n}$, that is, we have  $\Psi_{3,n}(\mathbf{X})< \Psi_{3,n}(\mathbf{X}')$.
\end{property}

Hence, when the goal is to decrease income inequality, these two properties say that well-off persons must transfer to struggling persons, and the index discourages  transfers between two well-off persons, or between two struggling ones, as the index views such transfers manipulative with no real consequences. Whether we agree with this or not determines whether or not we shall adopt the index $\Psi_{3,n}$ for measuring income inequality.

Having by now discussed the three indices and their properties, we next have a numerical example that illustrates the performance of the three indices side-by-side.

\begin{example}\label{ex-5.1}
Consider the six distributions of incomes specified in~\eqref{incomes-0}--\eqref{incomes-5} and visualized in Figure~\ref{transfers2}. Table~\ref{table-3indices}
\begin{table}[h!]
\centering
\begin{tabular}{ccccccc}
  \hline\hline
Indices  &  \eqref{incomes-0}  &  \eqref{incomes-1}  &  \eqref{incomes-2}  &  \eqref{incomes-3}  &  \eqref{incomes-4}  &  \eqref{incomes-5}\\
  \hline
$\Psi_{1,n}$  &  0.5714  &  0.5714  &  0.5714  &  0.5238  &  0.4286  &  0.2857\\
$\Psi_{2,n}$  &  0.8472  &  0.8472  &  0.8442  &  0.8296  &  0.7870  &  0.6640\\
$\Psi_{3,n}$  &  0.7694  &  0.7917  &  0.8046  &  0.7139  &  0.6713  &  0.6217\\
  \hline
\end{tabular}
\caption{The three indices for income distributions~\eqref{incomes-0}--\eqref{incomes-5}.}
\label{table-3indices}
\end{table}
contains the numerical values of the three indices for the six income distributions. We see from the values that the index $\Psi_{1,n}$ remains the same when transfers are only among well-off persons (distributions~\eqref{incomes-1} and~\eqref{incomes-2}) and diminishes in the case of transfers from well-off persons to struggling ones (distributions~\eqref{incomes-3}--\eqref{incomes-5}). The performance of the index $\Psi_{2,n}$ has already been amply discussed and so we move on to $\Psi_{3,n}$. Unlike $\Psi_{1,n}$, the index $\Psi_{3,n}$ increases when transfers are only among well-off persons (distributions~\eqref{incomes-1} and~\eqref{incomes-2}) but diminishes in the case of transfers from well-off persons to struggling ones (distributions~\eqref{incomes-3}--\eqref{incomes-5}). This concludes Example~\ref{ex-5.1}.
\end{example}

\section{Conclusion}
\label{conclusion}

In this paper we have introduced and explored three inequality indices that reflect three views of measuring income inequality:
\begin{enumerate}[label={(\arabic*)}]
\item\label{v.1} The median income of the poor is compared with the median income of the entire population.
\item\label{v.2}
 The median income of the poor is compared with the median income of those who are not poor.
\item\label{v.3}
  The median income of the poor is compared with the median of the same proportion of the richest.
\end{enumerate}
We have presented these inequality indices and their equality curves in two ways: one that is suitable for modeling populations parametrically, and the other one that is suitable for direct  data-focused computations. Several properties of the indices have been derived and discussed, most notably their behaviour with respect to income transfers. The indices and their curves have been illustrated using popular parametric models of income distributions, and also calculated and interpreted using real data. The new indices do not require any finite moment and, therefore, are suitable (mathematically) for analyzing all populations, including those that are ultra heavily tailed, that is, do not even have a finite first moment.

\appendix
\section{Technicalities}
\label{technicalities}

\begin{proof}[Justification of definitions~\eqref{def-psi1}--\eqref{def-psi3}]
The three empirical indices arise from formulas~\eqref{index-1}--\eqref{index-3} by  first replacing the population quantile function $Q$ by the empirical quantile function $Q_n$ in all the formulas. (We have asymptotically insignificantly modified the obtained expressions to facilitate their intuitive appeal.) In detail, with $F_n$ denoting the empirical cumulative distribution function based on $X_{1},\dots , X_{n} $, the empirical quantile function is given by equation~\ref{equantile function}.
Thus, for example, $Q_n(1/2)=X_{M :n}$, which is the empirical median used in the definition of $\Psi_{1,n}$. Note also that $\lfloor n/2 \rfloor + M = n $, and thus the definition of the index $\Psi_{2,n}$ does not go beyond the random variables $X_{1},\dots , X_{n} $.
\end{proof}

\begin{proof}[Proof of Property~\ref{prop42}]
The inequality holds because
$a/b\le (a+c)/(b+c)$ for all (positive) $a\le b$ and $c\ge 0$, and to have the strict inequality, we note that $a/b< (a+c)/(b+c)$ holds for all (positive) $a< b$ and $c> 0$.
\end{proof}

\begin{proof}[Proof of Property~\ref{prop43}]
The property follows from  $(a+c)/(b+c)\to 1$ when $c\to \infty $ irrespective of the values of (positive) $a\le b$.
\end{proof}

\begin{proof}[Details of definition~\eqref{hlc}]
\begin{align*}
X_{i:n}'&=X_{i:n} \quad \textrm{for} \quad 1\le i \le L-1,
\\
X_{L:n}'&=X_{L:n}+c ,
\\
X_{i:n}'&=X_{i:n} \quad \textrm{for} \quad L+1\le i \le H-1,
\\
X_{H:n}'&=X_{H:n}-c ,
\\
X_{i:n}'&=X_{i:n} \quad \textrm{for} \quad H+1\le i \le n,
\end{align*}
where $L$ and $H$ are integers such that $1\le L<H\le n$, and $c>0$ is any positive real number (i.e., the amount transferred from $H$ to $L$) such that the following ordering holds:
\begin{equation}\label{bounds-0}
X_{1:n}< \cdots
<
X_{L-1:n}< X_{L:n}+c< X_{L+1:n}
< \cdots
<
X_{H-1:n}< X_{H:n}-c < X_{H+1:n}
<  \cdots < X_{n:n}.
\end{equation}
When inequalities~\eqref{bounds-0} hold, we succinctly denote this transfer by $L \stackrel{c}{\longleftarrow} H $.
\end{proof}

\begin{proof}[Proof of Property~\ref{prop53}]
Since $L< M < H$, the increase in $L$'s income affects the index
\[
\Psi_{1,n}= 1- {1\over \lfloor n/2 \rfloor} \sum_{k=1}^{\lfloor n/2 \rfloor}
{X_{k:n} \over X_{M :n}}
\]
because
\[
{X_{L:n} \over X_{M :n} }
< {X_{L:n}+c \over X_{M :n}},
\]
whereas the decrease in $H$'s income does not affect $\Psi_{1,n}$ because $H$ is not among the terms making up the definition of the index. Hence,
$\Psi_{1,n}(\mathbf{X})> \Psi_{1,n}(\mathbf{X}')$.
\end{proof}

\begin{proof}[Proof of Property~\ref{prop51}]
When $M <L<H$, the index $\Psi_{1,n}$
is not affected by the transfer $L \stackrel{c}{\longleftarrow} H $ because, mathematically speaking, $L$ and $H$ are outside the summation range due to $\lfloor n/2 \rfloor \le M$ and, according to property~\eqref{bounds-0}, the transfer does not change the ordering of incomes. In other words, the median income and the incomes below it are not affected by the transfer, and  we therefore have $\Psi_{1,n}(\mathbf{X})= \Psi_{1,n}(\mathbf{X}')$.

When $L<H< M $, both $L$ and $H$ are among the terms in the sum making up the definition of $\Psi_{1,n}$. Since we have the equations
\begin{align*}
{X_{L:n} \over X_{M :n}}+ {X_{H:n} \over X_{M :n}}
&={X_{L:n}+c  \over X_{M :n}}+ {X_{H:n}-c  \over X_{M :n}}
\\
&= {X'_{L:n} \over X'_{M :n}}+ {X'_{H:n} \over X'_{M :n}}
\end{align*}
the value $\Psi_{1,n}$ is not affected by the transfer $L \stackrel{c}{\longleftarrow} H $. This implies
$\Psi_{1,n}(\mathbf{X})= \Psi_{1,n}(\mathbf{X}')$ and establishes Property~\ref{prop51}.
\end{proof}

\begin{proof}[Proof of Property~\ref{prop56}]
Consider first the case when $L< M < H$.
Since $L< M $, we have $L\le \lfloor n/2 \rfloor $, and so the index
\[
\Psi_{2,n}= 1- {1\over \lfloor n/2 \rfloor} \sum_{k=1}^{\lfloor n/2 \rfloor}
{X_{k:n} \over X_{M+k :n}} .
\]
is affected by the transfer $L \stackrel{c}{\longleftarrow} H $  because
\begin{align*}
{X_{L:n} \over X_{M  +L :n}}+{X_{H-M :n} \over X_{H:n}}
&< {X_{L:n}+c \over X_{M  +L :n}}+{X_{H-M :n} \over X_{H:n}-c },
\\
&= {X'_{L:n} \over X'_{M  +L :n}}+{X'_{H-M :n} \over X'_{H:n}}
\end{align*}
which implies $\Psi_{2,n}(\mathbf{X})> \Psi_{2,n}(\mathbf{X}')$.

When $L<H< M $, we have $L<H \le \lfloor n/2 \rfloor$, and so
\begin{align*}
{X_{L:n} \over X_{M  +L :n}} + {X_{H:n} \over X_{M  +H :n}}
&<
{X_{L:n}+c \over X_{M  +L :n}} + {X_{H:n}-c \over X_{M  +H :n}}
\\
&= {X'_{L:n} \over X'_{M  +L :n}} + {X'_{H:n} \over X'_{M  +H :n}},
\end{align*}
with the inequality holding because $X_{M  +L :n}<X_{M  +H :n}$. Hence, $\Psi_{2,n}(\mathbf{X})> \Psi_{2,n}(\mathbf{X}')$, thus concluding the proof of Property~\ref{prop56}.
\end{proof}

\begin{proof}[Proof of Property~\ref{prop55}]
Since  $M <L<H$, the incomes of $L$ and $H$ are above the median $X_{M :n}$, and so there are two $k$'s in the sum in the definition of $\Psi_{2,n}$
that give $M+k=L$ and $M+k=H$, respectively.
Consequently,
$\Psi_{2,n}(\mathbf{X})> \Psi_{2,n}(\mathbf{X}')$
holds if and only if the following inequality holds:
\begin{align*}
{X_{L-M :n} \over X_{L:n}}+{X_{H-M :n} \over X_{H:n}}
&<
{X_{L-M :n} \over X_{L:n}+c }+{X_{H-M :n} \over X_{H:n}-c },
\\
&= {X'_{L-M :n} \over X'_{L:n}}+{X'_{H-M :n} \over X'_{H:n}} .
\end{align*}
Simple algebra shows that the inequality is equivalent to $c> c_2$, where $c_2$ is defined by equation~\eqref{delt-1}.
This establishes Property~\ref{prop55}.
\end{proof}

\begin{proof}[Proof of Property~\ref{prop59}]
The transfer $L \stackrel{c}{\longleftarrow} H $  affects the index
\[
\Psi_{3,n}= 1- {1\over \lfloor n/2 \rfloor} \sum_{k=1}^{\lfloor n/2 \rfloor}
{X_{k:n} \over X_{n-k+1 :n}}
\]
via both $L$ and $H$ because
\begin{align*}
{X_{L:n} \over X_{n-L+1:n}}+ {X_{n-H+1:n} \over X_{H:n}}
&< {X_{L:n}+c \over X_{n-L+1:n}}+ {X_{n-H+1:n} \over X_{H:n}-c }
\\
&= {X'_{L:n} \over X'_{n-L+1:n}}+ {X'_{n-H+1:n} \over X'_{H:n}},
\end{align*}
which implies
$\Psi_{3,n}(\mathbf{X})> \Psi_{3,n}(\mathbf{X}')$ and establishes Property~\ref{prop59}.
\end{proof}

\begin{proof}[Proof of Property~\ref{prop57}]
Consider first the case of two well-off persons, that is, $M<L<H$.
Since $\lfloor n/2 \rfloor + \lceil n/2 \rceil=n$ and $M=\lceil n/2 \rceil$, we have $n-k+1>M$ for every $k\le \lfloor n/2 \rfloor$. Consequently, there are two $k$'s in the sum in the definition of $\Psi_{3,n}$ that give $n-k+1=L$ and $n-k+1=H$, respectively, because $M <L<H$. Hence, the inequality  $\Psi_{3,n}(\mathbf{X})\ge \Psi_{3,n}(\mathbf{X}')$ holds if and only if
\begin{align*}
{X_{n-L+1:n} \over X_{L:n}}+{X_{n-H+1:n} \over X_{H:n}}
&\le {X_{n-L+1:n} \over X_{L:n}+c }+{X_{n-H+1:n} \over X_{H:n}-c },
\\
&= {X'_{n-L+1:n} \over X'_{L:n}}+{X'_{n-H+1:n} \over X'_{H:n}}
\end{align*}
which is equivalent $c\ge c_3$, where
\[
c_3={ X_{n-L+1:n}X_{H:n}^2-X_{n-H+1:n}X_{L:n}^2\over X_{n-L+1:n}X_{H:n}+X_{n-H+1:n}X_{L:n}}.
\]
Recall now that the transfer  $L \stackrel{c}{\longleftarrow} H $ does not change the ordering of incomes, and thus we must have $X_{L:n}+c < X_{H:n}-c$, which is equivalent to $c<c_0$, where \[
c_0={ X_{H:n}-X_{L:n} \over 2}.
\]
Hence, to have $L \stackrel{c}{\longleftarrow} H $ for some $c>0$, we must have $c_3<c_0$, which is equivalent to
\[
2\big( X_{n-L+1:n}X_{H:n}^2-X_{n-H+1:n}X_{L:n}^2\big)< \big(X_{H:n}-X_{L:n}\big)\big( X_{n-L+1:n}X_{H:n}+X_{n-H+1:n}X_{L:n}\big),
\]
which simplifies to
\[
 X_{n-L+1:n}X_{H:n}^2-X_{n-H+1:n}X_{L:n}^2
 < \big( X_{n-H+1:n}  - X_{n-L+1:n}\big) X_{L:n}X_{H:n}.
\]
The latter inequality is impossible because $L<H$ implies $X_{H:n}>X_{L:n}$ and $X_{n-H+1:n} < X_{n-L+1:n}$. Consequently, it is impossible to have $c_3<c_0$ and so there is not a single $c>0$ that satisfies  $c\ge c_3$ and $c<c_0$ simultaneously. This shows that the only possibility that exists is $\Psi_{3,n}(\mathbf{X})< \Psi_{3,n}(\mathbf{X}')$.

Consider now the case of two struggling persons, that is, $L<H< M $. In this case we have $L<H\le \lfloor n/2 \rfloor $ and so $L \stackrel{c}{\longleftarrow} H $  affects $\Psi_{3,n}$ because of the inequality
\begin{align*}
{X_{L:n} \over X_{n-L+1:n}} + {X_{H:n} \over X_{n-H+1:n}}
&>
{X_{L:n}+c \over X_{n-L+1:n}} + {X_{H:n}-c \over X_{n-H+1 :n}}
\\
&= {X'_{L:n} \over X'_{n-L+1:n}} + {X'_{H:n} \over X'_{n-H+1:n}}
\end{align*}
that holds due to $X_{n-L+1:n} > X_{n-H+1:n}$. Hence,
$\Psi_{3,n}(\mathbf{X})< \Psi_{3,n}(\mathbf{X}')$, concluding the proof of Property~\ref{prop57}.
\end{proof}

\end{document}